\documentclass[11pt]{article}
\pdfoutput=1
\usepackage[utf8]{inputenc}
\usepackage{dsfont}
\usepackage{diagbox}
\usepackage{amsfonts}
\usepackage{mathtools}
\allowdisplaybreaks[4]        
\usepackage{amssymb}
\usepackage{euscript}     
\usepackage{braket}
\usepackage{starfont}
\usepackage{color,soul}         
\usepackage{braket}
\usepackage{tensor}        
\usepackage{amsthm}
\usepackage{graphicx}
\usepackage{slashed}
\usepackage{leftidx}
\usepackage{subfigure}
\usepackage{bbm}
\usepackage{indentfirst}

\definecolor{outerspace}{rgb}{0.25, 0.29, 0.3}
\definecolor{scarlet}{rgb}{1.0, 0.13, 0.0}
\usepackage[header,title,page,titletoc]{appendix}  
\definecolor{princetonorange}{rgb}{1.0, 0.56, 0.0}
\definecolor{WildStrawberry}{rgb}{1.0, 0.26, 0.64}
\definecolor{rossocorsa}{rgb}{0.83, 0.0, 0.0}
\definecolor{navyblue}{rgb}{0.0, 0.0, 0.5}
\usepackage[numbers,sort&compress]{natbib}  
\usepackage{float}
\usepackage[paper=a4paper,margin=1in]{geometry}
\usepackage{titlesec}

\pdfoutput=1
\usepackage{amsmath}
\usepackage{color}
\usepackage{amsfonts}
\usepackage{amssymb}
\usepackage{graphicx}
\usepackage{geometry}
\usepackage{amssymb,epsfig,subfigure}
\usepackage{amssymb}
\usepackage{hyperref}
\usepackage{comment}
\usepackage[font=footnotesize]{caption}
\usepackage[T1]{fontenc}
\usepackage[utf8]{inputenc}
\usepackage{latexsym}

\usepackage{cancel}

\usepackage[numbers,sort&compress]{natbib}  
\usepackage{float}

\usepackage{dsfont}
\usepackage{diagbox}

\usepackage{mathtools}
\allowdisplaybreaks[4]

\usepackage{euscript}     
\usepackage{braket}
\usepackage{starfont}
\usepackage{color,soul}         
\usepackage{tensor}        
\usepackage{amsthm}

\usepackage{slashed}
\usepackage{leftidx}
\usepackage{bbm}

\makeatletter
\renewcommand\section{\@startsection {section}{1}{\z@}%
                                 {-3.5ex \@plus -1ex \@minus -.2ex}
                                   {2.3ex \@plus.2ex}%
                                   {\normalfont\large\bfseries}}
\renewcommand\subsection{\@startsection{subsection}{2}{\z@}%
                                   {-3.25ex\@plus -1ex \@minus -.2ex}%
                                     {1.5ex \@plus .2ex}%
                                     {\normalfont\bfseries}}
\renewcommand\subsubsection{\@startsection{subsubsection}{3}{\z@}%
                                   {-3.25ex\@plus -1ex \@minus -.2ex}%
                                     {1.5ex \@plus .2ex}%
                                     {\normalfont\itshape}}
\makeatother

\def\pplogo{\vbox{\kern-\headheight\kern -29pt
\halign{##&##\hfil\cr&{\ppnumber}\cr\rule{0pt}{2.5ex}&\ppdate\cr}}}
\makeatletter
\def\ps@firstpage{\ps@empty \def\@oddhead{\hss\pplogo}%
  \let\@evenhead\@oddhead 
}
\def\maketitle{\par
 \begingroup
 \def\thefootnote{\fnsymbol{footnote}}
 \def\@makefnmark{\hbox{$^{\@thefnmark}$\hss}}
 \if@twocolumn
 \twocolumn[\@maketitle]
 \else \newpage
 \global\@topnum\z@ \@maketitle \fi\thispagestyle{firstpage}\@thanks
 \endgroup
 \setcounter{footnote}{0}
 \let\maketitle\relax
 \let\@maketitle\relax
 \gdef\@thanks{}\gdef\@author{}\gdef\@title{}\let\thanks\relax}
\makeatother

\numberwithin{equation}{section}
\newcommand\nn{\nonumber}
\newcommand\eea{\end{eqnarray}}
\newcommand\bea{\begin{eqnarray}}

\def\beq{\begin{equation}}
\def\eeq{\end{equation}}

\newcommand{\be}{\begin{equation}}
\newcommand{\ee}{\end{equation}}
\newcommand{\ba}{\begin{align}}
\newcommand{\ea}{\end{align}}
\newcommand{\bg}{\begin{gather}}
\newcommand{\eg}{\end{gather}}
\newcommand{\bseq}{\begin{subequations}}
\newcommand{\eseq}{\end{subequations}}

\usepackage{hyperref}
\hypersetup{
    colorlinks,
    citecolor=rossocorsa,
    filecolor=navyblue,
    linkcolor=navyblue,
    urlcolor=navyblue
}

\begin{document} 

\begin{titlepage}

\begin{center}

\phantom{ }
\vspace{3cm}

{\bf \Large{Entropic order parameters for the phases of QFT}}
\vskip 0.5cm
Horacio Casini${}^{*}$, Marina Huerta${}^{\dagger}$, Javier M. Mag\'an${}^{\ddagger}$, Diego Pontello${}^{\mathsection}$
\vskip 0.05in
\textit{Instituto Balseiro, Centro At\'omico Bariloche}
\vskip -.4cm
\textit{ 8400-S.C. de Bariloche, R\'io Negro, Argentina}

\begin{abstract}
We propose entropic order parameters that capture the physics of generalized symmetries and phases in QFT's. We do it through an analysis of simple properties (additivity and Haag duality) of the net of operator algebras attached to space-time regions. We observe that different types of symmetries are associated with the breaking of these properties in regions of different non-trivial topologies. When such topologies are connected, we show that the non locally generated operators generate an Abelian symmetry group, and their commutation relations are fixed. The existence of order parameters with area law, like the Wilson loop for the confinement phase, or the 't Hooft loop for the dual Higgs phase, is shown to imply the existence of more than one possible choice of algebras for the same underlying theory. A natural entropic order parameter arises by this non-uniqueness. We display aspects of the phases of theories with generalized symmetries in terms of these entropic order parameters. In particular, the connection between constant and area laws for dual order and disorder parameters is transparent in this approach, new constraints arising from conformal symmetry are revealed, and the algebraic origin of the Dirac quantization condition (and generalizations thereof) is described. A novel tool in this approach is the entropic certainty relation satisfied by dual relative entropies associated with complementary regions, which quantitatively relates the statistics of order and disorder parameters.

\end{abstract}
\end{center}

\small{\vspace{4.5 cm}\noindent${}^{\text{\text{*}}}$casini@cab.cnea.gov.ar\\
${}^{\ddagger}$marina.huerta@cab.cnea.gov.ar\\
${}^{\dagger}$javier.magan@cab.cnea.gov.ar\\
${}^{\mathsection}$diego.pontello@ib.edu.ar
}

\end{titlepage}

\setcounter{tocdepth}{2}

{\parskip = .4\baselineskip \tableofcontents}
\newpage

\section{Introduction}

Transcending the weak coupling regime has been a recurring theme in the context of QFT in the past decades. Many pressing reasons motivate this interest. We have the everlasting confinement problem in gauge theories \cite{Jaffe:2000ne,Greensite:2011zz}, examples of non-Fermi liquid behavior at low temperatures in condensed matter theory \cite{ong2001more}, electromagnetic dualities in QFT \cite{AlvarezGaume:1997ix}, and the holographic duality \cite{Aharony_2000}.

In the quest of understanding strong coupling phenomena, it is natural to seek sufficiently robust features that remain valid at any value of the coupling. This includes looking for alternative descriptions, or new structures, which may be studied in a controllable manner. The present article is framed within the Haag-Kastler algebraic approach to QFT \cite{Haag:1963dh,haag2012local}. This approach has been fruitful for progress at the conceptual level. As described below, it can be considered a minimalistic approach, that only assumes very general and basic properties about the way operator algebras are assigned to space-time regions. Moreover, it is the natural approach for the description of entanglement entropy and other statistical measures of states \cite{Hollands:2017dov}.

Structures that transcend the perturbative regime are generally connected to symmetries, whether space-time or internal ones. Examples are conformal symmetry, supersymmetry, global and local symmetries, and the recently introduced generalized global symmetries \cite{Gaiotto:2014kfa}.  However, most of the time the way these symmetries are considered is linked to the Lagrangian QFT definition, relying on a weak coupling regime. 

There are two notable exceptions. For the case of global symmetries, a first principle algebraic approach was carried out by Haag, Doplicher, and Roberts \cite{Doplicher:1969tk,Doplicher:1969kp,Doplicher:1971wk,Doplicher:1973at}. They studied the imprint of the symmetry already in the neutral (observable) sector of the theory. They found that the superselection sectors arising by including charged operators in the model were seen to be in correspondence with certain endomorphisms of the observable algebra. Having identified the imprint, one can try to reverse the logic. Given a structure of endomorphisms with certain defining properties, called in the literature DHR superselection sectors, one seeks to derive the symmetry group itself. This problem was completed leading to the reconstruction theorems \cite{Doplicher:1990pn}. For the case of conformal symmetries, a first principle approach started with the works of Polyakov, Ferrara, Grillo, and Gatto \cite{Polyakov:1974gs,Ferrara:1973yt}, known as the conformal bootstrap, and which is being used with great success at present \cite{Poland:2018epd}.

One would like to extend the algebraic approach to other kinds of symmetries, such as local ones. This extension turns out to be more complicated. The reason is that for local symmetries, the associated charged operators cannot be localized in a ball. One can measure their charge at arbitrarily long distances employing local operators only. An example is an electric charge which can be measured by the electric flux at infinity. Some modifications of the DHR formalism were proposed in this regard. One considers sectors which, instead of being localizable in balls, are localizable in cones that extend out to infinity \cite{Buchholz:1981fj,frohlich2006quantum,horuzhy2012introduction}. This approach departs from the local QFT philosophy it started with, due to the infinite cones. It would be better to understand all kinds of symmetries already in a bounded region of Minkowski space-time and keep aligned with the local QFT attitude. We would also like to include symmetries associated with higher dimensional cones. Presumably, these would be related to the generalized symmetries introduced more recently in \cite{Gaiotto:2014kfa}. But from the algebraic perspective, the higher dimensional cones would represent superselection charges with infinite energy in an infinite space, and they have been discarded in that regard.

In this article, we propose a unified approach to symmetries in QFT which is fundamentally local. We do not want to resort to a Lagrangian or any local current, and we want to be able to frame the description talking about the vacuum state on subregions of flat topologically trivial Minkowski spacetime. To connect with more conventional approaches, we seek to define order parameters that signal the presence and breaking of the symmetries, allowing a broad characterization of phases in QFT's. As will be clarified through the text, order parameters in this context are naturally defined using information theory, and we call them entropic order parameters. These can be related to operator order parameters, though not to the standard singular line operators that are usually considered. In particular, these operators cannot be renormalized arbitrarily.

Quite surprisingly, such a path to symmetries in QFT has a simple and geometrical starting point, based on causality.  In QFT, causality is enforced by the requirement of commutativity of operators at spatial distances. This is summarized by
\be \label{dualintro}
{\cal A}(R)\subseteq {\cal A}(R')'\,,
\ee
where ${\cal A}(R)$ is the algebra of operators localized in a certain region $R$, $R'$ is the set of points causally disconnected from $R$, and ${\cal A}'$ is the commutant of the algebra ${\cal A}$, that is, the set of all operators that commute with all the operators in ${\cal A}$. Naively, one could be inclined to believe that this relation might be saturated in a QFT, i.e, we would have the equality in \eqref{dualintro}. Such saturation is called Haag duality, or duality for short.\footnote{Haag duality should not be confused with the dualities relating different descriptions of the same QFT or linking different QFT's.} It turns out that something more interesting can happen. The previous inclusion does not need to be saturated. Indeed, as we will describe, it is precisely in the difference between both algebras where generalized symmetries appear. This difference consists of operators that cannot be locally generated in $R$ but are still commuting with operators in $R'$. Therefore, if we include them in the algebra of $R$ to restore duality, we introduce a violation of additivity, the property stating that operators in a region are generated as products of local operators inside the region. The tension between duality and additivity in these theories cannot be resolved. 

The observation that global symmetries entail violations of Haag duality was known a long time ago, see for example \cite{haag2012local}. The reason is that one can form observables out of the product of local charged operators. If one chooses a region $R$ which is disconnected, so that it has non-trivial homotopy group $\pi_0$, then Haag duality will not hold due to the existence of charge-anti-charge operators localized at different disconnected patches. These non-local operators are called intertwiners. This type of breaking of Haag duality was studied in full detail for two-dimensional conformal field theories in \cite{brunetti1993}, where the structure of the algebra was unraveled and shown to be controlled by the structure of superselection sectors. In higher dimensions, the analysis was complemented in \cite{Casini:2019kex}, by describing the breaking of duality in the region $R'$ complementary to $R$. This region has a non-trivial $\pi_{d-2}$ homotopy group, and the violation of duality is due to the existence of twist operators, that implement the symmetry locally. This will be described in more detail below.

While the relation between duality violation for topological non-trivial regions and global symmetries was appreciated, the starting point for the algebraic derivation of global symmetries was the DHR endomorphisms \cite{Doplicher:1969tk,Doplicher:1969kp,Doplicher:1971wk,Doplicher:1973at}. In this paper, we take the breaking of duality as the fundamental physical feature, from which the symmetries could be derived. This seemingly mild change of perspective eases the way to generalizations. 
We will be able to discuss symmetries by focusing on the ``kinematical'' properties of algebras and regions in the vacuum. For this purpose, we avoid studying superselection sectors, which may have a dynamical input, or may require infinite cones for their description.  We observe that different types of symmetries are related to the breaking of duality for regions of different topologies. While global symmetries entail the breaking of duality for regions with non-trivial $\pi_0$ or $\pi_{d-2}$, we observe that generalized symmetries arising from gauge symmetries appear for QFT's in which duality is broken for regions with non-trivial $\pi_1$ or $\pi_{d-3}$. Going up in the ladder, in QFT's with higher form-generalized symmetries, duality is broken for regions with non-trivial $\pi_i$ or $\pi_{d-2-i}$. We argue that for any $i\geq 1$, the symmetries are bound to form an Abelian symmetry group. Finally, we show that the breaking of the duality of the complementary regions $\pi_i$ and $\pi_{d-2-i}$ is due to the existence of non-local operators with specific commutation relations between themselves. Physically, these dual non-local operators correspond to order and disorder parameters, and their behavior characterizes the phases of the theory.

As a by-product of this analysis, it follows that the Dirac quantization condition nicely fits into the algebraic framework. It turns out to be simply originated when enforcing causality of the net of algebras. Although this might sound trivial, the causality of the net becomes threatened in situations where the inclusion \eqref{dualintro} is not saturated. Enforcing duality and causality directly provides the generalized quantization condition.

Having identified the connection between the failure of duality and generalized symmetries in QFT's, in the second part of the article we proceed to construct order parameters that sense their presence and their breaking. We start by showing that the non-local order-disorder operators that violate additivity are the only ones that can display area laws, typical of confinement of electric or magnetic charges in gauge theories. Equivalently, the breaking of duality in the appropriate region is seen as a necessity for the existence of order parameters with area law behavior, like the Wilson loop of the fundamental representation in pure gauge theories.

The choice of operator order parameters is not unique. Indeed there is an infinite number of possibilities. This is somewhat in contrast with the previous inclusion of algebras \eqref{dualintro}, which is robust and completely unambiguous. Natural order parameters should arise from such inclusions. To accomplish this, it seems more natural to us to resort to information theory. In fact, given an inclusion of the previous type, entropic order parameters can be defined as the relative entropy between the vacuum and a state in which we have sent to zero all expectation values of non-local operators. This relative entropy is a well-defined notion of uncertainty for the algebra of non-local operators, and it will play a central role in the article. The entropic approach to global symmetries recently developed in \cite{Casini:2019kex}, which in turn was inspired by the work \cite{Longo:2017mbg} concerning free fermions in two dimensions, is here generalized to regions of different topology.

On one hand, the choice of relative entropy is convenient because it is robust and standard. But more importantly, it allows us to quantitatively relate the physics of order and disorder parameters. This is due to a general property of relative entropies called certainty principle \cite{Magan:2020ake,Casini:2019kex}. In the present light, it relates the entropic order parameter with the entropic disorder parameter, for complementary geometries.  
In other words, quoting a specific example, the statistics of Wilson loops and t' Hooft loops in complementary regions, are precisely related to each other by the certainty relation.

We will compute the entropic order and disorder parameters for symmetries and phases in QFT's in several cases of interest. In some instances, we can check compatibility with the certainty principle, or use this relation to understand their behavior. We will start with QFT's with global symmetries, and consider scenarios with conformal symmetry and with spontaneous symmetry breaking. Both phases will be seen to be distinguished already at a qualitative level by the order parameters, as they should. Similarities with the phase structure of gauge theories that arise from the present approach will be highlighted. Interestingly, for scenarios with spontaneous symmetry breaking, the computations are related to the solitons/instantons of the theory, as could have been anticipated. We then move to the case of gauge theories. We will first analyze the case of the Maxwell field, which can be done in great detail, and where the match between the order and disorder approaches will be confirmed with surprisingly good accuracy. We then analyze several interesting constraints that appear in gauge theories with conformal symmetry in four dimensions. In this scenario, a specific relative entropy becomes enough constrained to be determined analytically. We finally move to the Higgs phase, which as explained by t' Hooft in \cite{tHooft:1977nqb}, is dual to the confinement scenario, and where semiclassical physics may be used to study the entropic order parameters.  

A final remark is in order. One of the initial motivations for this work was to understand issues about entanglement entropy in gauge theories. Several specific regularizations of entropy were proposed in the literature, which pointed to some UV ambiguities of entropy in gauge theories \cite{Donnelly:2014fua,Ghosh:2015iwa,Soni:2015yga,Soni:2016ogt,Huang:2014pfa}. As explained in \cite{Casini:2013rba,Casini:2019kex,Casini:2019nmu}, such ambiguities do not survive the continuum limit.   In this paper, we find that for specific QFT's (the ones with generalized symmetries), there is more than one possible algebra for a region of specific topology. These multiple choices are macroscopic and physical, and they pertain to the continuum model itself. They have no relation with regularization ambiguities, nor with the description in terms of gauge fields. Corresponding to the multiplicity of algebras there are multiple entropies for the same region. These entropies measure different quantities and therefore should not be understood as ambiguities. The relative entropy order parameters introduced in this paper are precisely well-defined notions of the differences between these entropies.

\section{Algebras, regions, and symmetries: additivity versus duality.}
\label{algebra-regions}

In the algebraic approach, a QFT is described by a net of von Neumann algebras. This is an assignation of an operator algebra to any open region of space-time. The particular QFT model is determined by how the algebras in the net relate to each other and with the state.

We will restrict to consider only causal regions, which will be typically denoted by $R$ below. Causal regions are the domain of dependence of subsets of a Cauchy surface.  In this paper, we will be interested in the properties of algebras assigned to causal regions based on the same (arbitrary) Cauchy surface ${\cal C}$. These regions will have in general non-trivial topologies whose properties are the same as the ones of subregions of ${\cal C}$ (typically the surface $t=0$)  in which they are based. Hence, we will often make no distinction between a $d-1$ dimensional subset of ${\cal C}$ and its causal $d$-dimensional completion. In this sense, our description of the structural properties of the net of algebras focuses on quite kinematical aspects. This description may also apply to non-relativistic theories, lattice theories, or finite volume models

The algebras ${\cal A}(R)$ attached to regions $R$ satisfy the basic relations of {\sl isotony} 
\be
{\cal A}(R_1)\subseteq  {\cal A}(R_2)\,,\hspace{1cm}R_1\subseteq R_2\,, \label{isotonia}
\ee
and {\sl causality}
\be
{\cal A}(R) \subseteq  ({\cal A}(R'))'\,,\label{causality}
\ee
where $R'$ is the causal complement of $R$, i.e. the space-time set of points spatially separated from $R$, and  ${\cal A}'$ is the algebra of all operators that commute with those of ${\cal A}$. For any von Neumann algebra we always have ${\cal A}''={\cal A}$. A (causal) {\sl net} is an assignation ${\cal A}(R)$ of algebras to regions satisfying \eqref{isotonia} and \eqref{causality}.

Extensions of these relations are expected to hold for sufficiently complete models but are not granted on general grounds. For example, \eqref{causality} could be extended to the relation of {\sl duality} (also called Haag's duality)
\be
 {\cal A}(R)= ({\cal A}(R'))'\,,   \label{duality}
\ee
and we could also expect a form of {\sl additivity}
\be
{\cal A}(R_1 \vee R_2)={\cal A}(R_1 )\vee{\cal A}( R_2)\,, \label{additivity}
\ee 
where $R_1 \vee R_2=(R_1\cup R_2)''$,  ${\cal A}_1 \vee{\cal A}_2=({\cal A}_1 \cup {\cal A}_2)''$ are the smallest causal regions and von Neumann algebras containing   $R_1, R_2$, and ${\cal A}_1, {\cal A}_2$ respectively. The relation \eqref{additivity} means the algebra of the bigger region is generated by the operators in the smaller ones. We will call a net {\sl complete} if it satisfies \eqref{duality} and \eqref{additivity} for all $R$ based on the same Cauchy surface. The main focus of the paper concerns nets that are not complete in this sense, and how this incompleteness is related to generalized symmetries in the QFT.

The de Morgan laws
\bea
({\cal A}_1\vee {\cal A}_2)'={\cal A}_1'\cap {\cal A}_2' \;,\nonumber \\
(R_1\vee R_2)'=R_1' \cap R_2' \,,
\eea
are universally valid for regions and algebras. 
From these relations it follows that if we have unrestricted validity of duality \eqref{duality} and additivity \eqref{additivity}, we have the {\sl intersection property}
\be
{\cal A}(R_1 \cap R_2)={\cal A}(R_1 )\cap {\cal A}( R_2)\,. \label{intersection} 
\ee
Conversely, additivity follows from unrestricted validity of duality and the intersection property. Therefore the intersection property is another aspect of duality and additivity.\footnote{Interestingly, the algebras (assumed to be factors) and causal regions both have the structure of orthocomplemented lattices in the order theoretical meaning, and the relations \eqref{duality}, \eqref{additivity}, and \eqref{intersection} for a complete theory can be interpreted as a homomorphism of lattices. See the discussion in \cite{Haag:1963dh}, section III.4.}

We are interested in studying algebra-region problems determined by the topology of the regions. Our starting point is a net ${\cal A}(R)$ where we assume additivity holds for topologically trivial regions whose union is also topologically trivial, i.e., ${\cal A}(R_1\vee R_2)={\cal A}(R_1)\vee {\cal A}(R_1)$, where $R_1,R_2,$ and $R_1\vee R_2$, all have the topology of a ball. This statement means that the algebra of $R$ is generated by the algebras of any collection of balls (of any size) included in $R$ and whose union is all $R$. This accounts for the idea that the operator content of the theory is formed by local degrees of freedom. This can be summarized by saying that any localized operator of the theory is locally generated.\footnote{A well-known counterexample is a conformal generalized free field with a two-point function $|x-y|^{-2 \Delta}$. This field appears in the large $N$ approximation of holographic theories, and it is equivalently described in terms of a free massive field in AdS. It is not difficult to see through this relation that algebras of many small overlapping balls will not generate the algebra of the causal union of the balls. See \cite{Duetsch:2002hc}.} However, a different question is whether any operator of a certain algebra  ${\cal A}(R)$ is locally generated inside $R$ itself when the region is topologically non-trivial. Below we will see several examples demonstrating that the existence of non locally generated operators in such ${\cal A}(R)$ is not an uncommon phenomenon.

Let us be more precise. Given a net, we can always construct an additive algebra for a region $R$ as
\be
{\cal A}_{\textrm{add}}(R)= \bigvee_{B \,\textrm{is a ball}, \,B\subseteq R} {\cal A}(B)\,. 
\ee
This provides to us a minimal algebra, in the sense that it contains all operators which must form part of the algebra because they are locally formed in $R$. The assignation of ${\cal A}_{\textrm{add}}(R)$ to any $R$  gives the minimal possible net 
and if ${\cal A}_{\textrm{add}}(R)\subsetneq {\cal A}(R)$ it follows that there are more than one net. 

In this freedom of choosing the operator content of different regions, the greatest possible algebra of operators that can be assigned to $R$ and still satisfies causality must correspond to a minimal one assigned to $R'$,
\be
{\cal A}_{\textrm{max}}(R)= ({\cal A}_{\textrm{add}}(R'))'\,.
\ee
We anticipate that, in general, the assignation ${\cal A}_{\textrm{max}}(R)$ does not form a (local) net since ${\cal A}_{\textrm{max}}(R)$ and ${\cal A}_{\textrm{max}}(R')$ may not commute. 
Also, it is evident that if ${\cal A}_{\textrm{add}}(R)\subsetneq {\cal A}_{\textrm{max}}(R)$, it follows that the additive net does not satisfy duality. In this situation one can enlarge the additive net by adding non locally generated operators, to generate a net satisfying duality \eqref{duality}. In general, this may be done in multiple ways. We will call such nets Haag-Dirac (HD) nets for reasons that will become apparent later on. By construction, Haag-Dirac nets satisfy duality
\be
{\cal A}_{\textrm{HD}}(R)=({\cal A}_{\textrm{HD}}(R'))'\,,
\ee
but in general, they will not satisfy additivity. Therefore, there is a tension between duality and additivity which cannot be resolved in these incomplete theories. Notice that for a global pure state the entropy of an algebra ${\cal A}$ is equal to the one of its algebraic complement ${\cal A}'$. The present discussion shows this does not translate to an equality of entropies for complementary regions, except for a HD net.

To be more concrete, let us call $a \in {\cal A}_{\textrm{max}}(R)$ to a  collection of non locally generated operators in $R$ such that 
\be
{\cal A}_{\textrm{max}}(R)=({\cal A}_{\textrm{add}}(R'))' = {\cal A}_{\textrm{add}}(R)\vee \{a\}\,.
\ee
In the same way we have operators $b \in {\cal A}_{\textrm{max}}(R')$ non locally generated in $R'$ such that
\be
{\cal A}_{\textrm{max}}(R')=({\cal A}_{\textrm{add}}(R))' = {\cal A}_{\textrm{add}}(R')\vee \{b\}\,.
\ee
Evidently, the {\sl dual} sets of operators $\{a\}$ and $\{b\}$ cannot commute with each other. 
Otherwise it would be  ${\cal A}_{\textrm{max}}(R) \subseteq ({\cal A}_{\textrm{max}}(R'))'= {\cal A}_{\textrm{add}}(R)$ and these operators would be locally generated. Given the existence of non locally generated operators $a$ in $R$, 
the necessity of the existence of dual complementary sets of non locally generated operators $b$ in $R'$  is  due to the fact that for two different algebra choices ${\cal A}_{1,2}$ for $R$ there are two different  choices ${\cal A}_{1,2}'$ associated with $R'$. The later cannot coincide because of the von Newman relation ${\cal A}''={\cal A}$.    
 
Since the dual non locally generated operators $\{a\}$ and $\{b\}$ do not commute, when constructing Haag-Dirac nets ${\cal A}_{\textrm{HD}}(R)$ satisfying duality, we have to sacrifice some operators of ${\cal A}_{\textrm{max}}(R)$ or ${\cal A}_{\textrm{max}}(R')$, to keep the net causal. The assignation ${\cal A}_{\textrm{max}}(R)$ for all $R$ does not form a net. 
A possible choice is  
${\cal A}_{\textrm{max}}(R)$ for $R$ and ${\cal A}_{\textrm{add}}(R')$ for $R'$ or vice-versa, and usually there are  some intermediate choices. In particular, if the topologies of $R$ and $R'$ are the same, both of these choices are not very natural and may break some spatial symmetries.\footnote{When referring to the topology of an infinite region, i.e. the complement of a bounded one, we will think that the full space $\mathbb{R}^{d-1}$ is compactified to a sphere $S^{d-1}$.}

An important remark is the following. Even if some non locally generated operator is excluded from the algebra of $R$, this does not mean it does not exist in the theory. All non locally generated operators that could be assigned to $R$ are always formed locally in a ball containing $R$ and thus its existence cannot be avoided. They will always belong to the algebra of this ball. In particular, the full operator content (which is generated by all the operators in all balls) of different nets is the same. The particularity of the theories that we are interested to describe in this paper is that they admit more than one possible (local) net constructed out of the same set of operators. For simplicity, throughout this article, we will often call the operators such as $a$ and $b$ as ``non-local operators'', meaning they are not locally generated (or not additively generated) in a specific region.

Notice also that the fact that the sets $\{a\}$ and $\{b\}$ form complementary sets of observables based on complementary regions does not imply a violation of causality. The reason is that to construct $\{a\}$ in a laboratory from microscopic operators we need to have access to a ball including $R$ which non trivially intersects $R'$.       
 
The operators $a$ may be chosen to form irreducible classes $[a]$ in ${\cal A}(R)$ under multiplication by locally generated operators.\footnote{The subset $[a]$ of ${\cal A}(R)$ is  the set generated as
$\sum_\lambda O_1^\lambda \,a\, O_2^\lambda$, with $O_1^\lambda$ and $O_2^\lambda$ locally generated operators from ${\cal A}_{\textrm{add}}(R)$. It is irreducible if there are no non trivial subspaces of $[a]$ invariant under the left and right action of the additive algebra. The class $[1]$ coincides with ${\cal A}_{\textrm{add}}(R)$ and $[1][a]=[a][1]=[a]$. We assume both ${\cal A}_{\textrm{add}}(R)$ and ${\cal A}_{\textrm{add}}(R')$ have no center (are factors), see \cite{bischoff2015tensor}. This is an expected property in the continuum QFT. A center would produce an irreducible sector unrelated to non locality.} With the class $[a]$ there is also the adjoint class $[\bar{a}]$. Because ${\cal A}(R)$ is an algebra, these classes must close a fusion algebra between themselves $[a][a']=\sum_{a''} [n]^{a''}_{a a'} [a'']$, with $[n]^{a''}_{a a'}=0,1$. These fusion rules simply indicate which classes appear in the decomposition.\footnote{Provided we can choose spatially separated representatives $a,a'$ in the same region $R$ (as in all examples in this paper) this algebra is commutative, namely $[n]^{a''}_{a a'}=[n]^{a''}_{a' a}$.} The same will happen with the complementary operators $[b][b']=\sum_{b''} [\tilde{n}]^{b''}_{b b'} [b'']$.  We will describe several specific examples below.

In the applications of this paper, these dual fusion rules are associated with group representations and their conjugacy classes. This brings in the idea of symmetries. In the specific models we analyze, duality is seen to fail when the algebras are constructed as the invariant operators under certain symmetries. Examples are orbifolds of a global symmetry and gauge-invariant operators for some gauge theories. We will see that the particular topology of $R$ where duality or additivity fails depends on the type of symmetry involved. Orbifolds show algebra-region ``problems'' when one of the homotopy groups $\pi_0(R)$ or $\pi_{d-2}(R)$ is non-trivial.\footnote{ Spontaneously broken global symmetries allow the construction of a net where duality fails for balls. We will describe this situation and its corresponding order parameter in section \ref{gs}.} The case of ordinary gauge symmetries might give problems for regions with non-trivial $\pi_1(R)$ or $\pi_{d-3}(R)$. Higher homotopy groups correspond to the case of gauge symmetries for higher forms gauge fields.  In these examples, the gauge symmetry plays an auxiliary role in the construction of the models, but it does not play a direct role in the final theory. However, the algebra of the non locally generated operators does play a fundamental role. It can be interpreted as a generalized symmetry in the sense of \cite{Gaiotto:2014kfa}.

\subsection{ Regions with non-trivial $\pi_0$ or $\pi_{d-2}$. Global symmetries.}
\label{global}

We consider the subalgebra ${\cal O}$ of a theory ${\cal F}$, consisting of operators invariant under a global symmetry group $G$ acting on ${\cal F}$. The theory ${\cal O}={\cal F}/G$ is called an orbifold. These models were treated in more detail in  \cite{Casini:2019kex}.  In this case, we take regions $R$ with non-trivial $\pi_0(R)$, that is, disconnected regions. The complement $R'$ will have non trivial $\pi_{d-2}(R')$.  The simplest example is two disjoint balls $R_1$ and $R_2$, and its complement $S=(R_1\cup R_2)'$, which is topologically a ``shell'' with the topology of $S^{d-2}\times \mathbb{R}$. In this section, we will focus on the case of an unbroken symmetry, where the Hilbert space generated out of the vacuum by invariant operators consists of invariant states. The discussion, in this case, can be done without appealing to the theory ${\cal F}$. We will deal with the modifications produced by a non-invariant vacuum state in section \ref{gs}.

\begin{figure}[t]
\begin{center}  
\includegraphics[width=0.55\textwidth]{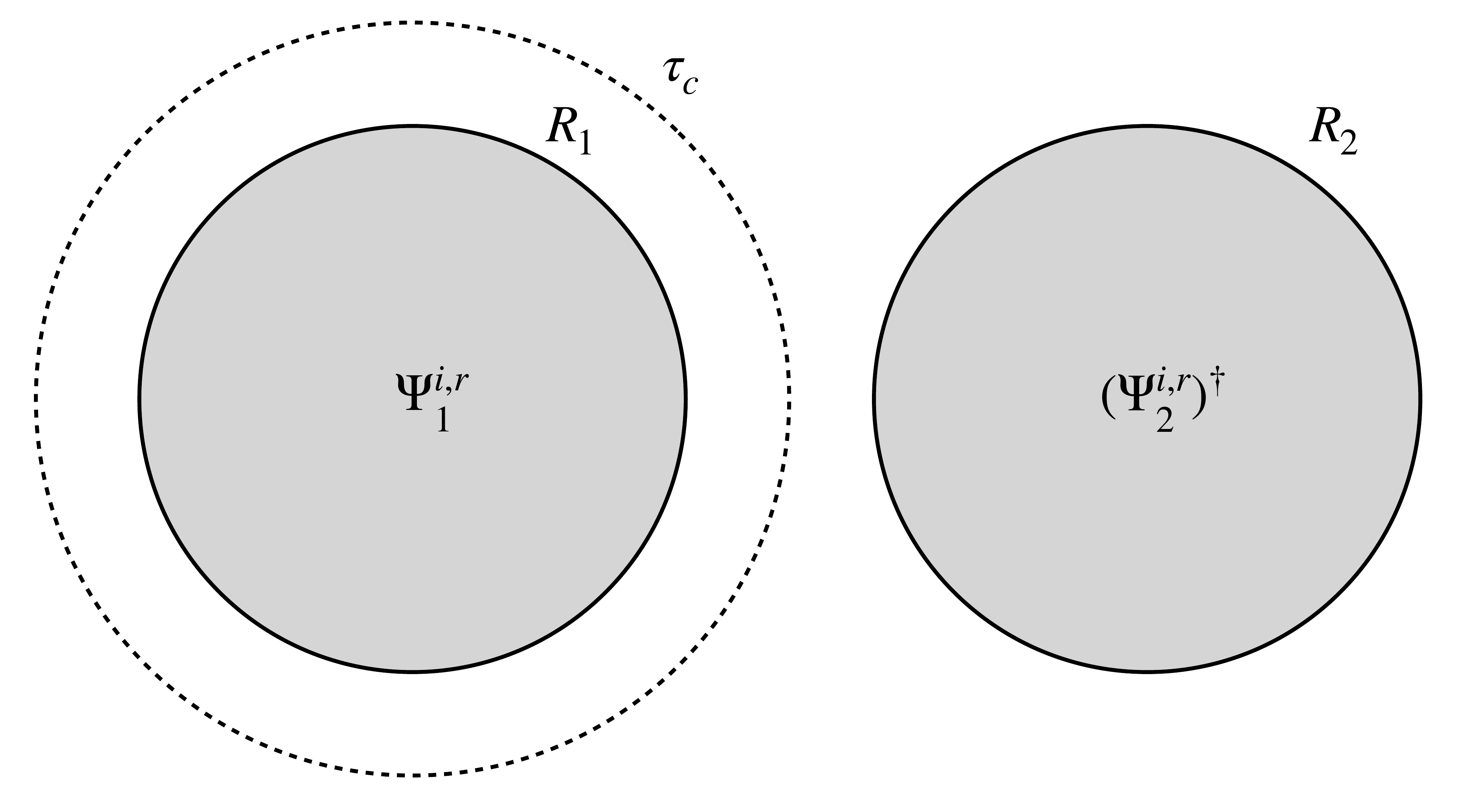}
\captionsetup{width=0.9\textwidth}
\caption{A region formed by two disjoint balls $R_1$ and $R_2$ (grey region) containing the intertwiner formed by a charge-anti-charge operator. In the complement $S$ of the two balls, which has a non-contractible $d-2$ dimensional surface, lives the twist operator.}
\label{int-twist}
\end{center}  
\end{figure}   

Let $\psi^{i,r}_1$, $\psi^{i,r}_2$ be charge creating operators in $R_1$ and $R_2$  in the theory ${\cal F}$, corresponding to the irreducible representation $r$, and where $i$ is an index of the representation.  The {\sl intertwiner} corresponding to this representation
\be\label{Ir}
{\cal I}_r= \sum_i \psi^{i,r}_1 (\psi^{i,r}_2)^\dagger\;,
\ee  
is invariant under global group transformations and belongs to the neutral theory ${\cal O}$, see figure \ref{int-twist}. It commutes with operators in ${\cal O}_{\textrm{add}}(S)$, but it cannot be generated additively by operators in the neutral algebras ${\cal O}(R_1)$ and ${\cal O}(R_2)$ since the charged operators $\psi^{i,r}$ belong to the field algebra ${\cal F}$ but not to ${\cal O}$.  

In a dual way, there are {\sl twist} operators $\tau_g$ implementing the group operations in $R_1$ and acting trivially in $R_2$. These commute with  ${\cal O}(R_1)$ and ${\cal O}(R_2)$, that is, uncharged operators in $R_1$ or $R_2$, but they do not commute with the intertwiners, which have charged operators in $R_1$. The twists can be chosen to satisfy\footnote{See \cite{Doplicher:1984zz,Doplicher:1983if,Bueno:2020vnx} for the construction of twist operators using the split property.}
\be\
\tau_g \tau_h=\tau_{gh}\,,\hspace{1cm}U(g) \tau_h U(g)^{-1}=\tau_{g h g^{-1}}\,, \label{tt}
\ee
where $U(g)$ is the unitary global symmetry operation. For a non-Abelian group, the twists are not invariant. The combinations of twist operators invariant under the global group
\be
\tau_{c}=\sum_{h\in c} \tau_{h} \,,\label{215}
\ee
are labeled by group conjugacy classes $c \subset G$, such that $g c g^{-1}=c$ for all $g\in G$. These operators belong to the neutral algebra ${\cal O}$.  While the full model ${\cal F}$, which includes the charge creating operators, satisfies duality and additivity, this is not the case for the neutral model ${\cal O}$. In fact, we have\footnote{Here and throughout this article, $R_1 R_2$ denotes de union $R_1 \cup R_2$.}
\bea
({\cal O}_{\textrm{add}}(R_1 R_2))' &=& {\cal O}_{\textrm{add}}(S)\vee \{\tau_{c}\} \;,\nonumber \\  
({\cal O}_{\textrm{add}}(S))' &=& {\cal O}_{\textrm{add}}(R_1 R_2) \vee \{{\cal I}_r\}\,.\label{copi2}
\eea
This shows explicitly that, retaining additivity, duality fails for the two-component region $R_1R_2$ and its complement $S$. The reason is the existence of operators (twists and intertwiners) in the model in these regions which cannot be additively generated inside the same regions by operators localized in small balls. However, the intertwiners and twists can be generated additively inside the model ${\cal O}$ in sufficiently big regions with trivial topology.

For finite groups, the number of independent twists coincides with the number of intertwiners. This is because the number $n_C$ of conjugacy classes of the group is equal to the number of irreducible representations. For Lie groups, there is an infinite number of irreducible representations, and the same occurs for conjugacy classes. In this case, as described in more detail below when discussing gauge theories, it is the duality between ``electric'' and ``magnetic'' weights the one ensuring that both sets of operators run over dual lattices.

As shown in appendix \ref{app2}, we can choose the intertwiners to satisfy a closed algebra. More concretely we get the fusion algebra
\be
 {\cal I}_{r_1} {\cal I}_{r_2}=\sum_{r_3} n_{r_1 r_2}^{r_3} {\cal I}_{r_3}\,,\hspace{1cm}{\cal I}_{\bar r}=({\cal I}_r)^\dagger\,,\hspace{1cm} {\cal I}_1=\mathbf{1}\,,\label{fusion}
\ee
where $\bar{r}$ is the representation conjugate to $r$, and  $n^{r_3}_{r_1 r_2}$ are the fusion matrices of the group representations
\be
[r_1]\otimes [r_1]=\oplus_{r_3} \,n^{r_3}_{r_1\, r_2} \, [r_3]\,,
\ee
providing the number of irreducible representations of type $r_3$ appearing in the decomposition of the tensor product of $r_1$ and $r_2$. Because $n_{r_1 r_2}^{r_3}=n_{r_2 r_1}^{r_3}$ the algebra \eqref{fusion} is Abelian. The same can be said of the twist algebra. From \eqref{tt} we get
\be\label{fusc}
\tau_{c_1} \tau_{c_2}= \sum_{c_3} m^{c_3}_{c_1 c_2}   \tau_{c_3}\,, 
\ee
with $ m^{c_3}_{c_1 c_2}$ the fusion coefficients of the conjugacy classes. 

The two Abelian algebras of twists and intertwiners do not commute with each other. For finite groups, they can be embedded in the non-Abelian matrix algebra of $|G|\times |G|$ matrices (see appendix \ref{app2}). A similar embedding works for Lie groups, but the embedding algebra needs to be infinite-dimensional. For Abelian symmetry groups, the commutation relations take a very simple form
\be\label{ccrel}
\tau_g \, {\cal I}_r = \chi_r(g) \, {\cal I}_r \, \tau_g\,,  
\ee 
where $\chi_r(g)$ is the group character. 

The DHR theory of ball localized superselection sectors gives examples of the failure of additivity-duality for regions with non-trivial $\pi_0, \pi_{d-2}$ for any dimension. The theory shows that under quite general conditions for these types of sectors and provided $d\ge 3$, the fusion algebras arise from a group, as described above \cite{Doplicher:1969tk,Doplicher:1969kp,Doplicher:1973at,Doplicher:1990pn}. More general fusion rules may appear in $d=2$ \cite{haag2012local,frohlich2006quantum,bischoff2015tensor}. As shown by the reconstruction theorem in such papers, starting with the model ${\cal O}$ with this type of duality failure a new theory ${\cal F}$ exists where charged operators cure these duality and additivity problems. The symmetry group is globally represented in ${\cal F}$ acting on the charged fields. It is important to remark that this reconstruction does not "modify" the subtheory ${\cal O}$ since the correlation functions of invariants operators do not change after the charged operators are included. This does not seem to have a transparent analog in gauge theories.

\subsection{Regions with non-trivial $\pi_1$ or $\pi_{d-3}$. Gauge theories. }

\begin{figure}[t]
\begin{center}  
\includegraphics[width=0.65\textwidth]{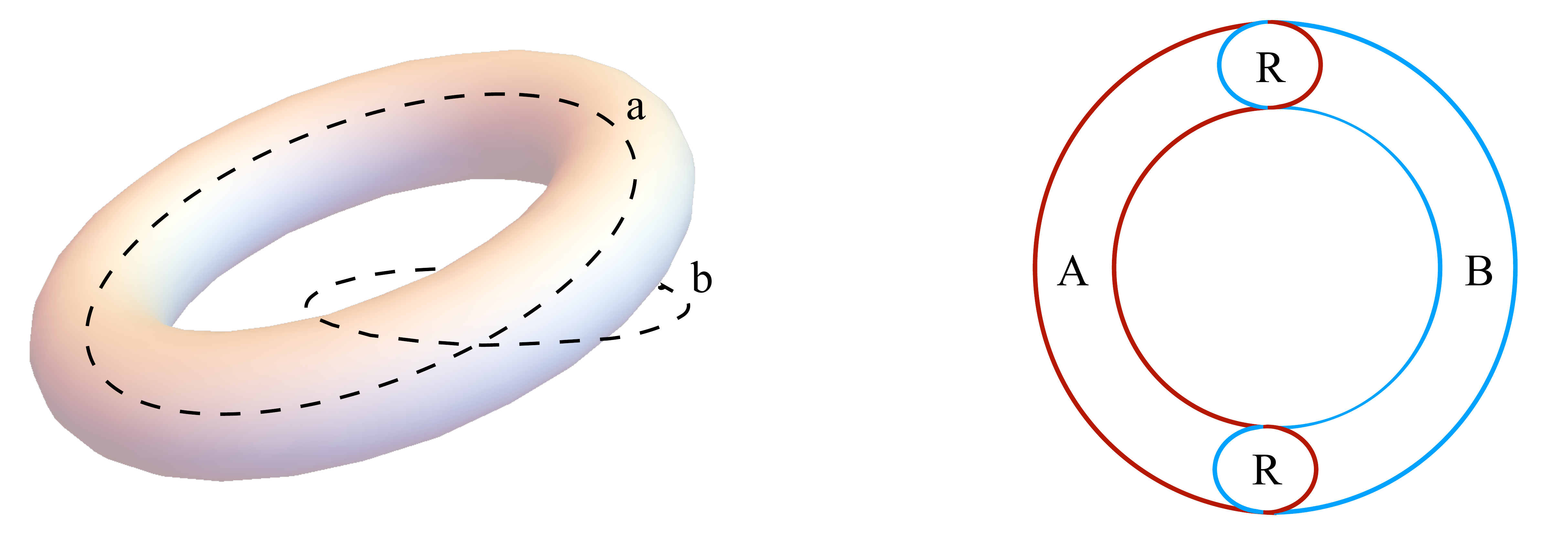}
\captionsetup{width=0.9\textwidth}
\caption{Left: Duality and additivity cannot be valid simultaneously for a ring-like region (solid torus) $R$. The operator $a$ is not additive in $R$. The interlocked operator $b$ is again not additive in the complement $R'$. $a$ and $b$ do not commute with each other. For commuting algebras attached to $R$ and $R'$, either $a$ or $b$ have to belong to the respective algebra (but not both of them at the same time), and additivity is lost. Right: Violation of the intersection property. The figure shows a section of two spherical cap regions $A$ and $B$  intersecting in the ring $R$ (here $d= 4$). A non-additive operator in $R$ is additive in both the topologically trivial regions $A$ and $B$. It then necessarily belongs to the intersection of the algebras of $A$ and $B$. This implies that additivity for $R$ cannot be maintained at the same time as the intersection property.}
\label{figufigu}
\end{center}  
\end{figure}

In this section, we move our focus towards theories that violate duality for regions having non-trivial $\pi_1(R)$. From the dual perspective, these theories will also show problems for regions with non-trivial $\pi_{d-3}(R')$. The failure of duality or additivity for these types of regions gives rise to a failure of the intersection property for topologically trivial regions $A$ and $B$, with an intersection $R$ or $R'$, see figure \ref{figufigu}.\footnote{In general we think in $d\ge 4$ since for $d= 3$ the breaking of additivity/duality in regions with non-trivial $\pi_1(R)$ and $\pi_{d-3}(R')=\pi_{0}(R')$ could arise from both global symmetries or gauge symmetries. This interesting feature makes the discussion less transparent. We will comment on it later. In any case, statements about gauge theories are valid for $d=3$ as well.}

The main working example in this situation will be that of gauge theories. However, before describing the specific non-local operators associated with gauge theories, we want to show how the structure arising from a failure of duality-additivity in these types of regions is rather fixed on general grounds, without referring to gauge fields. In particular, it is possible to show that the dual non-local operators form dual Abelian groups, and the commutation relations are fixed. 

 For gauge theories, these features appear when there is a subgroup of the center of the gauge group which leaves invariant all matter fields. For pure gauge theories, as we will show below, the non-local operators correspond to t' Hooft and Wilson loops associated respectively to the center $Z$ of the gauge group and its dual group $Z^*$, the group of its characters (which is isomorphic to $Z$). All other independent Wilson and t' Hooft loops are locally generated. Any finite Abelian group can be formed in this way with a gauge theory because the cyclic group $Z_n$ is the center of $SU(n)$ and any finite Abelian group is a product of cyclic groups. In $d=4$, $R$ and $R'$ have the same topology of $S^1$, and both the Wilson and t' Hooft loops are now non-local operators in the same ring $R$. For pure gauge fields, the group of non-local operators is then $Z\times Z^*$ in $d=4$. Adding matter fields, several subgroups of $Z\times Z^*$ can be realized. We describe the non-local operators for a Maxwell field and non-Abelian (pure) gauge theories. In the appendix \ref{apg} we compute explicitly the non-local operators for arbitrary gauge fields in a lattice.

\subsubsection{The non-local operators form Abelian groups}

Now we show that the dual algebras of non-local operators correspond to dual Abelian groups, and the structure of the commutation relations is fixed. We keep the discussion as simple as possible. A mathematically precise proof would follow the ideas of the DHR analysis for global symmetries, see \cite{haag2012local} and \cite{Longo:1994xe}. Some natural assumptions have to be made. Borrowing the terminology of that analysis, an underlying assumption is that the non-local operators are {\sl transportable}. This just states that the non-local sectors are preserved by deformations. More precisely, for any two (open) regions $R_1$ and $R_2$ with the same topology (in particular, they are homotopic to each other), we assume there is a one-to-one correspondence between the non local sectors $[a]_1$ and $[a]_2$ located in $R_1$ and $R_2$ respectively. This correspondence has two steps. First, any non local operator $a$ for a region $R$ is a non local operator for an homotopic region $\tilde{R}$ if $R\subseteq \tilde{R}$. Second, the tube of homotopy $R_{12}$ connecting $R_1$ and $R_2$ has the same topology of $R_1$ and $R_2$, and includes both of these regions. Therefore, non-local operators in either $R_1$ or $R_2$ give non-local operators in  $R_{12}$, and the classes can be matched.

\begin{figure}[t]
\begin{center}  
\includegraphics[width=0.7\textwidth]{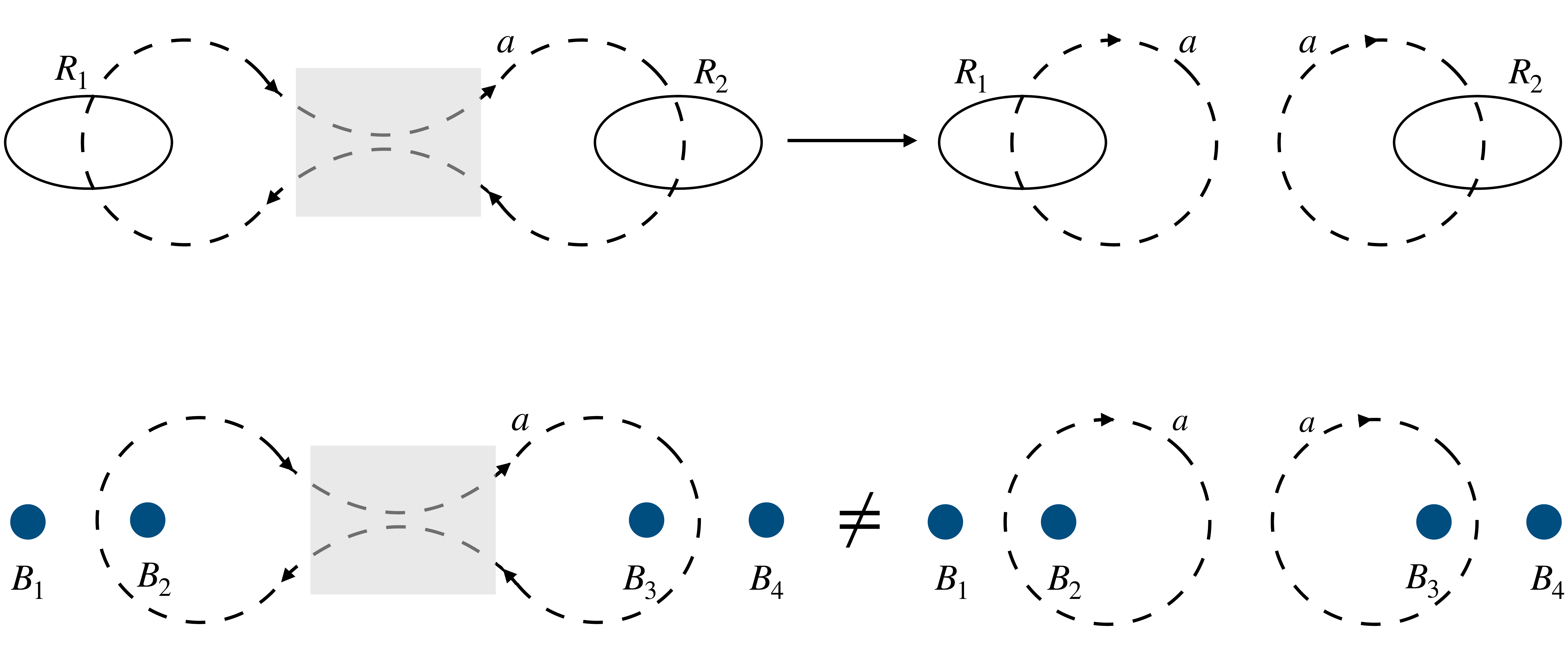}
\captionsetup{width=0.9\textwidth}
\caption{Upper panel: reconnecting a loop operator using local operators in the shaded region. Lower panel: this cannot be done for non Abelian twists. }
\label{2wl}
\end{center}  
\end{figure}  

A simple property is that given two arbitrary regions $R_1$ and $R_2$, if $R_1$ is included in a topologically trivial region $R_0$ disjoint from $R_2$, then any non locally generated operators based in $R_1$ and $R_2$ must commute with each other. This follows from the assumption that non locally generated operators in a region $R_1$ become locally generated in the topologically trivial region $R_0$ containing $R_1$. In this case, we say that $R_1$ and $R_2$ are not linked.  
 
To show the Abelianity of the non-local algebra, we first refer to the upper panel in figure \ref{2wl} (see appendix \ref{apg} for the explicit construction of these operations in lattice gauge theories). 
We take two non-linked loop regions $R_1$ and $R_2$, and take a 
 loop operator of class $a$ (dashed curve) in the complement of $R_1,R_2$, which is linked once with $R_1$ and $R_2$. The product of two disjoint loop operators of class $a$, each one linked once with just one of the two rings $R_1$ and $R_2$ (upper right panel of figure \ref{2wl}), belongs to the same class as the original single component loop of class $a$. This is because the algebra of non-local operators in the two rings $R_1$ and $R_2$ is the tensor product of the algebras of non-local operators in $R_1$ with the ones in $R_2$.\footnote{There may be interplays between symmetries related to different topological characteristics. We are not studying these scenarios in the paper and assume algebras-region problems for only one type of topology. In the present case, the non-local operators of $R_1 R_2$ are due to non-contractible loops. They are products of non-local operators in each of the rings.} It is not difficult to see that the original one-component loop of type $a$ based on $(R_1R_2)'$ has the same action on the non-local algebra of the region $R_1R_2$ as the product of the two independent loops of class $a$. Then, the single component loop and the two loops belong to the same class. They must be related by local operations in $(R_1R_2)'$. This is represented by the shaded region in the figure. This is an important step in showing that the non-local algebra is Abelian.  
 
The lower panel of figure \ref{2wl} shows why this fails for the case of twist operators in QFT's with non-Abelian global symmetries. What in the previous case were two spatially separated rings $R_1$ and $R_2$, in this case consists of four spatially separated balls $B_1$, $B_2$, $B_3$ and $B_4$. It is no longer the case that the algebra of non-local operators in the four balls is the tensor product of the non-local operators (intertwiners) in $B_1$ and  $B_2$ with the ones in $B_3$ and $B_4$. The reason is that we can cross intertwiners between $B_2$ and $B_3$. In the non-Abelian case, the twist on the lower-left panel does not have the same action on this algebra as the product of two twists on the right panel. The reason is that the invariant non-Abelian twists are sums of twists corresponding to a given conjugacy class of the group (see equation (2.14)) and the products of two of these twists generally decompose into a sum of invariant twists of different classes.

\begin{figure}[t]
\begin{center}  
\includegraphics[width=0.5\textwidth]{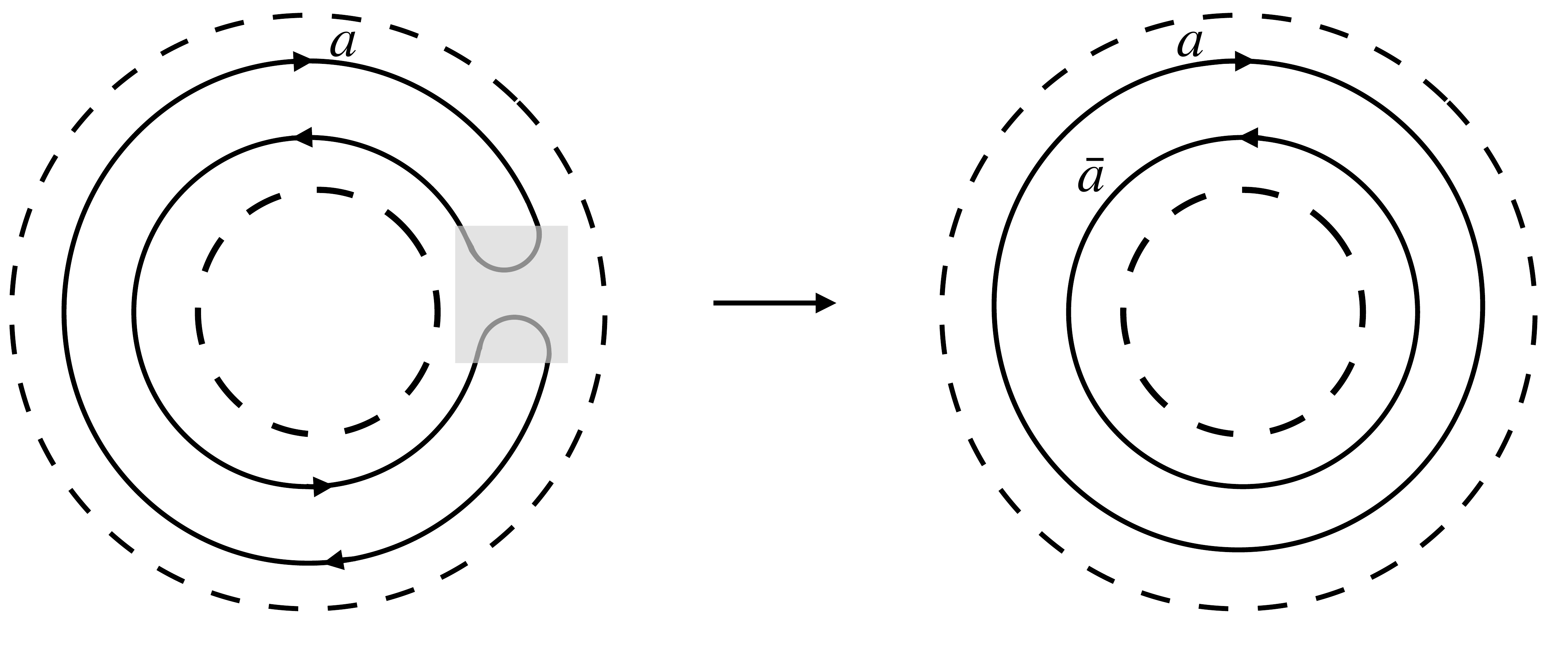}
\captionsetup{width=0.9\textwidth}
\caption{The operators of the form $a\bar{a}$ are locally generated inside a ring $R$, marked with the dashed line.}
\label{2wl2}
\end{center}  
\end{figure}

We conclude we can glue and split loops associated with the same class in this form. Now let us take a simple ring $R$ drawn with a dashed line in figure \ref{2wl2}. Inside the ring, we can place an elongated loop of class $a$. This is a folded version of the loop in the left upper panel of the figure \ref{2wl}. This loop is locally generated inside $R$ since its topology can be shrunk inside $R$. If we glue two extremes of this loop (the same operation as in figure \ref{2wl}), we obtain two single component loops inside $R$, as shown in the left panel of figure \ref{2wl2}. The product of these two loops must, therefore, be equivalent to the trivial class since it is locally generated. They have to correspond to conjugate classes in the ring $R$, which are the only ones that can contain the identity in their product. This gives
\be    
[n]_{a\bar{a}}^{a'}=0\,, \hspace{1cm}a'\neq 1\,,
\ee
and it implies that the fusion rules arise from an Abelian group. Let us see how this comes about. The product of classes is associative and commutative.  We already have the unit and the inverse, $[1][a]=[a]$ and $[a][\bar{a}]=[1]$, where $[1]$ is the class of locally generated operators. To realize the structure of an Abelian group, we further need to prove that the fusion of two arbitrary classes gives rise to only one class.  Such fusion takes the generic form
\be
[a_1][a_2]=\sum_{a_3} [n]_{a_1 a_2}^{a_3}[a_3]\,. \label{class}
\ee
Multiplying this expression by $[\bar{a}_1]$ we get the class $[a_2]$ on the left-hand side, which must be equal to the right-hand side. The right hand side results in the sum of the classes $[\bar{a}_1][a_3]$. These classes must then all be equal to the class $[a_2]$. Assume now there is more than one class, say $[a']$ and $[a'']$, in the right hand side of \eqref{class}. We must have $[a_2]=[\bar{a}_1][a']=[\bar{a}_1][a'']$. Multiplying by $[a_1]$ in this expression we get that in fact $[a']$ and $[a'']$ are equal. Therefore, for fixed $a_1$ and $a_2$, the coefficient $[n]_{a_1 a_2}^{a_3}$ can only be non zero for just one class $[a_3]$. This defines an Abelian group $G_a$ for the product of classes. The elements of the group are just the classes, which contain an inverse and an identity, and the product in the group is the product of classes. All this argument runs in the same way for the classes $[b]$ associated with the non-local operators in $R'$. These dual classes form a group $G_b$. Below we will show how to choose actual operators of the theory representing the abstract fusion of classes. In other words, we will find loop operators representing the Abelian symmetry groups.
 
This argument does not hold in this generality for regions with non-trivial $\pi_1(R)$ in $d=3$, as shown by the examples of global symmetries having non-Abelian groups discussed in the preceding section. As explained above, the reason is that in $d=3$ (two spatial dimensions) the operation of figure \ref{2wl} does not hold in general. Still, for pure gauge theories in $d=3$, the proof holds (see appendix \ref{apg}), and we have an Abelian group for the non-local sectors. 
    
The same proof of Abelianity should work for sectors corresponding to regions with the topology of spheres $S^{k}$ for $0<k<d-2$. The conclusion is that living aside the case of dimensions $0$ and $d-2$, which includes the case of global symmetries, in all other cases the product of a class $[a]$ and its inverse $[\bar{a}]$ is an operator that is locally generated on the appropriate region.

A slightly different chain of arguments is as follows. We can imagine we started with a different and bigger set $S$ of sectors $s$. These abstract sectors could run for example over all the irreducible representations of a certain non-Abelian group, whether of discrete or Lie type, as it is the case of Wilson loops for non-Abelian gauge theories. To run the argument we only assume these sectors satisfy some generic notion of fusion rules
\be 
s\ast s'=\sum\limits_{s''}\,N_{ss'}^{s''}\,s''\;.
\ee
Here the fusion coefficients might be associated with a non-Abelian symmetry group, or with a more general structure. We only ask the fusion algebra to be Abelian $N_{ss'}^{s''}=N_{s's}^{s''}$, which follows from the locality principle in QFT.

But crucially, not all the sectors $s\in S$ are non locally generated in the region $R$. All the sectors being produced in the fusion of arbitrary products of $s\bar{s}$ are locally generated, for the same reason as above. Let us call the set of sectors appearing in arbitrary products of $s\bar{s}$ by $S_{1}$. By construction, $S_{1}$ defines a subcategory of the category $S$. The true classes associated with the violation of Haag duality arise as the quotient of the whole set $S$ by the sectors in $S_{1}$. In the literature of tensor categories, see \cite{etingof2016tensor}, this is called the universal grading of $S$, and the associated group the universal grading group. Grading of a category $S$ by a group $G$ is a partition of $S$ of the form
\be 
S=\sqcup_{g\in G}S_{g}\;,
\ee
such that for any $s_{g}\in S_{g}$ and $s_{h}\in S_{h}$ the product $s_{g}\ast s_{h}$ belongs to $S_{gh}$. The universal grading, as its name suggests, can always be found, and it is associated with $S_{1}$ being formed by arbitrary products of $s\bar{s}$. For symmetric fusion rings, like the ones we are considering, the resulting universal braiding group $G$, shown to be associated with the breaking of Haag duality, is necessarily Abelian.

An analogous result holds for theories with $k$-form symmetries \cite{Gaiotto:2014kfa}. The proof of Abelianity in such work relies on the Euclidean continuation of the QFT, in particular the Euclidean continuation of the generators of the generalized global symmetry. Here we did not invoke a particular Hamiltonian and no relativistic symmetry was necessary for the argument. The Abelian nature just follows from the physical requirement that the true non-local classes should be not locally generated. This directly forces us to consider the universal grading of the original fusion rules above alluded, which is necessarily an Abelian group.

\subsubsection{Algebra of non-local operators}
\label{allgg}

We have shown that the classes of non locally generated operators in $R$ form an Abelian group. We now want to show we can take operator representatives of these classes providing the actual group operations. An Abelian group $G$ is a product of cyclic subgroups $Z_{n_{1}}\otimes Z_{n_{2}}\cdots$. If we can construct operators for any of the cyclic subgroups, then it is enough to take representatives for each of the factor cyclic subgroups in different spatially separated non linked rings (in order that they commute with each other) inside the region $R$ to get representatives for the full group.    
 
Then, let $C_a=\{[a^{k}]\}$, $k=0,\cdots ,n-1$ be a cyclic factor of order $n$ of the group $G_a$ associated with the class $[a]$ in $R$.  Let $\tilde{a}$ be a representative of $[a]$, an actual operator in the theory. Choosing $\tilde{a}$ such that $\tilde{a} \tilde{a}^\dagger$ is invertible, the unitary operator $\hat{a}=\tilde{a}/\sqrt{\tilde{a} \tilde{a}^\dagger}=\tilde{a}/|\tilde{a}|$ belongs to the same class $[a]$. We have $\hat{a}^n= U$, with $U$ unitary, commuting with $\hat{a}$, and $U\in[1]$. All the spectral projections of $U$ belong to the algebra of locally generated operators and commute with $\hat{a}$. Using the spectral decomposition we can construct $V=U^{-1/n}$ by taking the $n^{th}$ root of the eigenvalues, with the same spectral projections. With these observations there are now many choices for $V$. Any of them will do the work. Define $a=V \hat{a}$. We have $a^{k}$ belongs to the class $[a^{k}]$, and $a^0=a^n=1$. The operators $a^{k}$ then provide a representation of the cyclic subgroup $C_a$. The same can be done for the other cyclic subgroups in $R$ and  also for the operators $b\in[b]$ with the group operation laws of $G_b$ for the dual classes $[b]$ in $R'$.   

Having constructed the operator representatives of the symmetry, consider the unitary transformation $x \mapsto b x b^{-1}$. It maps ${\cal A}_{add}(R')$ into itself, since for $x$ in this additive algebra, $b x b^{-1}$ is in the identity class. It also maps its commutant ${\cal A}_{max}(R)={\cal A}(R)\vee \{a\}$ into itself.  We also observe that this automorphism of ${\cal A}_{max}(R)$ does not depend on the precise choice of representatives $b$. This is because any other choice arises from $b$ as products of locally generated operators in $R'$, and these operators commute with all $x\in {\cal A}_{\textrm{max}}(R)$.

It will be more useful to define the following maps of ${\cal A}_{max}(R)$, associated with each irreducible representation $r$ of $G_b$,
\be
E_r(x)= |G|^{-1}\, \sum_{b\in G_b} \chi_r^*(b)\, b\, x\, b^{-1}\,, \hspace{.6cm} x\in {\cal A}_{\textrm{max}}(R)\,,\hspace{1cm}  E_{r'}(E_r(x))=\delta_{r r'}\, E_r(x)\,. \label{sds}
\ee
The third equation just follows by direct evaluation. 

A not so transparent property of the previous map is that $E_r(a)\in [a]$. The reason is that we can choose a representative $a$ of $[a]$ in $R$, such that it is actually supported in a smaller ring $\tilde{R}\subset R$. Then, for the purpose of the action \eqref{sds} on $a$, we can replace the $b$ operators by new ones inside $R$ but outside $\tilde{R}$. Then, the map is composed by locally generated operators in $R$, and hence, they cannot change the class $[a]$. Finally, from the last equation in \eqref{sds}, it is clear that if $E_r(a)\neq 0$ for some $r$, then $E_{r'}(a)=0$ for all $r\neq r'$.

The previous observations imply there is a one to one correspondence between representations $r$ of $G_b$ and the non local classes $[a]$. It has to be one-to-one since otherwise there would be linear combinations of elements of different classes which vanish, or the $b$ operators would not be linearly independent. Therefore we can label the representations $r$ of $G_b$ by the class labels $a$, such that $E_r(a)=E_a(a)\in [a]\neq 0$. Further, we can show that $E_a(a_1)=a_1$, for any $a_1$ of $[a]$. First let us define $\tilde{a}=E_a(a)$, for which $E(\tilde{a})=\tilde{a}$ and $\tilde{a}\in [a]$. Now, any element $a_1$ of $[a]$ can be written by taking $\tilde{a}$ and multiplying by arbitrary products of locally generated operators. Therefore
\be
E_a(a_1)=E_a\left(\sum_\lambda O_1^\lambda\, \tilde{a}\, O_2^\lambda\right)= \sum_\lambda O_1^\lambda\, E(\tilde{a})  O_2^\lambda= a_1\,,
\ee
where $O^\lambda_1,O^\lambda_2$ are additive elements on $R$. In particular
\be\label{ia}
E_a(a)=a\,.
\ee  
Essentially, the intuition is that the previous map is a projection of ${\cal A}_{\textrm{max}}(R)$ into its different classes $[a]$. In the context of von Neumann algebras, projections are often associated with conditional expectations, which we will describe below in detail. In this case, $E_a$ is not a conditional expectation for $a\neq 1$. The reason is that for $a\neq 1$ the target space is not actually an algebra since the non-trivial classes do not contain the identity by construction. The map is better seen as a projection in a vector space.

In any case, using both \eqref{sds} and \eqref{ia} it follows that
\be
b \, a= b \, E_a(a)=\chi_a (b) \, a \, b\,,\label{ccrr}
\ee
or equivalently
\be
a \, b= \chi_b(a) \, b \, a\,,\label{ccrr1}
\ee
with $\chi_a(b)=(\chi_b(a))^{-1}=(\chi_b(a))^{*}$. Since all operators in $[a]$ and $[b]$ are constructed by multiplying the representatives $a$ and $b$ by arbitrary products of locally generated operators in $R$ and $R'$ respectively, and these commute between each other, it follows that the same commutation relation holds for all other elements of $[a]$ and $[b]$.

Finally, in order to construct a maximal causal net satisfying duality, we have to take subsets of dual operators $\{a\}$ and $\{b\}$, such that they satisfy causality and close under fusion. This is equivalent to take maximal sets of pairs of non-local operators $M=\{(a_i,b_j)\}$ such that
\be
 (a_{i_1},b_{j_1})\in M\,, (a_{i_2},b_{j_2})\in M\rightarrow  (a_{i_1}a_{i_2},b_{j_1}b_{j_2})\in M \,\,\textrm{and}\, \,\chi_{a_i}(b_j)=\chi_{b_j}(a_i)=1\,. \label{ggg}
\ee
These maximal causal nets were called Haag-Dirac nets in the introduction exactly for this reason. The generalized Dirac quantization condition $\chi_{a}(b)=1$ arises in the local algebraic approach by requiring Haag duality and causality.

To summarize, we conclude that the number of elements in $\{a\}$ and $\{b\}$ is the same. Besides,  $\{a\}$ is the group of characters of $\{b\}$, and the other way around. The dual Abelian groups arising from the breaking of Haag duality are Pontryagin duals of each other. The commutation relations are fixed to be \eqref{ccrr}, and the phases $\chi_a(b)$ in this relation form the table of characters of the symmetry group. The Dirac quantization condition arises by enforcing causality of the net. Remarkably, these features are simply inescapable consequences of the violation of Haag duality for regions with non-trivial $\pi_1$ and $\pi_{d-3}$ in local QFT. In particular, we have not defined the dual operators, say the $b$'s, by their commutation relations with the $a$'s, as it is usually done since 't Hooft's original work \cite{tHooft:1977nqb}. Also, we have not assumed any symmetry group structure and charged operators to start with.

\subsubsection{Standard non-local operators}
\label{standard}
Interestingly, given a region $R$ with non-local operators, there is a standard way to obtain representatives of the non-local operators. The construction generalizes the Doplicher-Longo construction of standard twists \cite{Doplicher:1983if,Doplicher:1984zz}.\footnote{The Doplicher-Longo construction is however associated with a type I factor that splits the algebra of two balls. Here this split is not needed.} These standard operators are uniquely defined by the condition 
\be
J_R\, a \, J_R =a\,,\label{pol}
\ee
where $J_R$ is the vacuum Tomita-Takesaki reflection corresponding to ${\cal A}_{\textrm{add}}(R)$. 

The existence of these operators is a simple consequence of a theorem that states that any automorphism of a von Neumann algebra with a cyclic and separating vector is implementable by a unitary operator, and one can choose the unitary to be invariant under the conjugation $J$ (see \cite{haag2012local} theorem 2.2.4). In the present case, the algebra is $({\cal A}_{\textrm{add}}(R))'$, the automorphism is the one induced by the non-local operators of type $a$ (which is independent of the representative), and the vector state is the vacuum. Then we get a unitary $a$ invariant under the modular conjugation of $({\cal A}_{\textrm{add}}(R))'$ which is the same as the modular conjugation of ${\cal A}_{\textrm{add}}(R)$, and hence \eqref{pol}. By construction, the algebra of the standard operators $a$ and $b$ is the expected one. By the same reason $a$ belongs to $({\cal A}_{\textrm{add}}(R'))'$ but not to ${\cal A}_{\textrm{add}}(R)$, and it is a non local operator in $R$. 

Further interesting properties follow from the fact that the standard operator leaves the natural cone ${\cal P}$ of vectors invariant. This cone is defined as the one generated by all vectors of the form $O J O |0\rangle$ for $O$ in the algebra \cite{haag2012local}. The important point here is that vectors in the natural cone include the vacuum and have a positive scalar product. If follows that $a_i |0\rangle \in {\cal P}$ and $\langle 0 | a_i |0\rangle > 0$. This last equation also entails $\langle 0 | a_i a_j |0\rangle > 0$.
 
This interesting construction gives, for example, standard smeared non-local Wilson and t' Hoof loops (for the center of the gauge groups) defined exclusively by the vacuum and the geometry of the chosen region. In particular, they enjoy all the symmetries that these regions and the vacuum state may have. 

\subsubsection{Maxwell field}

A simple example of these scenarios is the Maxwell field in $d=4$. This is the Gaussian theory of the electric and magnetic fields satisfying the equal-time commutation relations
\be
[E^i(\vec{x}),B^j(\vec{y})]=i \varepsilon^{ijk}\, \partial_k \delta^3(\vec{x}-\vec{y})\,.
\ee
Equivalently, the theory can be described by the normal oriented electric and magnetic fluxes $\Phi_E$ and $\Phi_B$, defined on two-dimensional surfaces with boundaries $\Gamma_E$ and $\Gamma_B$.  For such fluxes, we have a commutator proportional to the linking number of $\Gamma_E$ and $\Gamma_B$,
\be
[\Phi_E,\Phi_B]=\frac{i}{4\pi}\int_{\Gamma_E}\int_{\Gamma_B} \frac{\vec{x}_1-\vec{x}_2}{|\vec{x}_1-\vec{x}_2|^3}\, d\vec{x}_1\times d\vec{x}_2  \,.\label{deci}
\ee  
We will always assume these fluxes to be smeared over positions of $\Gamma_E$ and $\Gamma_B$ such that the flux operators are well-defined linear operators and not operator-valued distributions. If the smearing region for $\Gamma_E$ and $\Gamma_B$ lies respectively inside a region with the topology of a ring $R$ and its complement $R'$, and the integral of the smearing function adds up to one (which we will also assume in the following), equation \eqref{deci} still holds for the smeared fluxes. In $d=4$ the topology of $R'$ is the same as the topology of $R$. It is $S^1\times \mathbb{R}^2$ and it has non-trivial $\pi_1(R)$. 

Because $\nabla E=\nabla B=0$ the fluxes are conserved. The surface over which they are defined can be deformed, keeping the boundary fixed, without modifying the operator. In turn, by deforming the surface of the flux we can take it away from some local operator lying in the original surface. Therefore, one concludes that the fluxes will commute with the locally generated operators associated with the complementary ring.   

We can write a bounded  electric flux operator (t' Hooft loop) $T^{g}=e^{i g \Phi_E}$, and a magnetic flux operator (Wilson loop) $W^{e}=e^{i q \Phi_B}$, for any $g,q\in \mathbb{R}$.  When they are linked, the commutation relations between them follows from \eqref{deci}
\be
T^{g}W^{q}= e^{i\, q\, g } \, W^q T^g\,.\label{cr}
\ee 
This non-commutativity implies these operators cannot be locally generated in the rings in which they are based. For example, if $T^g$ were locally generated in $R$ (where its boundary lies) this would imply, by the arguments given above, it necessarily commutes with $W^q$ based on the complementary ring. But this is not possible according to \eqref{cr}. Notice this is an explicit example of relation \eqref{ccrr}.    

Therefore, the algebra of a ring $R$ and its complement $R'$ (also a ring) cannot be taken additive without violating duality. The reason is that the commutant of the additive algebra of the ring contains both the electric and magnetic loops of any charge based on $R'$, and this is not additive. We have
\bea
{\cal A}_{\textrm{max}} (R')\equiv ({\cal A}_{\textrm{add}}(R))'={\cal A}_{\textrm{add}}(R')\vee \{W^q_{R'} T^g_{R'}\}_{q,g \in \mathbb{R}}\,,
\eea  
and analogously by interchanging $R\leftrightarrow R'$. Here we have denoted $W^q_{R'}$ and $T^g_{R'}$  for the Wilson and t' Hooft loops based on $R'$. 

One can repair duality at the expense of additivity by defining the ring algebras to contain, on top of the locally generated operators, some particular set of non locally generated ones. To form a (local) net, such choice has to respect causality. A natural condition is to add operators with electric and magnetic charges $(q,g)$ to all rings. This choice does not ruin translation and rotation invariance. Given two dyons $(q,g)$ and $(q',g')$ in the same ring, the one formed by their product $(q+q', g+g')$, and the conjugates $(-q,-g)$ and $(-q',-g')$, should also be present to close an algebra. Therefore, the set of all dyons should be an additive subgroup of the plane, giving a lattice 
\be
(q,g)=n (q_1,g_1) + m (q_2,g_2)\,,\label{lat}
\ee
where $n,m\in \mathbb{Z}$, and $(q_1,g_1),\,(q_2,g_2) \in \mathbb{R}^2$ are the generating vectors of the lattice.
Locality between a would-be dyon $T^{g}W^{q}$ with charges $(q,g)$ in $R$ and another one $(\tilde{q},\tilde{g})$ in $R'$ (i.e. the vanishing of the phase in \eqref{cr}) results in the Dirac quantization condition\footnote{The Dirac quantization condition is typically a statement that arises when we include charges in the model, as we comment below. But indeed, it is more naturally originated in a setup without charges when studying causal nets of the form described here.}
\be
q \tilde{g}-\tilde{q} g= 2\pi k\,,\label{dcc}
\ee
for an integer $k$.  This is compatible with \eqref{lat} provided that $q_1 \,g_2-g_1\,q_2\in 2 \pi \mathbb{Z}$.
If we want to construct a Haag-Dirac net, we need to take a maximal set of charges that satisfy \eqref{dcc}. This forces us to choose
\be
q_1 \,g_2-g_1\,q_2=2 \pi\,.\label{sis}
\ee
This is the most general condition for a $U(1)$ symmetry. However, for the case of the relativistic Maxwell field, in solving for the space of solutions of the previous equation, we need to take into account that there is a duality symmetry (see, for example, \cite{AlvarezGaume:1997ix})
\be
(E+i B)\rightarrow e^{i \phi} (E+i B)\,,
\hspace{.7cm}
(q+i g)\rightarrow e^{i\phi} (q+ i g)\,. 
\ee
Then, there is a hidden free parameter in the solution of \eqref{sis} that moves us between isomorphic Haag-Dirac nets. This freedom can be eliminated by writing the different solutions as
\be
(q,g)= \left(q_0 \left(n_e+\frac{\theta}{2 \pi} n_m \right), g_0 n_m \right)\,, \label{buscar}
\ee
where $q_0>0$ and $\theta\in [0,2\pi)$ are parameters, $g_0=2\pi/q_0$, and $n_e, n_m\in \mathbb{Z}$. \footnote{We can also consider the limiting cases when $q\rightarrow0$ ($g\rightarrow\infty$) and $q\rightarrow \infty$ ($g\rightarrow0$). In the first case, the HD net is formed by adding all the Wilson loops for a ring-like regions and none of the t' Hooft loops. The second case is the opposite.} Writing the two real parameters as a single complex one $\tau=\theta/(2 \pi)+2 \pi i /q_0^2$, the Haag-Dirac nets verifying duality and causality are determined by this parameter, ${\cal A}_{\textrm{HD}}(\tau)$. In this parametrization there is a residual duality symmetry, since nets with $\tau'=\tau+1$, $\tau'=-1/\tau$ are isomorphic.

Nets with $\theta \neq 0,\pi$ are not time reflection symmetric. Notice that in a specific model describing electric charges and monopoles, when adding a topological $\theta$ term to the Lagrangian (or equivalently considering the $\theta$ vacua), we change the lattice of charges according to the Witten effect \cite{Witten:1979ey}. We see such a parameter here, as arising from the previous freedom we encountered in describing the lattice of charges.    

The nets constructed in this way will satisfy duality, but they will not satisfy additivity. Additivity can be recovered if we couple the theory to charged fields. For example, if we have a field $\psi$ of electric charge $q$, we can now consider Wilson line operators of the form
\be
\psi(x)e^{i q \int_x^y dx^\mu A_\mu } \psi^\dagger(y) \,.
\ee  
Taking products of consecutive Wilson lines, and allowing for the fusion of the fields with opposite charges at the extremes of the lines we want to join, the Wilson loop $W^{q}$ in $R$ (with the specific charge $q$), becomes an operator in the additive algebra of $R$. In the same way, if we have magnetic charges $g$, the 't Hooft loops $T^{g}$ corresponding to this charge should be additive in $R$, and with a dyon $(q,g)$ we can break the operators $T^{g}W^{q}$. For the theory to still satisfy locality the charges have to satisfy \eqref{dcc}. This is now converted into the Dirac-Schwinger-Zwanziger (DSZ) quantization condition for the charges. As mentioned before, this condition is seen here as a consequence of causality in the net of algebras for the theory without charges. In this way, by adding a full set of charged fields with charges corresponding to a HD net, we can make the theory  ``complete'' in the sense of both satisfying duality and additivity. If we do not add charged operators for a full lattice,  there still will be some problems between algebras and regions, which can be studied by taking a quotient by the new locally generated loops.

Let us close this section with an important remark. In the presence of charged fields, the flux operators $T^g W^q$ continue to exist even if $(q,g)$ does not belong to the lattice. But they now depend on a surface rather than a closed curve. Since $\nabla E=\nabla B=0$ is modified, the flux on a given surface cannot be deformed to other surfaces with the same boundary. Then, in this scenario, the operator belongs to a topologically trivial region and cannot be associated with a ring.

\subsubsection{Non-Abelian Lie groups}
\label{nonabelian}

In this section, we consider the case of non-Abelian Lie groups, whose features can be described directly in the continuum limit.\footnote{For a detailed analysis of the failure of duality and/or additivity of gauge theories we refer to Appendix \ref{apg}, where we give explicit lattice constructions of all the involved operators.} We start with pure gauge theories, without charged matter. Later we will consider the effect of adding matter fields. The objective is again to understand the failure of duality and additivity for these theories.

For generic pure gauge theories, as for the Maxwell field, the set of gauge-invariant non-local operators, with the potential of being non additively generated, is given by the Wilson and t' Hooft loops \cite{PhysRevD.10.2445,tHooft:1977nqb,PhysRevD.74.025005}. Although Wilson loops are one-dimensional in all dimensions, 't Hooft loops are only one-dimensional objects in four spacetime dimensions, where they were at first defined. In other spacetime dimensions, t' Hooft operators are defined for $d-3$ dimensional surfaces (see Appendix \ref{apg} for an explicit construction). This suggests that Wilson loops are the right candidates to violate additivity in regions with non-trivial $\pi_1 $, while the dual 't Hooft operators are the right candidates to violate additivity in regions with non-trivial $\pi_{d-3} $. In $d=4$, both operators potentially contribute to the violation of duality in ring-shaped regions.

Let us start with the Wilson loops. These are defined for each representation $r$ as
\be
W_{r}:= \textrm{Tr}_{r} \mathcal{P} \,e^{i\oint_C dx^{\mu} A^r_{\mu}}\;,
\ee
where $C$ is a loop in space-time and $\mathcal{P}$ is the path order. There is one independent Wilson loop per irreducible representation of the gauge group. As shown in Appendix \ref{apg}, they can be chosen in order to satisfy the fusion rules of the (gauge) group representations
\be
W_{r}W_{r'}=\sum\limits_{r''}N_{rr'}^{r''}W_{r''}\;.
\ee
 
We now seek to know whether Wilson loops are unbreakable or not. A Wilson loop of representation $r$ can be certainly broken into pieces if there are charged fields $\phi_{r}$ in the model transforming under the representation $r$. With such charged field, we can construct Wilson lines
\be 
\phi_{r}(x)\,\mathcal{P} e^{i \int_x^y  dx^\sigma A_\sigma}\,\phi^\dagger_{r}(y)\;.
\ee
These lines decompose the Wilson loop of representation $r$ into a product of operators localized in segments. Although we are considering pure gauge theories without charges, we cannot escape the fact that, for non-Abelian gauge fields, the gluons are charged themselves. They are charged under the adjoint representation. Indeed, we can form the following Wilson line, terminated by curvatures
\be
F_{\mu\nu}(x) \,\mathcal{P} e^{i \int_x^y  dx^\sigma A_\sigma}\, F_{\alpha\beta(y)}\,,\label{indices}
\ee
where all fields are in the adjoint representation of the Lie algebra. We conclude that a loop operator in the adjoint representation can be generated locally by multiplying several of these lines along a loop. Since the adjoint Wilson loop is locally generated, the same can be said for all representations generated in the fusion of an arbitrary number of adjoint representations. Therefore, the `truly'' non-local Wilson loops, those violating Haag duality, are labeled by the equivalence classes that arise when we quotient the set of irreducible representations by the set of representations generated from the adjoint.\footnote{This is the gauge theory analog of the general discussion in the previous section concerning the universal grading of the set of representations.}

To understand in precise terms what we mean by the last statement we need to invoke several notions from the theory of representations of Lie groups. Since introducing and describing them in detail would take some time and space, and it will certainly interrupt the flow of the presentation, we will assume here knowledge of such topic, and refer to the references \cite{brocker2003representations,Cornwell:1997ke,Costa:2012zz,hamermesh1989group,sternberg2009lie,
zee2016group,carter_macdonald_segal_taylor_1995,roman2011fundamentals} for more details. For the present context, the most important notions we need are the weight and root lattices. For a Lie algebra $\mathfrak{g}$, a Cartan subalgebra $\mathfrak{h}$ is a maximal Abelian subalgebra. If $\mathfrak{h}$ is generated by $l$ elements, the Lie algebra is said to have rank $l$. Since $\mathfrak{h}$ is Abelian, it can be diagonalized in every irreducible representation of the algebra. A weight associated with a certain eigenvector in certain irrep is defined as the $l$-component vector formed by the eigenvalues of the Cartan subalgebra generators. It turns out that the weights form a lattice
\be 
\Lambda_{\omega}:= \left\{ \sum\limits_{i=1}^{l} a_{i}\omega^{(i)}\,\,\,\,\,\,\textrm{with}\,\,\,\,\, a_{i}\in \mathds{Z} \right\} \;,
\ee
generated by arbitrary linear combinations with integer coefficients of a set of fundamental weights $\omega^{(i)}$. The number of fundamental weights is equal to the rank. Physically, this lattice contains the information of all the representations of the algebra. In this lattice, each irreducible representation is labeled by a dominant weight. In the weight lattice, such dominant weights are in one-to-one correspondence with orbits of the Weyl group, and then we have
\be 
\Lambda_{\textrm{dom}}\sim \Lambda_{\omega}/W\;.
\ee
These equivalence classes label all the (inequivalent) irreducible representations, and therefore, all (inequivalent) Wilson loops. Furthermore, for every Lie group, there is a universal representation called the adjoint representation. It is the representations in which the Lie algebra transforms into itself. The weights of the adjoint representation are called roots. The roots also form a lattice, called the root lattice
\be 
\Lambda_{\textrm{root}}:= \left\{\sum\limits_{i=1}^{l} a_{i}\alpha^{(i)}\,\,\,\,\,\,\textrm{with}\,\,\,\,\, a_{i}\in \mathds{Z} \right\}\,.
\ee
It is generated from a set of $l$ fundamental roots $\alpha^{(i)}$. Physically, while the weight lattice contains all possible weights, and therefore all weights appearing in arbitrary products of fundamental representations, the root lattice contains all weights appearing in arbitrary products of the adjoint representation.

The dominant weights appearing in the root lattice can be isolated in the same way as before, employing the Weyl group
\be 
\Lambda_{\textrm{root-dom}}\sim \Lambda_{\textrm{root}}/W\;.
\ee
The non locally generated classes of Wilson loops are then labeled by
\be 
WL_{\textrm{non-local}}\sim \Lambda_{\textrm{dom}}/\Lambda_{\textrm{root-dom}}\sim (\Lambda_{\omega}/W)/(\Lambda_{\textrm{root}}/W)\sim \Lambda_{\omega}/\Lambda_{\textrm{root}}\sim \Lambda_{Z}\;,
\ee
where $\Lambda_{Z}$ is equivalent to $Z^*$, the group of representations of the center $Z$ of the gauge group $G$. These representations form the dual of the Abelian group $Z$, which is isomorphic with $Z$.\footnote{In the original work \cite{tHooft:1977nqb}, 't Hooft loops were defined only in correspondence with the center of the gauge group. A natural question arose as to why we have so many more Wilson loops (for Lie groups an infinite number of them), and so few 't Hooft loops. This was clarified in \cite{PhysRevD.74.025005} by enlarging the set t' Hooft loops. It was noticed there that t' Hooft loops can be defined for any dominant magnetic weight. Here we have taken a complementary approach for the clarification of such an issue. From the present perspective, the only important Wilson loops are the non-locally generated ones. These are in one-to-one correspondence with the dual $Z^*$ of the center $Z$ of the gauge group, which is isomorphic to the center $Z$ itself. The equality in number from the Wilson loops and the 't Hooft loops arises here by this drastic reduction of significant Wilson loops.}

One can construct actual representatives of such non-additive classes using the generic construction described in the previous section. We conclude that we can find a set of non additively generated operators in a ring satisfying the algebra of the characters of the center of the group. This quotient is an example of the universal grading alluded to in the previous section.

A similar discussion goes for t' Hooft loops, when one starts with the dual description in terms of the dual GNO group \cite{Goddard:1976qe,PhysRevD.74.025005}. Then, this results in a non additively generated t' Hooft loop (violating duality for regions with non-trivial $\pi_{d-3}$) per element of the center of the gauge group, as they were originally defined in \cite{tHooft:1977nqb}. A construction of such non-additive 't Hooft loops, which do not use the dual description, is provided in Appendix \ref{apg}. Such construction also allows constructing the non-additive Wilson loops using the dual group. One can also label the t' Hooft loops by the conjugacy classes of the gauge group. These conjugacy classes are in one-to-one correspondence with orbits of the Cartan subalgebra under the Weyl group \cite{brocker2003representations}, as it is the case for magnetic monopoles \cite{Goddard:1976qe}. But again, labeled in this way, not all 't Hooft loops are non-locally generated.

We conclude that the physical symmetry group violating Haag duality in pure gauge theories is $Z^*\times Z$, where $Z^*$ is generated by the non-breakable Wilson loops and $Z$ by the non-breakable 't Hooft loops. We thus find the algebraic origin of the generalized global symmetries described in \cite{Gaiotto:2014kfa}. Haag-Dirac nets can be constructed by enforcing duality and causality to the net. These conditions were studied in \cite{Aharony:2013hda}, and the lattices found there are seen here as labeling HD nets.\footnote{We want to remark a possible source of confusion. In the literature, see, for example, the mentioned \cite{Aharony:2013hda} and the lecture notes written by Tong \cite{TongLectures}, it is sometimes stated that the solution of the appropriate Dirac quantization condition implies that some theories have some loop operators and not others. For example, in a pure gauge theory, if we have all Wilson loops, we cannot have 't Hooft loops transforming under the center of the dual group. We clarify here that this statement only applies to the \emph{net}, not to the full content of the QFT. In a ball-shaped region, we always have all Wilson and 't Hooft loops. It is only the assignation of algebras to regions with non-trivial topology, namely the specification of the net of algebras, which is constrained by the Dirac quantization condition. This was transparently seen in the Maxwell field scenario described earlier, where the loop operators are simply electric and magnetic fluxes. See \cite{Harlow:2018tng} for a discussion on this point from a different perspective with similar conclusions.} 

Finally, let us mention how these features change when one includes matter. For $d>4$, matter fields will break non-local operators only if they are charged under the center $Z$ of the group (electrically charged fields) or the dual $Z^*$ of the center (magnetically charges fields).  Let us call $M_e\subseteq Z^*$ and $M_m\subseteq Z$ to the electric and magnetic charges with respect to $Z$ and $Z^*$ respectively. These can fuse, and then $M_e$ and $ M_m$ are subgroups. These charges have to satisfy the generalized Dirac quantization condition \eqref{ggg} by causality.
 All operators in $Z$, which do not commute with $M_e$, can no longer be considered operators that live in a ring, and are now only operators that exist in balls. Then, the remaining t 'Hooft loops in the ring are given by $(Z^*/M_e)^*\subseteq Z$, which is a subgroup of $Z$. $M_m$ is included in this subgroup due to the Dirac quantization condition, and the loops in $M_m$ are now locally generated, broken by magnetic charges. Hence, the remaining non locally generated t' Hooft loops in the ring are given by $(Z^*/M_e)^*/M_m$. Analogously, the non local Wilson loops will be $(Z/M_m)^*/M_e$. In a complete theory, these isomorphic groups should be trivial. For $d=4$, the group of non locally generated operators for a pure gauge theory is $Z\times Z^*$, which is now naturally isomorphic to $Z^*\times Z$. Let the subgroup of dyons be $D\subseteq Z\times Z^*$ and its isomorphic image $D^*\subseteq Z^*\times Z$. We have that the group of non-local operators is given by $((Z^*\times Z)/D^*)^*/D$.

\subsection{Generalizations}

The same arguments apply for QFT's in which the violation of Haag duality appears for regions $R$ with non-trivial $\pi_{n}$, whose complementary regions $R'$ have non-trivial $\pi_{d-n-2}$. There are particular instances that need to be taken with special care. But in general, a violation of duality due to some operators which can be localized in regions with non-trivial $\pi_{n}$, for $n\geq 1$, $d-n-2\ge 1$, will give rise to an Abelian group. The reason is the same as before. A region with such properties is connected. Therefore, if an operator of a given representation $r$ lives in a region $R$ (with such topological properties), then the representation $rr^{*}$ can be additively generated inside the region. This implies that the set of non additively generated operators corresponds to the universal grading of the associated tensor category. This universal grading results in an Abelian group.

Examples of these types of symmetries should come from p-form gauge fields $A_{\mu_1\mu_2\cdots}$, but we will not consider explicit examples in this paper. This construction again connects with the generalized global symmetries described in \cite{Gaiotto:2014kfa}. We remark that the Abelianity of the sectors we have discussed is rooted in the analysis of Haag duality and it can be proven without the necessity of going to Euclidean space.

Among the zoo of possible situations, there are some specially interesting cases in which both $R$ and $R'$ are ``ring'' like regions sharing the same topology. This occurs for $n=(d-2)/2$, in which both regions have non trivial $\pi_{(d-2)/2}$. This possibility only appears in even dimensions. In particular, in $d=4$, we have both Wilson and 't Hooft loops violating duality of the ring $R$ and its complement $R'$, which is also a ring. In this case, the groups $G_a$ and $G_b$ of complementary (simple linked) regions are not only dual to each other, but there is also a natural isomorphism arising from transporting the non-local operators from $R$ to $R'$ by deformations.
These situations have the additional interest that, for some geometries, one can construct conformal transformations mapping the complementary regions as we discuss further below.

It is an interesting program to understand what kind of non-local algebras could more generally appear in different topologies, under some simple assumptions such as that the local algebras have trivial centers and the sectors are homotopically transportable. For example,  regions with knots would not be necessarily equivalent to other topologically equivalent ones without them. This general analysis may reveal interesting new cases depending on the assumptions. 

A different simple example that is not covered by ordinary gauge theory is the case of higher helicity fields. The free (linearised) graviton is described by a field $h_{\mu\nu}$ with gauge invariance $h_{\mu\nu}\rightarrow h_{\mu\nu}+\partial_\mu \xi_\nu+\partial_\nu \xi_\mu$. Gauge invariant operators are generated by the curvature tensor $R_{\alpha\beta\gamma\delta}$, which is conserved in all its indices. This conservation should give rise to flux operators across two-dimensional surfaces that are not locally generated operators on the one-dimensional boundary. However, in contrast to the gauge theories described above, the non-local operators are indexed with space-time indices. We might anticipate from this observation a breaking of the Lorentz symmetry for a HD net.

\section{Entropic order parameters}

In this section, we seek to construct entropic order parameters that capture the physics of generalized symmetries. In other words, we want to find natural entropic order parameters that can distinguish, from a unified perspective, different phases of QFT's. In particular, these entropic order parameters should capture the confinement, Higgs, and massless phases in gauge theories.

Taking as a starting motivation the confinement phase, it is well-known that the Wilson loop of a fundamental representation was initially devised as an order parameter for it \cite{PhysRevD.10.2445}. The expectation value of such a Wilson loop can decay exponentially fast with the area of the loop. This behavior is indicative of confinement since it implies a linear quark-antiquark potential. On the other hand, a perimeter law scaling of the Wilson loop excludes the possibility of confinement.

However, in theories such as QCD, whose matter content includes charged fields in the fundamental representation, the Wilson loop has a perimeter law even if quarks are confined. Moreover, even in the absence of charged matter fields, the same holds for the Wilson loops in the adjoint representation. It seems no coincidence that these two examples concern precisely line operators that are locally generated in the ring.

These observations trigger the following hypothesis. The right order parameters in QFT's, characterizing the phase of some generalized symmetry, should be the appropriate non-additive operators discussed in the previous section. These are the operators that violate Haag duality in the appropriate region. In turn, the right entropic order parameters should be those able to capture the physics of the non-additive operators. The objective of this section is to build on this hypothesis, define the right entropic order parameters, and study them in different phases of several systems.

We start by setting the idea that non-additive operators are the right order parameters on firmer ground. To do so, we argue that for any general QFT it is not possible to construct a loop order parameter displaying an area law by employing only operators that are locally generated in the ring. We can wave only a sub-perimeter law behavior (perimeter law, or even a constant law). This implies that the existence of a confinement order parameter requires a non locally generated operator, with the associated failure of the additivity property for ring-like regions.

Associated with this failure of additivity, and as discussed in the previous section, there will be multiple choices of nets of algebras. We will use this multiplicity to define natural ``blind'' entropic order parameters, which do not rely on a particular operator, but just on the algebraic structure of the net of algebras. We will show that such entropic order parameters can be defined both for order parameters, such as intertwiners and Wilson loops, and for disorder parameters, such as twists and 't Hooft loops. It turns out that both perspectives, order vs. disorder, are related through the entropic certainty relation \cite{Magan:2020ake}.

We will finally use all these tools to analyze different known phases in QFT's, such as spontaneous symmetry breaking scenarios, Higgs and confinement phases, and conformal ones.

\subsection{An area law needs non locally generated operators} \label{confina-additivo}
     
\begin{figure}[t]
\begin{center}  
\includegraphics[width=0.4\textwidth]{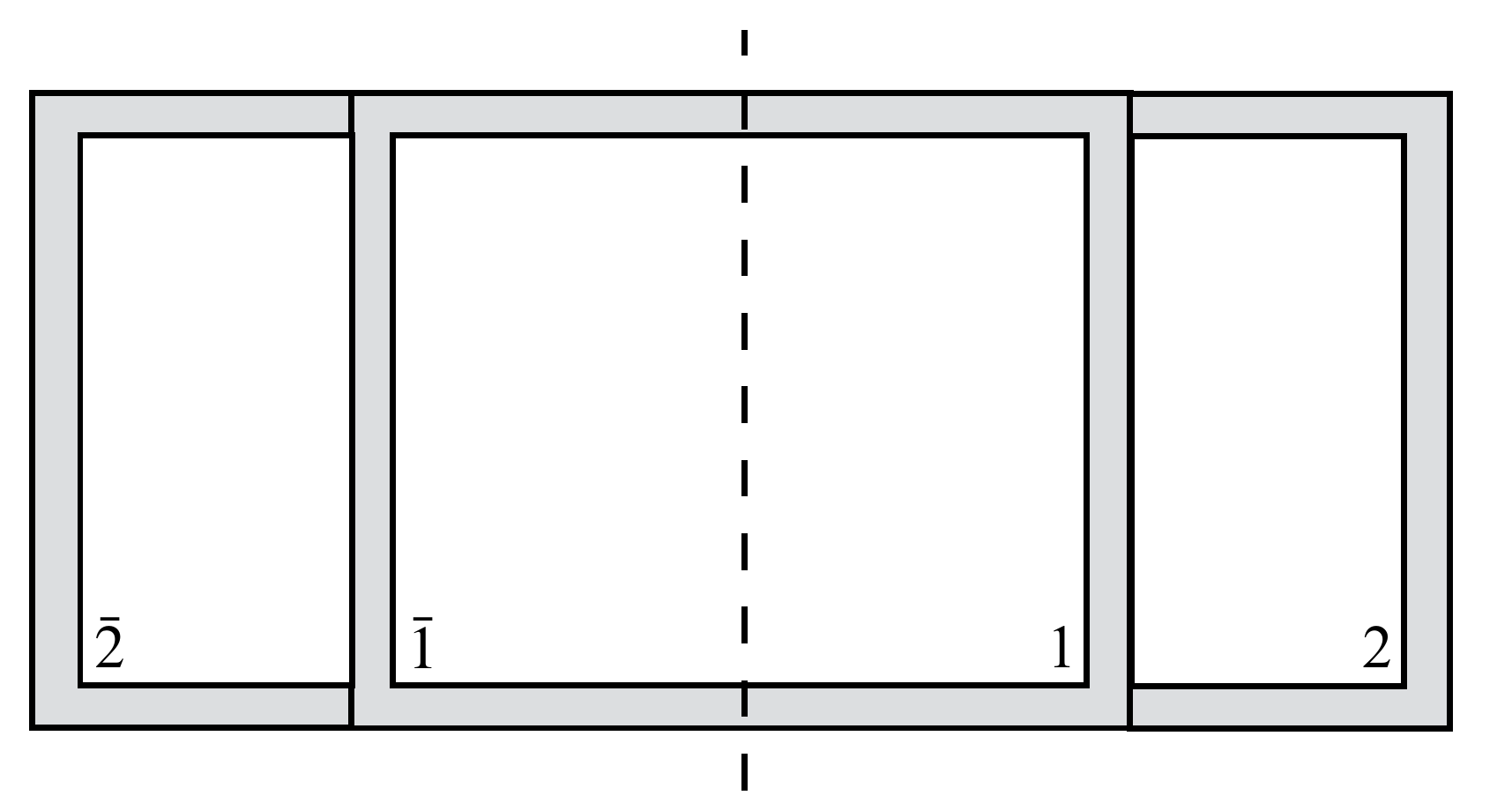}
\captionsetup{width=0.9\textwidth}
\caption{The construction of Bachas that shows the convexity of the quark-antiquark potential \cite{Bachas:1985xs}.}
\label{fig-bachas}
\end{center}  
\end{figure}

Let us first recall that the exponential decay of the expectation value of an (appropriately smeared) line operator is always bounded from below by an area law \cite{Bachas:1985xs}. To explain this, we refer to figure \ref{fig-bachas}, which shows an arrangement of four partially superimposed rectangular loop type operators. We call these loops $W_{1\bar{1}}$, $W_{2\bar{2}}$, $W_{1 \bar{2}}$ and $W_{2\bar{1}}$. These are formed by products of two half-loops on the right half plane (labeled $W_1$ and $W_2$) reaching just to the plane of reflection drawn with the dashed line,  and their reflected CRT images (labeled $W_{\bar{1}}$ and $W_{\bar{2}}$ respectively). The application of reflection positivity in the Euclidean version, or CRT positivity in real time,\footnote{CRT positivity is associated with the CRT (or CPT) symmetry of QFT \cite{Witten:2018lha}.  It is also known as wedge reflection positivity \cite{Casini:2010bf}, or Rindler positivity. It follows from Tomita-Takesaki theory (see \cite{haag2012local}), but it holds more generally in any purification of a quantum system \cite{Casini:2010nn}. In the present case, this is the purification of the vacuum state in the Rindler wedge by the global vacuum state in the whole space. These inequalities are manifestations of the positivity of the Hilbert space scalar product.} leads to the inequality
\be
\langle (a W_1+b W_2)(a^* W_{\bar{1}}+b^* W_{\bar{2}})\rangle \ge 0\,,
\ee 
for any constants $a$ and $b$. In particular, the determinant of the matrix of coefficients for this quadratic expression is positive
\be
\langle W_{1\bar{2}}\rangle \langle W_{2\bar{1}}\rangle \le   \langle W_{1\bar{1}}\rangle \langle W_{2\bar{2}}\rangle \,. \label{calcu}
\ee
Writing $\langle W\rangle = e^{-V(x,y)}$, with $x,y$ the two sides of the rectangle, it follows from this relation  and the analogous one produced by reflecting in the $x$ axis that the potential $V(x,y)$ must be concave in the two variables
\be\label{conc}
\partial_x^2 V(x,y)\le 0\,,\hspace{.8cm} \partial_y^2 V(x,y)\le 0\,. 
\ee
Then, the slopes $\partial_x V(x,y)$ and $\partial_y V(x,y)$ never increase. As these slopes cannot become negative (hence making the loop expectation value increase without bound), they will converge to a fixed non-negative value in the limit of large size. If the coefficient of $x y$ in $V(x,y)$ for large size is non zero we have an area law. If it is zero, we have a sub-area law behavior. No loop operator expectation value can go to zero faster than an area law $\langle W\rangle\sim e^{-c\,A}$ as the size tends to infinity. This calculation holds for any loop, whether locally or non locally generated in the ring, provided they are locally generated in the plane, and they are CRT reflection symmetric.\footnote{For non locally generated loops, the half loops have to close in the plane of reflection.} The derivation can be justified more rigorously in a lattice model \cite{Bachas:1985xs}.

Now we focus on loop operators formed additively in a ring. We want to show the expectation values of these operators cannot decay faster than a perimeter law $\langle W\rangle \ge c \, e^{-\mu R}$, where $R$ is the loop radius. The presentation will be rather sketchy. In appendix \ref{perimeter} we expand on how these arguments could be made mathematically precise.
 
It is more convenient to use circular loops for our present purposes. As the loops are locally generated, we can imagine forming a partial operator $W(l_1,l_2)$ in an arc $(l_1,l_2)$ of the ring of longitudinal size $l=l_2-l_1$. The idea is that we now construct a loop of a certain size, not by increasing the size of a smaller loop as above, but by increasing the size of an operator in an arc until the arc closes into a ring.

\begin{figure}[t]
\begin{center}  
\includegraphics[width=0.25\textwidth]{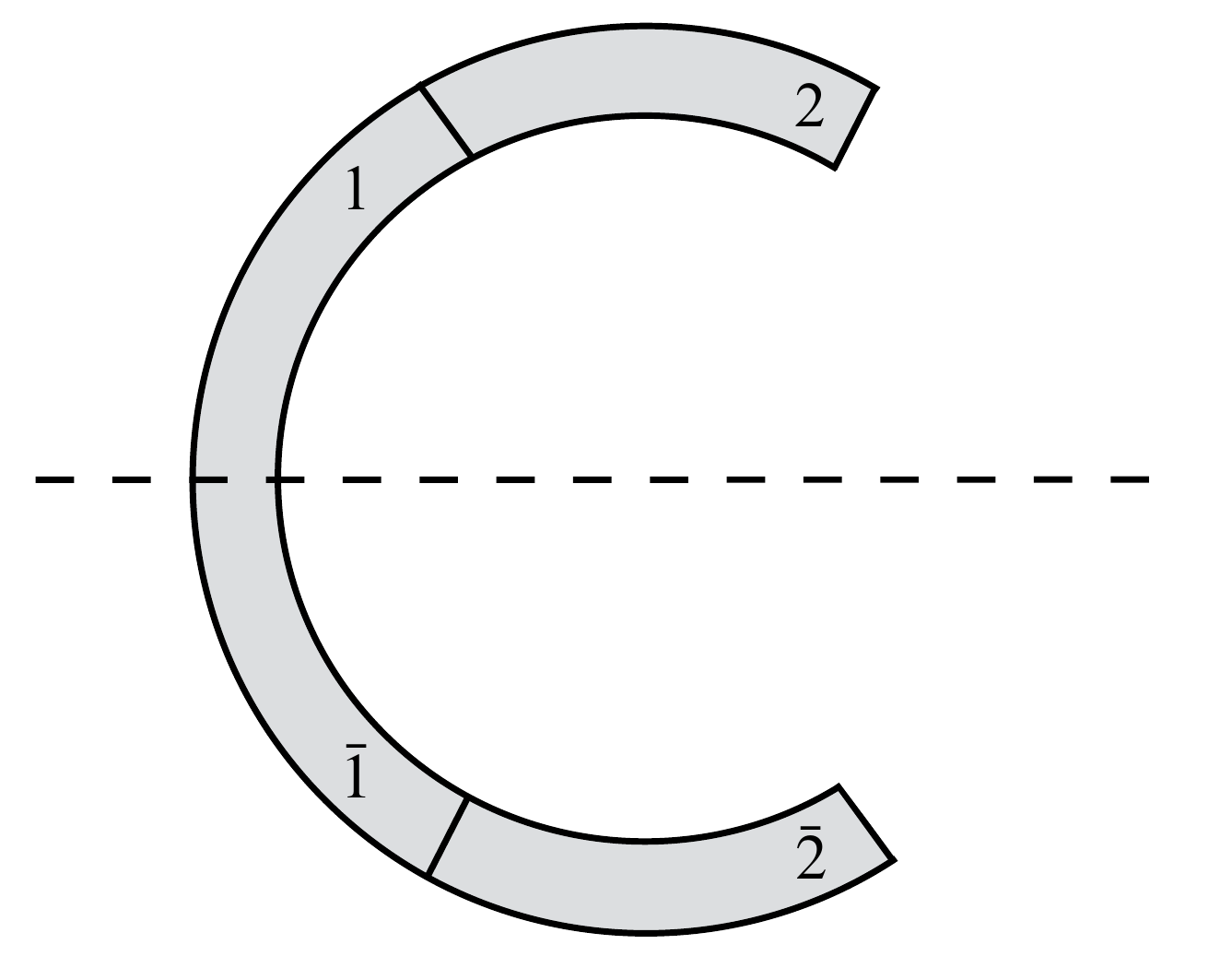}
\captionsetup{width=0.9\textwidth}
\caption{Reflection positivity applied to line operators along angular portions of a ring. Operators $1$ and $2$ reach to the reflection plane, and $\bar{1}$ and $\bar{2}$ are their CRT reflections.}
\label{fig-sectors}
\end{center}  
\end{figure}

We assume rotational invariance and define the potential
\be
\langle W(l_1,l_2)\rangle =e^{-V(l)}\,.\label{hh}
\ee
We can use CRT positivity again in this case, as shown in figure \ref{fig-sectors}. The result is
\be 
 V''(l)\le 0\,.
\ee   
Therefore, the slope of $V'(l)$ is non increasing and
\be
\langle W \rangle \ge e^{-2 \pi R \,V'(0)-V(0)} \,. \label{boundito}
\ee
If the loops are formed as products of small pieces in a rotationally symmetric way, we can form loops of larger radius starting with the same cross-section. For such a sequence of loops of different radius, we have the same value $V'(0)$, independently of the radius.  Equation \eqref{boundito} gives us a perimeter law, or more precisely a sub-perimeter law behavior. In particular, this excludes the possibility of an area law or any other law where the potential increases faster than linearly in the perimeter.  

Applying the same idea to the case of non locally generated loop operators in the ring fails. The reason is that we cannot define the partial (non-closed) line operators. Using a non-gauge-invariant Wilson line introduces several problems when some gauge fixing is chosen. If we do not fix the gauge, the expectation value of this line operator is zero, and the potential infinity. This prevents the calculation to give any useful bound. When we have charged operators to define the relevant Wilson line, and consequently the loop operator can be broken into pieces, its expectation value cannot decay faster than a perimeter law.

We expect similar results to hold for spherical shells of different dimensions $k$, based on the same arguments. General operators should have an area or sub-area law behavior $V\lesssim R^{k+1}$, generalizing \eqref{conc}. Besides, additive operators will not be able to display ``area law'' ($V\sim R^{k+1}$). Their expectation values will be restricted to have a sub-perimeter law behavior (here $V\lesssim R^{k}$), and they will not be appropriate order parameters. The argument leading to this statement should parallel the one deriving \eqref{boundito}.

\subsection{Definition and general properties of entropic order parameters}

Let's now move towards constructing sensible entropic order parameters signaling the presence or absence of generalized symmetries in QFT's. As discussed until now, these generalized symmetries manifest themselves through the violation of duality and additivity for different regions with different topologies. In all the cases considered, these violations have been related to definite non-additive ``order'' operators $\{a\}$ (such as intertwiners, Wilson loops, or high dimensional generalizations), and definite non-additive ``disorder'' operators $\{b\}$ (such as twists and 't Hooft loops). These operators close specific algebras, typically given by fusion algebras associated with a certain group of generalized symmetries.

Since both order and disorder operators, when properly chosen, generate self-consistent von Neumann algebras $\{a\}$ and $\{b\}$, the first obvious information theoretic notion coming to our minds is the von Neumann entropy
\bea\label{VN}
S_{\{a\}}(\omega):=-\textrm{Tr}_{\{a\}} \,\omega\log\omega \nonumber\\
S_{\{b\}}(\omega):= -\textrm{Tr}_{\{b\}} \,\omega\log\omega \;,
\eea
where for a given algebra $\mathcal{M}$, the symbol $\textrm{Tr}_{\mathcal{M}}$ instructs us to take the canonical trace associated with it \cite{ohya2004quantum}. The symbol $\omega$ on the right-hand-side denotes the reduced density matrix of the state $\omega$ in the corresponding algebra. Although we will also implicitly study these quantities, it turns out there are better suited entropic order parameters for the characterization of symmetries. In particular, we notice that the quantities \eqref{VN} have an unpleasant dependence on the choice of the non-local operators. 

To motivate new order parameters, we follow the logic described in \cite{Magan:2020ake}. Let us first remind that for a finite $d$-dimensional Hilbert space, the von Neumann entropy can be written equivalently as
\be
S_{\mathcal{M}} (\omega) := -\mathrm{Tr}_{\mathcal{M}}\left(\omega\log\omega\right) =\log d - S_{\mathcal{M}}\left(\omega\mid \tau\right)\,, \label{sre}
\ee
where $\tau={\bf 1}/d$ is the maximally mixed density matrix, and $S_{\mathcal{M}}\left(\omega\mid \tau\right)$ is a quantity known as relative entropy. The relative entropy is defined for two quantum states in the same algebra and, in the finite dimensional scenario, is defined by
\be
S_{\mathcal{M}}\left(\omega\mid\varphi\right):=\mathrm{Tr}_{\mathcal{M}}\left(\omega\left(\log\omega-\log\varphi\right)\right)\,.\label{re_def}
\ee
Intuitively, relative entropy is a measure of distinguishability between the two underlying states. Relation \eqref{sre} expresses that the uncertainty measured by the von Neumann entropy is also measured by the distance between the given state and the state with maximal uncertainty. Since there is a minus sign in the previous relation, the higher the relative entropy in \eqref{sre}, the smaller the uncertainty of $\omega$ on $\mathcal{M}$.

Using relative entropy has several advantages. First, relative entropy displays monotonicity under general quantum channels and restrictions onto subalgebras \cite{ohya2004quantum}. We will use this property frequently in the applications below. Second, relative entropy is well-defined across different types of algebras, including type-III von Neumann algebras appearing in QFT. This will ease the application to QFT, as it avoids many potential issues just from the start.

Finally, using relative entropy suggests certain generalizations. Notice that in \eqref{sre}, the maximally mixed state $\tau$ can be equivalently written as the composition\footnote{A state over an algebra $\mathcal{A}$ is a linear positive normalized functional $\omega: \mathcal{A} \rightarrow \mathbb{C}$ from the operators to the complex numbers, giving expectation values. In this sense, we use the terminology of states as functions on operators.} of $\omega$ with a map $E:\mathcal{M}\rightarrow \mathds{1}$, defined by $E (m) := \frac{1}{d}\textrm{Tr} (m)\mathbf{1}$. Rewriting the relative entropy in \eqref{sre} as
\be 
S_{\mathcal{M}}(\omega\vert\omega\circ E)
\ee
suggests a couple of generalizations to this notion of uncertainty. First, the map $E (m) := \frac{1}{d}\textrm{Tr} (m)\mathbf{1}$ is one example of a whole space of maps of the type $E:\mathcal{M}\rightarrow \mathds{1}$, as we describe below. Second, instead of using the identity as the target algebra, we could choose any subalgebra $\mathcal{N}\subset\mathcal{M}$. The relevant maps $E : \mathcal{M}\rightarrow \mathcal{N}$ are called conditional expectations \cite{ohya2004quantum,takesaki}. They are positive, linear, and unital maps from an algebra $\mathcal{M}$ to a subalgebra $\mathcal{N}$. They leave the target algebra invariant, and they further satisfy the following bimodule property
\be
\hspace{-1mm} E\left(n_{1}\,m\,n_{2}\right)=n_{1}E\left(m\right)n_{2}\,,\hspace{3mm} \forall m\in\mathcal{M},\,\forall n_{1},n_{2}\in\mathcal{N}.\label{ce_def_prop}
\ee
These maps are the mathematical definition of what restricting our observational abilities means (see \cite{ohya2004quantum} for an extensive review). Examples of conditional expectations are tracing out part of the system
\be 
\mathcal{F}:=\mathcal{O}\otimes\mathcal{A}\,,\hspace{7mm} E(O\otimes A):= \frac{\textrm{Tr}(A)}{d_{\cal A}}\, O\otimes 1_{\mathcal{A}}\;,
\ee
or retaining the neutral part ${\cal O}$ of an algebra ${\cal F}$ under the action of a certain symmetry group $G$
\be 
E(F):= \frac{1}{G}\sum\limits_{g\in G}\,\tau_g \,F\,\tau_{g}^{-1}\,, \hspace{5mm} F \in \mathcal{F}\,.
\ee
From a general standpoint, if $\mathcal{M}= \mathcal{N}\vee \mathcal{Q}$ is the algebra generated by $\mathcal{N}$ and certain subalgebra $\mathcal{Q}$, we say the conditional expectation ``kills'' $\mathcal{Q}$.

For our purposes, one of the most important properties of these maps, which we will use continuously, is that they can be used to lift a state in the subalgebra $\mathcal{N}$ to a state in the larger algebra $\mathcal{M}$. This lift is defined as
\be 
 \omega_{\mathcal{N}}\mapsto \omega_{\mathcal{N}}\circ E \,.
\ee
The generalization we are seeking is thus
\be 
S_{\mathcal{M}}(\omega \mid \omega\circ E)\,\hspace{7mm} E : \mathcal{M}\rightarrow \mathcal{N}\,.
\ee
If $\mathcal{M}= \mathcal{N}\vee \mathcal{Q}$, this quantity measures the uncertainty of $\mathcal{Q}$ in the state $\omega$, given the knowledge of $\mathcal{N}$. The fact that side correlations with the algebra $\mathcal{N}$ are taken into account in this quantity will be very important for QFT applications.

We now define the order-disorder entropic parameters in the following manner. If the algebra of non-additive operators $\{a\}$ lives in a certain region $R$, this provides us with a natural inclusion of algebras
\be 
 {\cal A}_{\textrm{add}}(R)\subseteq  {\cal A}_{\textrm{max}}(R)={\cal A}_{\textrm{add}}(R)\vee \{a\}\;.
\ee
Associated with this inclusion, we should have a space of conditional expectations
\be 
E : {\cal A}_{\textrm{max}}(R)\rightarrow {\cal A}_{\textrm{add}}(R)\,,
\ee
leading to the following entropic order parameter
\be 
S_{{\cal A}_{\textrm{max}}(R)}(\omega \mid \omega\circ E)\,.
\ee
This entropic order parameter was considered in \cite{Casini:2019kex} for the case of global symmetries, inspired by ideas in Ref. \cite{Longo:2017mbg}. See also \cite{Furuya:2020wxf}.

A parallel story works for the "disorder" operators $\{b\}$. We remind that they belong to the complementary region $R'$, and they provide us with the following inclusion of algebras
\be 
{\cal A}_{\textrm{add}}(R')\subseteq  {\cal A}_{\textrm{max}}(R')={\cal A}_{\textrm{add}}(R')\vee \{b\}\;,
\ee
with its associated space of conditional expectations
\be 
E' :  {\cal A}_{\textrm{max}}(R')\rightarrow {\cal A}_{\textrm{add}}(R')\,,
\ee
and the following entropic disorder parameter
\be 
S_{{\cal A}_{\textrm{max}}(R')}(\omega \mid \omega\circ E')\,.
\ee
An important difference between these order parameters and the von Neumann entropies \eqref{VN} is that these relative entropies are purely geometric objects depending only on the region $R$ and the state $\omega$ (typically the vacuum state in our applications). They do not depend on a particular choice of operators in $R$. The other difference is that they include the side correlations between the order-disorder operators with the appropriate additive algebras. This turns out to be important, as we now describe.

The order-disorder algebras do not commute between themselves. These commutation relations are completely fixed, as shown in equation \eqref{ccrel}. These commutation relations imply fundamental uncertainty principle type bounds between the two algebras. Such implications, which arise from quantum complementarity, can be accommodated in the entropic formulation. This problem was considered in detail in \cite{Magan:2020ake}, inspired by the result in \cite{Casini:2019kex}. To analyze it, we notice there is a natural way to understand quantum complementarity in this context.

Given a generic inclusion of von Neumann algebras $\mathcal{N}\subseteq \mathcal{M}$ and a conditional expectation $E:\mathcal{M}\rightarrow\mathcal{N}$, there is a natural complementarity diagram
\bea\label{cdia}
\mathcal{M} & \overset{E}{\longrightarrow} & \mathcal{N}\nonumber \\
\updownarrow\prime\! &  & \:\updownarrow\prime\\ \label{ecr_diagr}
\mathcal{M}' & \overset{E'}{\longleftarrow} & \mathcal{N}'\,.\nonumber 
\eea
In this diagram, going vertically takes the algebras ${\cal M}$ and ${\cal N}$ to its commutants ${\cal M}'$ and ${\cal N}'$ respectively. Going horizontally in the arrow direction means restricting to the target subalgebra. If ${\cal M}={\cal N}\vee {\cal Q}$ and ${\cal N}'={\cal M}'\vee \tilde{{\cal Q}}$, then $E$ kills  $\mathcal{Q}\subset\mathcal{M}$, and the \textit{dual conditional expectation} $E '$ kills  $\tilde{\mathcal{Q}}\subset\mathcal{N}'$.

Notice that while $\mathcal{N}$ commutes with ${\cal M}'$, the algebras $\mathcal{M}$ and $\mathcal{N}'$ do not commute with each other.  The only operators that do not commute with each other are the ones in $\mathcal{Q}$ and $\tilde{\mathcal{Q}}$. These are the ones killed by the appropriate conditional expectations. These algebras $\mathcal{Q}$ and $\tilde{\mathcal{Q}}$ are called complementary observable algebras (COA) \cite{Magan:2020ake}. They generalize the notion of complementary operators to operator algebras.

As a simple example, take $\mathcal{M}$ as the Abelian algebra $\mathcal{X}$ generated by the position operator. Then choose a conditional expectation that kills the full $\mathcal{M}=\mathcal{Q} = \mathcal{X}$. In other words $E: \mathcal{X}\rightarrow\mathds{1}$. The complementarity diagram becomes in this case 
\bea
\mathcal{X} & \overset{E}{\longrightarrow} & \mathds{1}\nonumber \\
\updownarrow\prime \!\! &  & \,\updownarrow\prime\\
\mathcal{X} & \overset{E'}{\longleftarrow} & \mathcal{X}\vee\mathcal{P}\,.\nonumber 
\eea
where $\mathcal{P}$ is the algebra generated by the momentum operator. As expected, we conclude that the algebras $\mathcal{X}$ and $\mathcal{P}$ forms a COA.

The case of interest to us, which will be a recurring theme in the following sections, concerns the one associated with order-disorder parameters in QFT. This is
\bea\label{cdiaor}
{\cal A}_{\textrm{add}}(R)\vee \{a\} & \overset{E}{\longrightarrow} &{\cal A}_{\textrm{add}}(R)\nonumber \\
\updownarrow\prime \!\! &  & \,\updownarrow\prime\\
{\cal A}_{\textrm{add}}(R')& \overset{E'}{\longleftarrow} & {\cal A}_{\textrm{add}}(R')\vee \{b\}\,.\nonumber 
\eea

Let us now continue with the general case \eqref{cdia}. Associated with such a diagram, we have an entropic order parameter for the upper side, namely $S_{{\cal M}}(\omega\vert \omega\circ E)$, and an entropic order parameter for the lower side, namely $S_{{\cal N}'}(\omega\vert \omega\circ E')$. In \cite{Magan:2020ake} the following relation between those was derived\footnote{After the publication of this paper in the arXiv database, this relation was derived in the context of type III von Neumann algebras \cite{hollands2020variational}, setting the certainty relation in a firm mathematical ground for QFT. We have also become aware of the work \cite{xu2020relative}, dealing with a similar type of ideas in a more mathematically oriented context.} for a pure global state $\omega$
\be\label{gcer}
S_{\mathcal{M}}\left(\omega|\omega\circ E\right)+S_{\mathcal{N}'}\left(\omega|\omega\circ E'\right)=\log\lambda\;.
\ee
The constant $\lambda \geq 1$ is a certain fixed number called the algebraic index of the conditional expectation $E$, which is equal to the one corresponding to the dual conditional expectation $E'$ and independent of the state $\omega$. In the examples of this paper $\lambda=|G|$, the order of a finite symmetry group $G$. This relation was called entropic certainty relation in \cite{Casini:2019kex}, where it was first derived for the case of global symmetry groups.  
The original references defining the algebraic index are \cite{Jones1983,KOSAKI1986123,longo1989}. On the study of the index in a generic inclusion of finite-dimensional algebras, see \cite{teruya,giorlongo}.

In the proof of the previous relation, a fundamental step is to understand the space of conditional expectations $E$ in a  generic inclusion of algebras $\mathcal{N}\subset\mathcal{M}$. The study of this space has been carried out in different scenarios. To our knowledge, the first references studying it were \cite{umegaki1,umegaki2,umegaki3,umegaki4}. In the context of the inclusion of factors in type III algebras, it was studied in \cite{longo1989}. In \cite{Magan:2020ake}, it was recently analyzed from a somewhat more physical perspective. Intuitively, the result is the following. Let us denote the space of conditional expectations from ${\cal M}$ to ${\cal N}$ as $C({\cal M},{\cal N})$. If the target algebra ${\cal N}$ has a center spanned by projectors $P_j^{{\cal N}}$, then any $E\in C({\cal M},{\cal N}) $ is of the form
\be 
E (m)=\bigoplus_{j=1}^{z_{\mathcal{N}}}E_{j}(m_{j})\,, \hspace{5mm} m_{j}:=P_{j}^{\mathcal{N}}mP_{j}^{\mathcal{N}} \in \mathcal{M}_j\;,
\ee
where $E_{j}\in C(\mathcal{M}_j,\mathcal{N}_j)$. But now the inclusion $\mathcal{N}_j\subset\mathcal{M}_j$ has a trivial center in the target algebra. Then one can prove that for such inclusions, the space $C(\mathcal{M}_j,\mathcal{N}_j)$ is isomorphic to the space of states in the relative commutant $\mathcal{M}_j\cap \mathcal{N}'_j$. Notice that if the original target algebra is a factor (it has a trivial center) and the relative commutant is also trivial, the space $C(\mathcal{M}_j,\mathcal{N}_j)$ contains only one element.

This is typically the case for the applications in continuum QFT, where the algebras have trivial centers, and the relative commutant (for the cases we are considering in this paper) is also trivial.\footnote{The argument for this triviality is that, in order to commute with the full local algebra of the region, an operator has to be localized in its boundary, but they do not exist operators localizable in a $d-2$ dimensional surface. There is not enough room in the boundary of the region to smear a field operator (operator-valued distributions) such as to construct a well-defined linear operator in the Hilbert space.} The conditional expectations are unique in this context and can be computed using the non-local operators themselves. We will show the explicit form of these conditional expectations in the next sections. In the lattice approximation of the continuum QFT, there might be several choices of conditional expectations. However, it is expected that the relative entropies for fixed states should approach in the continuum limit to the same values independently of the particular choice.

Returning to the general case, it can also be proven, for finite von Neumann algebras \cite{umegaki1,umegaki2,umegaki3,umegaki4}, that there always exists a conditional expectation preserving the trace. In this case
\be
\textrm{tr}(m)=\textrm{tr}(E(m))\;.
\ee
For these trace preserving conditional expectations, we have that \cite{Casini:2019kex}
\be 
S_{\mathcal{M}}\left(\omega|\omega\circ E\right)=S_{\mathcal{M}}\left(\omega\circ E\right)-S_{\mathcal{M}}\left(\omega\right)\;,
\ee
where on the right-hand side we have von Neumann entropies. Important examples of trace preserving conditional expectations are group averages. They will play a central role below, although the framework is more general. 
Therefore, one can think that the relative entropy order parameter, in the continuum limit of a QFT, is a well-defined version of the subtraction of two cutoff entropies. It also teaches us that this subtraction is monotonic with the region. Implied by this same monotonicity, the continuum limit in a cutoff theory is independent of the details of the cutoff.

Applying the generic certainty relation \eqref{gcer} to the case of entropic order-disorder parameters, characterized by the diagram \eqref{cdiaor}, we obtain for a pure global state $\omega$
\be
S_{{\cal A}_{\textrm{add}}(R)\vee \{a\}}\left(\omega|\omega\circ E\right)+S_{{\cal A}_{\textrm{add}}(R')\vee \{b\}}\left(\omega|\omega\circ E'\right)=\log\lambda\;.\label{ede}
\ee
From this expression and the positivity of relative entropy, we obtain the individual bounds
\be 
S_{{\cal A}_{\textrm{add}}(R)\vee \{a\}}\left(\omega|\omega\circ E\right)\leq\log\lambda\,, \hspace{.6cm}
S_{{\cal A}_{\textrm{add}}(R')\vee \{b\}}\left(\omega|\omega\circ E'\right)\leq\log\lambda\;.\label{uubb}
\ee
Moreover, from the entropic certainty relation \eqref{ede}, other bounds can be obtained using monotonicity of relative entropy under quantum channels or algebra restrictions. For example, for the applications in QFT described below, we will typically use that the certainty relation, together with monotonicity of relative entropy, implies
\be
S_{ \{a\}}\left(\omega|\omega\circ E\right)\leq S_{{\cal A}_{\textrm{add}}(R)\vee \{a\}}\left(\omega|\omega\circ E\right)\leq\log\lambda -S_{ \{b\}}\left(\omega|\omega\circ E'\right)\;,\label{boce1}
\ee
and similarly
\be 
S_{ \{b\}}\left(\omega|\omega\circ E'\right)\leq S_{{\cal B}_{\textrm{add}}(R)\vee \{b\}}\left(\omega|\omega\circ E'\right)\leq\log\lambda -S_{ \{a\}}\left(\omega|\omega\circ E\right)\;.\label{boce2}
\ee
From the physical point of view, the certainty relation includes uncertainty relations for the non-commuting and non-local operators $\{a\}$ and $\{b\}$. This is the reason why the two upper bounds \eqref{uubb} cannot be realized at the same time, which would imply maximal expectation values for these operators simultaneously. But instead of being an inequality, it is an equality. In this sense, it can be thought of as a generalization of the equality of entropies for complementary algebras in a global pure state to the case of non-commuting algebras \cite{Magan:2020ake}. 

How the certainty relation is realized is easily guessed in certain limits. If the expectation values of the operators $\{a\}$ tend to zero (for example when the region $R$ is very thin and fluctuations of $a$ are large), the states $\omega$ and $\omega\circ E$ will not be easily distinguished. These states differ precisely in that $\omega\circ E$ assigns exactly zero expectation value for any non-local operator $a$. The order parameter, in this case, goes to zero while the dual disorder one saturates the bound
\be
S_{{\cal A}_{\textrm{add}}(R)\vee \{a\}}\left(\omega|\omega\circ E\right)\rightarrow 0\,,\hspace{.7cm}S_{{\cal A}_{\textrm{add}}(R')\vee \{b\}}\left(\omega|\omega\circ E'\right)\rightarrow \log\lambda\,.
\ee
Analogously, when the expectation values of $b$ goes to zero we have 
\be
S_{{\cal A}_{\textrm{add}}(R)\vee \{a\}}\left(\omega|\omega\circ E\right)\rightarrow \log \lambda\,,\hspace{.7cm}S_{{\cal A}_{\textrm{add}}(R')\vee \{b\}}\left(\omega|\omega\circ E'\right)\rightarrow 0\,.
\ee
Interesting physical information about the phase of the theory can be learned from the geometric setup in which these limits are achieved, and from the subleading terms in these expressions.    

Summarizing, symmetries are associated with the appearance of two different algebras for the same region, the additive algebra ${\cal A}_{\textrm{add}}$ and the maximal algebra ${\cal A}_{\textrm{max}}$. Entropic order parameters of these symmetries are then naturally suggested from the fact that two different states ($\omega$ and $\omega\circ E$) can be produced out of the vacuum for the same algebra ${\cal A}_{\textrm{max}}$. Furthermore, the relative entropy between these states satisfies a remarkable relation that ties the statistics of complementary dual non-local operators of complementary regions. Another natural geometric order parameter would be produced by expectation values of the standard non-local operators described in section \ref{standard}. Indeed, the definition of these operators involves the full algebra in the region $R$. We will not study these standard operators further in the present paper. 

\subsection{Improving bounds by including additive operators}
\label{improving}

In the previous section, with the help of the relative entropy, we have defined entropic order parameters associated with a generic region $R$ and its complement $R'$. The relative entropy is a measure of distinguishability, and the order parameter essentially compares the vacuum state on the additive and non-additive algebras. If we restrict attention to just one arbitrary set of non-local operators, the relative entropy is smaller than the optimal, and the certainty relation is not saturated. However, a natural way to approximate the values of the entropic order parameters is to start from the knowledge of the expectation values of some algebraically closed set of non-local operators. From the certainty relation and the monotonicity of relative entropy we have
\be
S_{\{a\}}\left(\omega|\omega\circ E\right)\le  S_{{\cal A}_{\textrm{add}}(R)\vee \{a\}}\left(\omega|\omega\circ E\right) \le \log \lambda - S_{\{b\}}\left(\omega|\omega\circ E'\right),\label{tiyo}
\ee
for particular algebras of non local operators $\{a\}$ in $R$ and $\{b\}$ in $R'$. These upper and lower bounds are simple functions of the expectation values of the non-local operators. The bounds \eqref{tiyo} can be improved by searching for the best non-local operators in each region, that is, the ones with higher expectation values. 
Including the information on the additive algebra, or, equivalently, on the multiple possible choices of non-local operators in $R$, should also improve the bounds obtained from \eqref{ede}.
In this section, before treating specific QFT examples, we want to get more intuition about these features. Concretely, we will look at how including larger sets of approximately decoupled non-local operators in a region (which is a particular way of including additive operators) improves the bounds in a characteristic way. In this scenario, the bound approximates saturation exponentially fast in the number of decoupled operators.

\begin{figure}[t]
\begin{center}  
\includegraphics[width=0.35\textwidth]{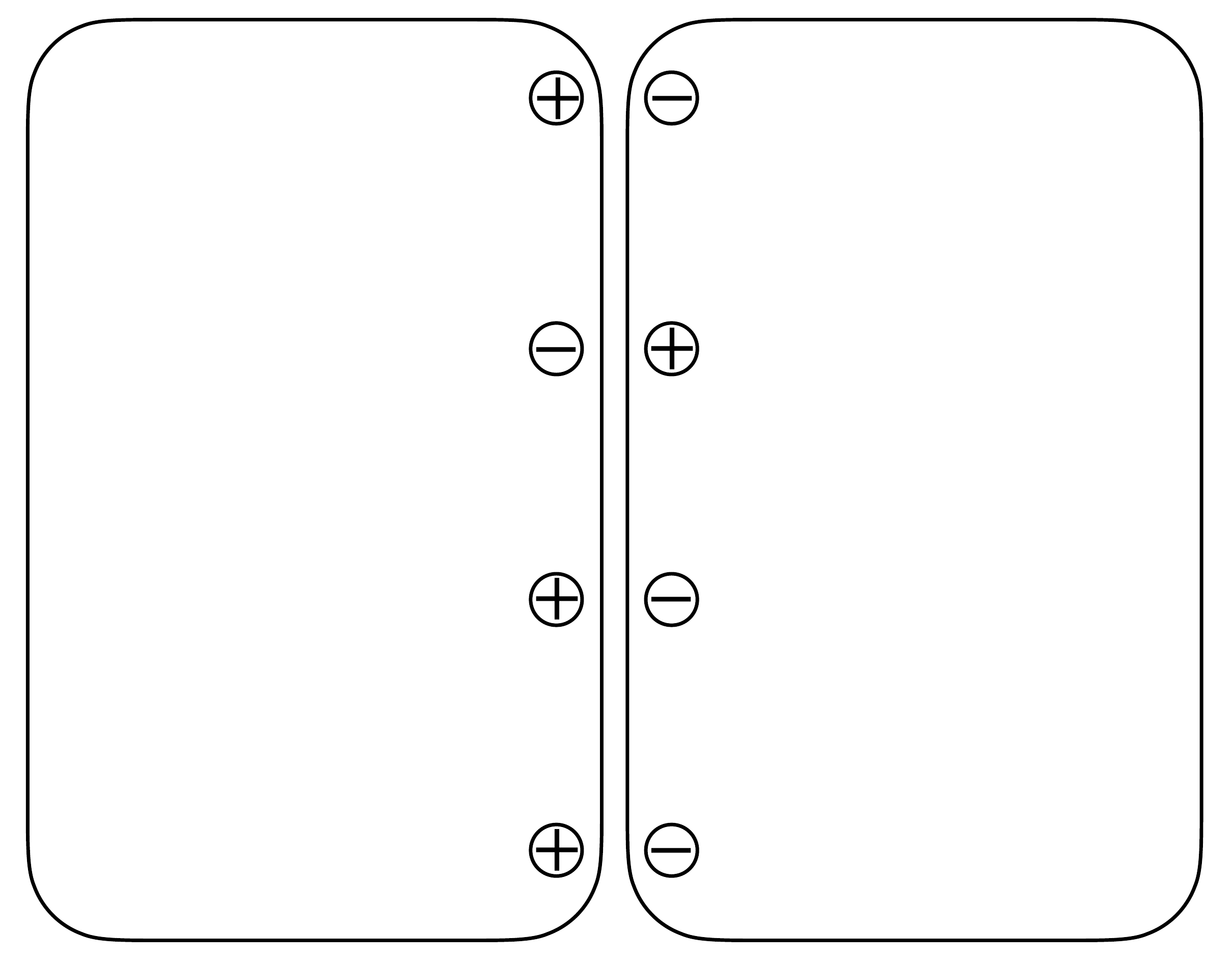}
\captionsetup{width=0.9\textwidth}
\caption{Approximately decoupled intertwiners formed by charge-anti-charge operator pairs along the boundaries of the two nearby regions.}
\label{hojales}
\end{center}  
\end{figure}

To motivate the calculation below, consider the order parameter for the case of global symmetry. We have a region formed by two single component regions as in figure \ref{hojales}, and we consider the case in which these regions are very near to each other. In the limit where they touch each other, the non-trivial twists are squeezed between the two regions and their expectation values go to zero. Then, we expect that the relative entropy over the intertwiners tends to the maximal value $\log |G|$ \cite{Casini:2019kex}. We want to have a handle on how this limit is approached. To produce a lower bound to this relative entropy, we can compute it in the algebra of any of the (unitary) intertwiners drawn along the surface in figure \ref{hojales}. These, however, will have some specific expectation values which may differ significantly from the maximal one $\langle {\cal I}\rangle=1$. Large expectation values $\langle {\cal I}\rangle \lesssim 1$ are needed to achieve a maximal relative entropy $\log |G|$. To find such improved thick intertwiners may be a complex task. The idea is then to use an algebra of many intertwiners along the surface, that can be taken to be uncorrelated to each other in good approximation, to improve the bound. We must remark that we are not enlarging the number of independent intertwiners, which is indeed impossible. There is only one independent intertwiner per irreducible representation of the symmetry group. In other words, all these ``new'' intertwiners that we are adding can be written, in a non-trivial way, in terms of the original intertwiners and some particularly chosen additive operators of the two regions. It is the contribution of the additive operators which greatly improves the result of the computation.
 
More concretely, let us take dual sets of non-local operators belonging to dual Abelian groups $G_a$ and $G_b$. 
They obey $b a b^{-1}=\chi_a(b) a$, where $\chi_a$ is the character of the representation $a$. As we will describe in detail below, the corresponding conditional expectation killing the operators $a$ is $E(x)=|G|^{-1}\sum_b b x b^{-1}$. We are interested in understanding the sub-leading terms in the approach of the relative entropy $S_{{\cal A}_{\textrm{add}}(R)\vee\{a\}}(\omega|\omega\circ E)$  to the saturation limit $\log |G|$.

With the knowledge of the expectation values of the operators $\{a\}$, we can produce a lower bound just restricting the calculation of the relative entropy to this subalgebra. As explained in more detail in the next section, it is convenient to use the orthogonal projectors 
\be
P_b:=\frac{1}{|G|}\sum_{a\in G_a} \chi_a(b)^* a\,, 
\ee
which are labelled by the elements $b\in G_b$. The expectation values of these projectors give a probability distribution $p_b:=\langle P_b\rangle$. The action of the conditional expectation is $E(P_b)=\mathbf{1}/|G|$. The relative entropy over the subalgebra $\{a\}$ is then the classical relative entropy between the probability distribution $(p_b)_{b \in G_b}$ and the uniform distribution, i.e.
\be
S_{\{a\}}(\omega|\omega\circ E)=H(p\,|\,|G|^{-1})=\log\,|G|-H(p)\,.
\ee
  
Let us now suppose that we have in ${\cal A}_{\textrm{add}}\vee\{a\}$ a sequence of $N$ commuting subalgebras of non local $\{a^i\}$, $i=1,\cdots,N$, and assume that they have identically distributed uncorrelated expectation values. That is, we have the Abelian algebra $G_a^{\otimes N}$ generated by the projectors $P_b^i$ with 
\be
\omega( P_{b_1}^1 \cdots P_{b_N}^N)= p_{b_1}\cdots p_{b_N}\,. 
\ee
Since $h^{-1} P_b h=P_{h b_1}$ for all $h \in G_b$, the state $\omega\circ E$ in this algebra is given by the mixture of $N$ states $\omega_h$ ($h\in G_b$) as
\bea
\omega\circ E= \frac{1}{|G|}\sum_{h\in G} \omega_h\,,\label{parte}\\
 \omega_h( P_{b_1}^1 \cdots P_{b_N}^N)= p_{h b_1}\cdots p_{h b_N}\,.
\eea
Then, each of the states $\omega_h$ just permutes the probability distributions $(p_b)\mapsto (p_{hb})$.

We want to understand the relative entropy in the limit of large $N$. We can reason using the operational interpretation of the classical Shannon relative entropy \cite{Vedral}.  Let us call $\beta=P_{b_1}^1 \cdots P_{b_N}^N$ to a generic sequence of projectors labelled by the sequence $(b_1,\cdots , b_N)$. The state $\omega$  has a probability distribution that is highly peaked around the set of projectors where $\beta$ contains each $P_b$ a number $\sim p_b N$ of times. Let us call $\beta_p$ to this set of projectors where the fraction of each $P_b$ is determined by the probability distribution $p(b)$.  
According to Shannon's theorem, this set of projectors form a fraction $\sim e^{N H(p)}$ of the total number $e^{N\log|G|}$ of projectors, and they all have the same probability $\omega(\beta_p)\sim e^{-N H(p)}$. On the other hand, consider the set of projectors $\beta_q$ corresponding to sequences having a different fraction of elementary projectors determined by the probability distribution $q(b)$. These are $\sim e^{N H(q)}$ projectors, and they have a probability
\be
\omega(\beta_q)\sim e^{-N H(q)} e^{-N H(q,p)}\,,
\ee    
where $H(q,p)$ is the classical relative entropy between the ``one site'' probability distributions \cite{Vedral}. The probability of the $\beta_q$ sequences is exponentially suppressed compared to $\beta_p$ by the relative entropy between these distributions. As a consequence, of all states in the mixture \eqref{parte} only the one with $h=1$ will have significant overlap with $\omega$, and we expect that the relative entropy converges to the maximum value $\log |G|$ exponentially fast.

To see this in more detail, we have to compute
\be
S_{\{a\}^{\otimes N}} (\omega|\omega\circ E)=\sum_\beta \omega(\beta) \log\left(\frac{\omega(\beta)}{|G|^{-1}\sum_{h\in G} \omega_{h}(\beta)}\right)=\log |G|-\sum_\beta \omega(\beta) \log\left(1+\sum_{h\neq 1} \frac{\omega_{h}(\beta)}{\omega(\beta)}\right)\,.
\ee 
For $h\in G_b$ and $p$ a probability distribution over  the group elements, we call $hp$ to the distribution $hp(b)=p(hb)$. Separating the sum in the different probability distributions $q$ we have 
\be
\log |G|-S_{\{a\}^{\otimes N}} (\omega|\omega\circ E)\sim \sum_q e^{-N H(q,p)} \log\left(1+\sum_{h\neq 1} e^{-N(H(q,h p)-H(q,p))}\right)\,.
\ee
This already tells us that, assuming $h p\neq p$ for all $h$, we will have an exponential decay with $N$. If $hp=p$ for some $h$, the $a$ operator expectation values do not break all the $b$ symmetries, and cannot provide the maximum value $\log|G|$ asymptotically. We exclude this case.

Evidently, in the large $N$ limit, for each $q$, the sum inside the logarithm will be dominated by a particular $h$ such that $H(q,h p)=H(h^{-1} q,p)$ is minimal. Replacing this sum by the best $h$, the saddle point approximation gives 
\be\label{qprobs}
\tilde{q}^{(p,h)}(b)=\frac{\sqrt{p(b) p(h b)}}{\sum_b \sqrt{p(b) p(h b)}}\,.
\ee
This distribution satisfies $H(\tilde{q}^{(p,h)},p)=H(\tilde{q}^{(p,h)},h p)$, and therefore we have   
\be
\log |G|-S_{\{a\}^{\otimes N}} (\omega|\omega\circ E)\sim   e^{-N \,\textrm{min}_{h\neq 1} H(\tilde{q}^{(p,h)},p)} \,.\label{saddle}
\ee
If we have many uncorrelated algebras $\{a^i\}$ with different expectation values, we expect a sum over the single classical relative entropies in the exponent,
\be
\log |G|-S_{\{a\}^{\otimes N}} (\omega|\omega\circ E)\sim   e^{-\sum_i \,\textrm{min}_{h\neq 1} H(\tilde{q_i}^{(p_i,h)},p_i)} \,.\label{saddle11}
\ee

Let us check the previous calculation in the simplest scenario of $G_b=\mathbb{Z}_{2}$. In this case we have only one non local operator $a\neq \mathbf{1}$, and the orthogonal projectors are simply $P_\pm=(\mathbf{1}\pm a)/2$. Then we have   
\be  
\langle a\rangle =1-2 p\,, \hspace{.5cm} \langle P_-\rangle=p\,,\hspace{.5cm} \langle P_+\rangle=1-p  \,.\label{exp_value_int}
\ee
A collection of uncorrelated equally distributed $a_i $ gives us the state over the multiple projectors
\be \label{state1}
p_{s_{1}\cdots s_{N}}^{\omega}=\frac{1}{2^{N}}\prod_{\alpha=1}^{N}\left(1+s_{\alpha}(1-2 p)\right)\,,
\ee
where the subindices $s_\alpha$ takes the values $  \pm $. The conditional expectation $E$ acts by averaging a projector $P_+^1 P^2_-\cdots$ with the one where the indices have changed signs. Hence the state transformed with the conditional expectation is
\be \label{state2}
p_{s_{1}\cdots s_{N}}^{\omega\circ E}=\frac{1}{2^{N+1}}\left[\prod_{\alpha=1}^{N}\left(1+s_{\alpha}(1-2 p)\right)+\prod_{\alpha=1}^{N}\left(1-s_{\alpha}(1-2 p)\right)\right]\,.
\ee
The relative entropy is 
 \be 
S(\omega\mid\omega\circ E)=S(\omega\circ E)-S(\omega)= -\sum_{k=0}^{N}\dbinom{N}{k}r_{k}\log(r_{k})- N H(p)\,,
\label{for}\ee
with
\bea
H(p)\!\!\! &=& \!\!\!-\left( 1-p\right) \log \left( 1-p\right) -p\log p\,,\\ 
r_k \!\!\! &=& \!\!\! \frac{1}{2} \left( 1-p\right) ^{k} p^{N-k}+\frac{1}{2}\left( 1-p\right) ^{N-k} p^{k}\,.
\eea
The sum cannot be done analytically. Numerically, the formula \eqref{for} agrees with the saddle point calculation \eqref{saddle} to leading order, and gives a sub-leading logarithm term,
\be
S(\omega\mid\omega\circ E)\sim \log(2)- k \,e^{- N  H\left((\frac{1}{2},\frac{1}{2})| (p,1-p)\right)-\frac{1}{2} \log N}\,.\label{vvvvv}
\ee
  
There is an interesting corollary of this calculation. For a gapped theory with topological contributions to the entropy, the topological term in the mutual information appears when the distance between the regions is smaller than the correlation length. However, we can also take regions separated from each other more than the correlation length, and the topological term will still appear in this case if the regions are exponentially large. This is because we can achieve saturation if we can take a sufficiently large number of uncorrelated intertwiners even if the intertwiner's expectation values are exponentially small.

\subsection{Global symmetry} \label{gs}

In this section, we study entropic order parameters for theories with global symmetries.  The algebraic structure of such theories was described previously in section \eqref{global} for the case of an unbroken symmetry. Summarizing that discussion, in these theories we focus on the global symmetry invariant algebra ${\cal O}$. There is a breaking of Haag duality in a pair of disconnected balls due to the existence of intertwiners \eqref{Ir}. These are neutral operators formed by a charged operator on one region and a compensatory anti-charge operator on the other region. There is one independent intertwiner per irreducible representation. In the complementary region, which has the topology of a spherical shell $S^{d-1} \times \mathbb{R}$, there is also a breaking of duality due to the existence of twists operators \eqref{tt}, which implement the symmetry group locally and do not commute with the intertwiners \eqref{ccrel}. 

We take two disconnected ball like regions $R_1$ and $R_2$ and their complement, the ``shell''  $S=(R_1 R_2)'$. As we described above, there are two choices for the algebra of $R_1 R_2$, namely, the additive algebra $\mathcal{O}(R_1R_2)$ and the additive algebra plus the intertwiners $\mathcal{O}(R_1R_2)\vee \mathcal\{{\cal I}\}$. Similarly, we have two choices of algebras for $S$, the additive one $\mathcal{O}(S)$ and the additive one plus the (invariant) twists $\mathcal{O}(S)\vee \{\tau\}$. The quantum complementarity diagram reads
\bea \label{cuadro}
\mathcal{O}(R_1R_2)\vee \{\mathcal{I}\} & \overset{E_{\cal I}}{\longrightarrow} & \mathcal{O}(R_1R_2)\nonumber \\
\updownarrow\prime\! &  & \:\updownarrow\prime\\
\mathcal{O}(S) & \overset{E_{\tau}}{\longleftarrow} & \mathcal{O}(S)\vee \{\tau\}\,.\nonumber 
\eea
We have called $E_{\cal I}$ and $E_{\tau}$ to the dual conditional expectations to emphasize they "kill" the intertwiners and twists, respectively. The associated entropic certainty relation relating the dual order/disorder parameters is \cite{Casini:2019kex}
\be \label{cglobal}
S_{\mathcal{O}(R_1R_2)\vee \{\mathcal{I}\}}\left(\omega|\omega\circ E_\mathcal{I}\right)+S_{\mathcal{O}(S)\vee \{\tau\}}\left(\omega|\omega\circ E_\tau\right)=\log \vert G\vert\;.
\ee
In this case, the algebraic index is $\vert G\vert$, the order of the group. With the help of the enlarged theory, ${\cal F}$ that includes the charged operators, the intertwiners relative entropy can also be written as \cite{Casini:2019kex}
\be
 S_{\mathcal{O}(R_1R_2)\vee \{\mathcal{I}\}}\left(\omega|\omega\circ E_\mathcal{I}\right)=I_{{\cal F}}(R_1,R_2)-I_{\cal O}(R_1,R_2)\,,
\label{mutualdif}\ee 
where on the right-hand-side appears the difference between the mutual information on the two models. 

These order parameters were studied at length in \cite{Casini:2019kex}, with a focus on the topological contributions to the entropy. This corresponds to the limit in which $R_2$ is nearly complementary to $R_1$. In this case, $S_{\mathcal{O}(R_1R_2)\vee \{\mathcal{I}\}}\left(\omega|\omega\circ E_\mathcal{I}\right)$ can be understood as the difference between regularized von Neumann entropies between the model with charges ${\cal F}$ and the orbifold ${\cal O}$ in a single ball $R_1$.   Here, our focus is on how these relative entropies behave as order/disorder parameters for phases of the theory. To understand this, the opposite geometry of far separated balls will also be useful, as well as the understanding of the subleading terms at saturation, which are important in distinguishing phases.

We recall that for global symmetries, the orbifold theory ${\cal O}$ comes together with the theory ${\cal F}$ containing charged operators. If the state $\omega$ is not invariant under the symmetry (whether the case of spontaneous symmetry breaking (SSB) or simply any non-invariant state, even in a compact space), acting with the algebra ${\cal O}$ on such a state produces a Hilbert space representation containing charged states and charged operators. If the symmetry is completely broken, the representation includes all charged operators in ${\cal F}$ as well. For simplicity, we will treat the case where the symmetry is completely broken. Partial symmetry breaking can be dealt in a similar way.

For the theory ${\cal F}$, as well as for its subalgebra ${\cal O}$, there are several different possibilities of algebra inclusions to play with in addition to  $\mathcal{O}(R_1R_1)\subseteq \mathcal{O}(R_1R_2)\vee \{\mathcal{I}\}$. In particular, the subnet ${\cal O}$ does not satisfy duality for single component regions $R$ since the commutant includes charged operators in the complement $R'$. Instead, starting from the inclusion ${\cal O}(R)\subset {\cal F}(R)$ for a ball $R$ we get the complementarity diagram
 \bea \label{dlp}
{\cal F}(R) & \overset{E_{ \psi}}{\longrightarrow} & \mathcal{O}(R)\nonumber \\
\updownarrow\prime\! &  & \:\updownarrow\prime\\
\mathcal{F}(R') & \overset{E_{\tau}}{\longleftarrow} & \mathcal{F}(R')\vee G\,.\nonumber 
\eea
Notice that in this equation the elements of the global group $G$ could be replaced by twists implementing the group operations on the sphere $R$. We have called $E_\psi$ to the conditional expectation (implemented by the twists), which eliminates the charged operators $\psi$ in the ball. Dually, $E_{\tau}$ is implemented by the charged operators, and its effect is to set to zero the twists (and the elements of the global group $G$).  The relevant entropic order parameters satisfy the certainty relation
\be
S_{{\cal F}(R)}(\omega|\omega\circ E_{{\psi}})+S_{{\cal F}(R')\vee G}(\omega|\omega\circ E_{\tau}) =\log|G|\,. \label{cr12}
\ee

The relative entropy $S_{{\cal F}(R)}(\omega|\omega\circ E_{{\psi}})$ is an order parameter for symmetry breaking. It goes to zero for regions much smaller than the symmetry breaking scale, and to $\log|G|$ for regions much larger than this scale \cite{Casini:2019kex}. It vanishes if $\omega$ is invariant under the group, i.e., an unbroken symmetry vacuum. This is because the two states that it compares are identical in this case. Then, according to \eqref{cr12}, we must have $S_{{\cal F}(R')\vee G}(\omega|\omega\circ E_{\tau}) =\log|G|$  for an invariant state $\omega$. It is a consequence of the non-local operators $U(g) \in G$ having maximal expectation values $\langle U(g)\rangle=1$ in this case.
 
In the rest of this section we will focus on the relative entropy for disconnected regions $S_{\mathcal{O}(R_1R_2)\vee \{\mathcal{I}\}}\left(\omega|\omega\circ E_\mathcal{I}\right)$ as an order parameter for symmetry breaking. This is the one having a natural generalization to higher form symmetries. Depending on the correlation functions of charged operators, rather than on the one-point functions, the information it contains is of a more local nature than the single ball order parameter. We will revisit this issue in section \ref{remarks}.
 
The generalization of \eqref{cuadro} for a scenario that may display symmetry breaking is obtained by thinking of all algebras as subalgebras of the theory ${\cal F}$. This gives us
 \bea \label{cuadro1}
\mathcal{O}(R_1R_2)\vee \{\mathcal{I}\} & \overset{E_{\cal I}}{\longrightarrow} & \mathcal{O}(R_1R_2)\nonumber \\
\updownarrow\prime\! &  & \:\updownarrow\prime\\
\mathcal{F}(S)\vee G & \overset{E_{\tau}}{\longleftarrow} & \mathcal{F}(S)\vee \{\tau\}\vee G\,,\nonumber 
\eea
and certainty relation
 \be \label{cglobal11}
S_{\mathcal{O}(R_1R_2)\vee \{\mathcal{I}\}}\left(\omega|\omega\circ E_\mathcal{I}\right)+S_{\mathcal{F}(S)\vee \{\tau\}\vee G}\left(\omega|\omega\circ E_\tau\right)=\log \vert G\vert\;.
\ee
This equation generalizes  \eqref{cglobal} to general pure states $\omega$ on ${\cal F}$, and coincides with \eqref{cglobal} when $\omega$ is a globally invariant state.

We want to understand the behavior of these relative entropies in different phases. To this end, we start by explaining how to put bounds on the entropic order parameters. These bounds come from entropies on the algebras of non-local operators, and they depend on their expectation values. These bounds follow by applying the generic relations~(\ref{boce1}) and~(\ref{boce2}) to the appropriate complementarity diagram. For example, for the diagram~(\ref{cglobal}) they become
\be
S_{ \{\mathcal{I}\}}\left(\omega|\omega\circ E\right)\leq S_{\mathcal{O}(R_1R_2)\vee \{\mathcal{I}\}}\left(\omega|\omega\circ E\right)\leq\log\vert G\vert -S_{ \{\tau\}}\left(\omega|\omega\circ E'\right)\;,\label{boceg1}
\ee
and
\be 
S_{ \{\tau\}}\left(\omega|\omega\circ E'\right)\leq S_{\mathcal{O}(S)\vee \{\tau\}}\left(\omega|\omega\circ E'\right)\leq\log\vert G\vert -S_{ \{\mathcal{I}\}}\left(\omega|\omega\circ E\right)\;.\label{boceg2}
\ee
After describing the generic formulas for the bounds, we will analyze the order parameters in symmetric and broken symmetry phases.
 
\subsubsection{Bounds from operator expectation values}
 
The algebra of a fixed set of twists was described in section \ref{global}. We have
\be
\tau_g \tau_h=\tau_{gh}\,,\hspace{1cm}U(g) \tau_h U(g)^{-1}=\tau_{g h g^{-1}}\,,
\ee
where $U(g)$ are the global symmetry operators. For non-Abelian groups, the twists are not (in general) observables since they are not invariant under the symmetry group. We can produce invariant combinations by averaging over the conjugacy classes\footnote{We will be writing the formulas for a finite group, but similar formulas appear for Lie groups. As usual, we just need to change the sums by integrals with a measure given by the Haar measure on the Lie group manifold.}
\be
\tau_{c}=\sum_{h\in c} \tau_{h}\;. 
\ee
These invariant twists generate a closed algebra with fusion coefficients given by the fusion of conjugacy classes (see equation \eqref{fusc}). This is an Abelian algebra.  For computing the entropy in this Abelian algebra, it is convenient to choose a diagonal basis for the fusion.  Define the projectors
\be
P_r=\frac{d_r}{|G|} \sum_g \chi_r^*(g) \tau_g \,.\label{centry}
\ee
They are labeled by the set of inequivalent irreducible representations. We have 
\be
P_r P_{r'}=\delta_{r,r'}\, P_{r}\,, \hspace{1cm}\sum_ r P_r=\mathbf{1}\, . 
\ee
We are interested in the conditional expectation $E_{\tau}$ in this algebra, which is dual to the conditional expectation associated with the intertwiners. This kills all twists except the identity \cite{Casini:2019kex}
\be 
E_{\tau}(\tau_g)=\delta_{g,1} \mathbf{1}\;. \label{tw_ce}
\ee
Therefore, from \eqref{centry} we get
\be
E_{\tau}(P_r)=\frac{d_r^2}{|G|}\, \mathbf{1}\,.
\ee
Any state $\omega$ in this Abelian algebra is determined  by the probabilities on the different projectors 
\be
q_r:=\langle P_r \rangle\,. 
\ee
Then, the relative entropy becomes
\be  \label{entryt}
S_{\{\tau\}} (\omega| \omega\circ E_\tau)= \log|G|+\sum_r q_r \log q_r -\sum_r q_r \log d_r^2\,.
\ee
This is the formula for the twists relative entropy presented in \cite{Casini:2019kex}. We have $S_{\{\tau_c\}}(\omega| \omega\circ E_\tau)\in [0,\log |G|]$, which follows by taking into account $|G|=\sum_r d_r^2$. If the group is Abelian, this is equal to $\log|G|-S_{\{\tau_c\}}$, where $S_{\{\tau_c\}}$ is the von Neumann entropy on the (invariant) twist algebra.

We want now to compute the relative entropy in the algebra generated by the intertwiners. In \cite{Casini:2019kex}, this computation was approached by enlarging the theory to include charged operators. Although the quantitative final result is bound to be the same, it proves useful for later use in gauge theories to have an approach based only on neutral operators. We should then work only with the algebra of intertwiners. As shown in appendix \ref{app2}, this is again an Abelian algebra where the intertwiner ${\cal I}_r$ corresponding to the irreducible representation $r$ is represented by the fusion matrix $N^{(r)}$, whose entries are given by the coefficients of the fusion of representations $N^{(r)}_{r'r''}=n_{r r'}^{r''}$. These matrices commute with each other because of the commutativity of the fusion algebra. Therefore they can be simultaneously diagonalized. It is convenient to label the basis vectors, where the matrices are simultaneously diagonalized, by the conjugacy classes $c$ of the group. The number of conjugacy classes is the same as the number of irreducible representations.  The expression of the matrices on this basis is given by
\be
N^{(r)}\equiv \delta_{c,c'}\, \chi_r(c) = \textrm{diag}(\chi_r(c)) \,.  
\ee
This results in the right algebra because the product of characters satisfy
\be 
\chi_r(c)\chi_{r'}(c)=\sum_{r''} n^{r''}_{r r'} \chi_{r'}(c)\;.
\ee
From this formula and the orthogonality of characters, it is simple to derive
\be
N^{(r)}_{r_1 r_2}=\sum_{c} \frac{d_{c}}{|G|}\, \chi_{r_1}(c) \chi_{r}(c)  \chi_{r_2}(c)^*\,,
\ee
where $d_{c}$ is the number of elements in the conjugacy class $c$. The matrix $S$ that diagonalizes the matrices $N^{(r)}$, namely $S^{-1} N^{(r)} S=\textrm{diag}(\chi_r(c))$, is 
\be
S_{r_1, c}=  \sqrt{\frac{d_{c}}{|G|}}\, \chi_{r_1}(c)\,,\hspace{1cm} S^{-1}_{ c,r_1}=  \sqrt{\frac{d_{c}}{|G|}}\, \chi_{r_1}(c)^*\,.
\ee
This matrix is unitary. In terms of these matrices\footnote{There is a dual version of these formulas in the basis of representations, where twists are diagonal while the intertwiners are non-diagonal.  These are based on the fusion rules of the conjugacy classes of the group \eqref{fusc} (see \cite{francesco2012conformal} page 404).} 
\be
N^{(r)}_{r_1 r_2}=\sum_{c}  S_{r_1,c} \, \frac{S_{r,c}}{S_{1,c}} \, S_{r_2,c}^*\,,
\ee
which is Verlinde's formula for group representations. The projectors over the diagonal are now
\be\label{proyI}
P_{c}=\sum_r  \frac{d_{c}}{|G|}\, \chi_r^*(c)   {\cal I}^{(r)}\equiv\sum_r  \frac{d_{c}}{|G|}\, \chi_r^*(c)   N^{(r)}\,,
\ee
as follows from the orthogonality of the characters. The conditional expectation kills all non-trivial intertwiners
\be
E_\mathcal{I} (N^{(r)})=\delta_{r,1}\, \mathbf{1}\,,
\ee
and therefore
\be
E_\mathcal{I}(P_{c})=\frac{d_{c}}{|G|} \, \mathbf{1}\,.
\ee
Defining the probabilities of the minimal projectors
\be
q_{c}=\langle P_{c}\rangle\,,
\ee
the associated relative entropy becomes
\be
S_\mathcal{I}(\omega|\omega\circ E_\mathcal{I})=\log|G|+\sum_{c} q_{c} \log q_{c} -\sum_{c} q_{c}\, \log d_{c}\,. \label{esta1}
\ee
The interest of formulas \eqref{entryt} and \eqref{esta1} is that they relate entropic quantities with operator expectation values. 

The certainty relation plus monotonicity of relative entropy imply the bounds~(\ref{boceg1}) and~(\ref{boceg2}). These can be used now to constrain the order parameters by using the operator expectation values. For example we have that
\be
\log|G|+\sum_{c} q_{c} \log q_{c} -\sum_{c} q_{c}\, \log d_{c}\le S_{\mathcal{O}(R_1R_2)\vee \{\mathcal{I}\}}\left(\omega|\omega\circ E_\mathcal{I}\right)\le -\sum_r q_r \log q_r +\sum_r q_r \log d_r^2\,.
\ee 

\subsubsection{Symmetric phase}
Below we will only consider finite groups. For Lie groups see \cite{Casini:2019kex}. Let us consider the case of a CFT and two nearly complementary regions given by a ball $R_1$ of radius $R$ and the complement $R_2$ of a ball of radius $R+\epsilon$.  The two regions are separated by a thin spherical shell of width $\epsilon$. 
As argued in  \cite{Casini:2019kex}, the twist expectation values in a thin shell in a symmetric phase will be exponentially small in the area $A$ of the shell
\be
\langle \tau_c \rangle \sim e^{-c_0 \frac{A}{\epsilon^{d-2}}}\,, \hspace{7mm} \tau_c \neq \mathbf{1}\,.
\ee
The twist does not change the vacuum state in the bulk of the region, and only a local contribution from the boundary arises. The constant $c_0$ depends on the twist and its precise form (smearing).

With this information, we can use the formulas in the previous section to put bounds to the entropic order parameters. We obtain
\be
q_r=\langle P_r\rangle =\frac{d_r}{|G|} \sum_g \chi_r^*(g) \langle\tau_g\rangle = \frac{d_r^2}{G}+ k_r\, e^{-c_0 \frac{A}{\epsilon^{d-2}}}\equiv \bar{q}_r+\delta \bar{q}_r\,,\label{centry1}
\ee
where $k_r$ is some constant that does not play an important role in what follows. We have defined $\bar{q}_r := \frac{d_r^2}{G}$, and from the normalization of probability we have that $\sum_{r}\delta \bar{q}_r =0$. This implies the last term in \eqref{entryt}, coming from the non-Abelian nature of the group, does not contribute to the correction. Introducing such probabilities in \eqref{entryt} and expanding in $\delta \bar{q}_r$, one finds that the first order correction vanishes.  The first correction appears at second order in $\delta \bar{q}_r$
\be 
S_{\{\tau_c\}} (\omega| \omega\circ E_\tau) \sim  \, e^{- 2 c_0 \frac{A}{\epsilon^{d-2}}}\;.\label{h1}
\ee

To understand the origin of this formula in terms of the intertwiners, we can apply the ideas of section \ref{improving}. To have such an exponential approach to saturation, one possibility is to find intertwiners with expectation values exponentially close to $1$. Instead of that, we just need to locate many independent intertwiners along the surface of the shell. The number of almost uncorrelated intertwiners will be proportional to the area. In the limit of small separation $\epsilon$, they can be separated enough between themselves to have small cross-correlations. Using the results of the previous section \ref{improving} this implies
\be
S_{\{\mathcal{I}\}}(\omega|\omega\circ E_\mathcal{I})\sim \log |G|- k\, e^{-c_1 \, \frac{A}{\epsilon^{d-2}}}\,,\label{h2}
\ee
for some constant $k$. Using (\ref{h1}-\ref{h2}) and the bounds~(\ref{boceg1}) and~(\ref{boce2}) that follow from the certainty relation, we get for the shell order parameter 
\be
S_{\mathcal{O}(S)\vee \{\tau_c\}}\left(\omega|\omega\circ E_\tau\right) \sim e^{-c \frac{A}{\epsilon^{d-2}}}\,,
\ee
where $ 2 c_0<c< c_1$.  The exact value of the dimensionless coefficient $c$ depends on the theory, which, in general, is not easy to compute.

To match the terminology of line operators in gauge theories, we will call this a {\sl perimeter law} for the shell order parameter, although it seems strange in the present case. We are taking the convention that for a non-local operator associated with a region of topology $S^k\times \mathbb{R}^{d-k-1}$, with a fixed width and large radius $R$ of the sphere $S^k$, we will call perimeter law to the case in which the expectation value decays with the exponential of $R^k$. As we have discussed above, this is the maximal rate of decay for locally generated operators. We will call area law if it decays exponentially with $R^{k+1}$. This is the maximal possible decay rate of non-local operators.

From this result, the certainty relation \eqref{cglobal}, by means of~(\ref{boceg1}) and~(\ref{boce2}), gives for the intertwiner order parameter the form
\be\label{sir}
S_{\mathcal{O}(R_1R_2)\vee \{\mathcal{I}\}}\left(\omega|\omega\circ E_\mathcal{I}\right)\sim \log |G|- k \, e^{- c\frac{A}{\epsilon^{d-2}}}  \;,
\ee
for some constant $k$. This is the dual version of the perimeter law. 

In the vein of adding more intertwiners to improve the lower bound to the intertwiner relative entropy, one could ask why not locate intertwiners all over the region, and not only along the boundary. Doing this we obtain a number of intertwiners proportional to the volume of the region $N\sim V/\epsilon^{d-1}$. However, for intertwiners localized all over the region, with one charged operator on each side of the shell, the expectation values decay fast as we get further from the shell. In fact, these expectation values will decay at least with a certain power, which depends on the conformal dimension of the charged operator.
\be 
\langle {\cal I}_r\rangle\sim \frac{\epsilon^{2 \Delta}}{R^{2\Delta}}\;.
\ee
The relative entropies over each individual intertwiner will go as $\sim \frac{\epsilon^{4 \Delta}}{R^{4\Delta}}$. We can make a rough estimate considering these intertwiners uncorrelated (which is hardly the case in a CFT) using \eqref{saddle11}. To overcome the area term, we need $\Delta< 1/4$, which is beyond the unitarity bound for $d\ge 3$. Therefore, the right scaling already arises just by considering a set of intertwiners close to the entangling surface, consistent with the twist result.

In the opposite geometric limit, in which the distance $L$ between the regions $R_1$ and $R_2$ is large compared to their size $R$, the roles of intertwiners and twists are qualitatively interchanged. In this scenario, the intertwiners decay as
\be 
\langle {\cal I}\rangle\sim \frac{R^{2 \Delta}}{L^{2\Delta}},
\ee
where $\Delta$ is the scaling dimension of the lowest dimensional operator charged under the group. This gives us a lower bound to the intertwiner order parameter. Noticing that corrections to saturation on the relative entropy only come at second order, this will scale as $(R/L)^{4 \Delta}$. However, from \eqref{mutualdif} and the results about the mutual information for well-separated regions \cite{Cardy.esferaslejanas} in a CFT, we get that this is the correct scaling, 
\be  
S_{\mathcal{O}(R_1R_2)\vee \{\mathcal{I}\}}\left(\omega|\omega\circ E_\mathcal{I}\right)\sim  \frac{R^{4 \Delta}}{L^{4\Delta}}=e^{- 4 \Delta\log (L/R)}\,.
\ee
This is a {\sl logarithmic law}. We will also say this is a {\sl sub-area law}, while the area law, in this case, would correspond to an exponent linear in $L$. 

This enforces the following behavior for the twists
\be
S_{\mathcal{O}(S)\vee \{\tau_c\}}\left(\omega|\omega\circ E_\tau\right) \sim \log |G|- k\, e^{- 4 \Delta\log (L/R)}\;,
\ee
and it implies the potentially useful fact that, in principle, it is possible to obtain the leading charged conformal dimension of the theory from the behavior of the expectation value of the best wide twists.  

For two balls in a CFT, these relative entropies are functions of the cross-ratio $\eta$, and we have quite different unrelated behavior in the two limits of $\eta\rightarrow 0$ and $\eta\rightarrow 1$. There is (apparently) nothing that connects the behavior in the two limits.

In the massive case, the changes are quite obvious. The area law for thin shells is the same except if we take $\epsilon m\gg 1$. In this case, the intertwiner contribution is exponentially suppressed, and we get
\be
S_{\mathcal{O}(R_1R_2)\vee \{\mathcal{I}\}}\left(\omega|\omega\circ E_\mathcal{I}\right)\sim 0\,,\hspace{.7cm} 
S_{\mathcal{O}(S)\vee \{\tau_c\}}\left(\omega|\omega\circ E_\tau\right)\sim \log |G|\,.
\ee 
If $\epsilon m\gg 1$, the twist operators can be chosen such that $\langle \tau\rangle\sim 1$, even if the size of the shell is large enough compared to $\epsilon$. This is a {\sl constant law} for the twist parameter. On the other hand, for very separated balls, we get
\be
S_{\mathcal{O}(R_1R_2)\vee \{\mathcal{I}\}}\left(\omega|\omega\circ E_\mathcal{I}\right)\sim e^{-2\,m\, L}\,,\hspace{.7cm} 
S_{\mathcal{O}(S)\vee \{\tau_c\}}\left(\omega|\omega\circ E_\tau\right)\sim \log |G|-c\, e^{-2\,m\, L}\,.
\ee
This corresponds to an {\sl area law} for the intertwiner parameter. This area law, simple as it is, can also be understood from a dual point of view, by noticing that we can insert many uncorrelated twists with approximately constant expectation values in between the line that separates the two charged operators.

Summarizing, we have described the characteristic pairs area vs. constant laws, and perimeter vs. sub-area laws, for the dual order parameters of thin regions. These dual behaviors have a rational explanation in terms of the certainty relation.

\subsubsection{Spontaneous symmetry breaking} \label{ssb}

From the previous discussion, one can anticipate that something qualitatively different is going to happen for scenarios with spontaneous symmetry breaking. In these cases, the correlation functions of intertwiners do not go to zero at large distances. The reason is that the one-point functions of charged operators in the vacuum do not vanish. For $R_1 $ and $R_2$ of fixed size and large separation, we get a constant law for the order parameter
\be
S_{\mathcal{O}(R_1R_2)\vee \{\mathcal{I}\}}\left(\omega|\omega\circ E_\mathcal{I}\right)\sim \textrm{const}\,.\label{11100}
\ee
Choosing charged operators bigger than the SSB scale, we can have large expectation values $\langle{\cal I}\rangle\sim 1$. In this limit, we obtain
\be
S_{\mathcal{O}(R_1R_2)\vee \{\mathcal{I}\}}\left(\omega|\omega\circ E_\mathcal{I}\right)\sim \log|G|\,.\label{111}
\ee

For the dual order parameter, it is interesting to look at the geometry of a thin shell. In the limit of large radius of the shell, we can now set a volume worth of different and approximately uncorrelated intertwiners crossing the shell, with constant expectation values. Using this large set of intertwiners and the certainty relation it follows, from the results of section \ref{improving}, that we must have
\be\label{sir2}
S_{\mathcal{F}(S)\vee \{\tau\}\vee G}\left(\omega|\omega\circ E_\tau\right)\sim  \, e^{-c \, V}  \;.
\ee
In other words, we expect the approach to saturation to be exponentially fast in the volume $V$ of the region enclosed by the shell. This is an ``area law'' for the shell order parameter (in the terminology adapted to the loop operators). Again we have area vs. constant laws for dual order parameters, coming from the certainty relation and the decoupling of the relevant dual non-local operators.

To study these features in more detail, we consider the simple case of a $\mathbb{Z}_2$ broken symmetry $\phi\rightarrow -\phi$ associated with real scalar field with a double-well potential $V(\phi)$. Call the two vacua $\vert \pm v\rangle$. Starting with the (non-trivial) twist, we should first find its expectation value. Taking a large ball $R$, the expectation value is the path integral in Euclidean space with a boundary condition
\be  
\langle \tau \rangle= \frac{Z_\tau}{Z}=\frac{\int_{\phi(0^-,\vec{x})=-\phi(0^+,\vec{x})\,, \vec{r}\in R} \,{\cal D} \phi\, e^{- S}}{\int {\cal D} \phi\, e^{- S}}\,.
\ee
When analyzing this expectation value, there are subtleties coming from the regularization at the borders. The operator has to be smeared there. But this smearing will contribute with a term proportional to the boundary area in the effective action. This will be superseded by the volume contribution, as long as the region is sufficiently large. Accordingly, we will think about the large volume limit and neglect boundary terms.

\begin{figure}[t]
\begin{center}  
\includegraphics[width=0.65\textwidth]{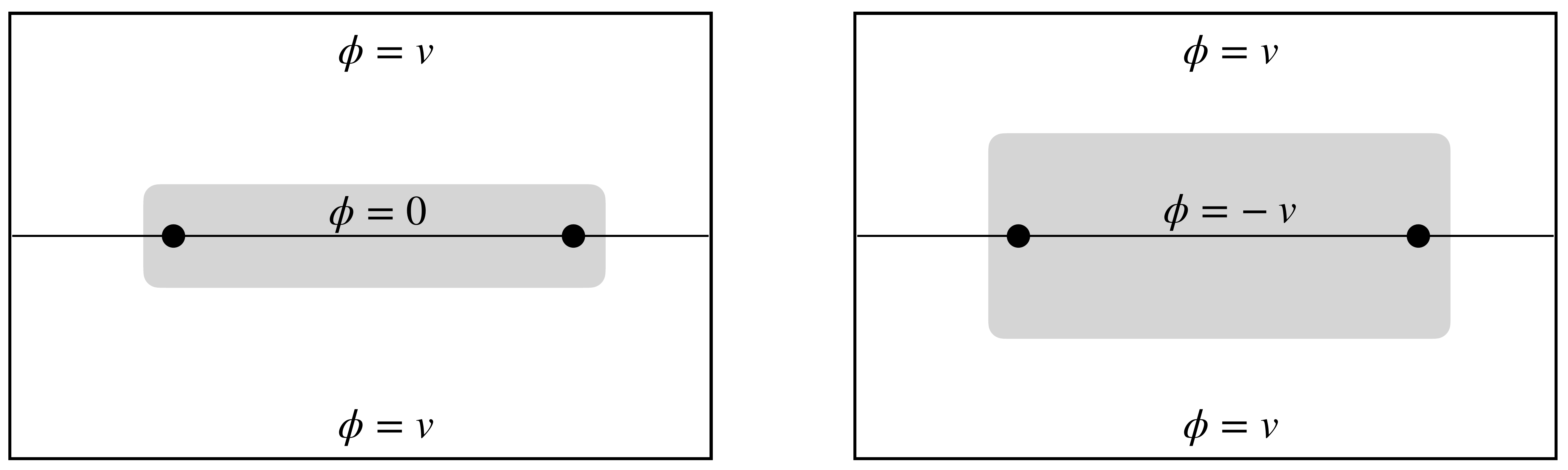}
\captionsetup{width=0.9\textwidth}
\caption{Left panel: path integral calculation of the expectation value of the twist inserted at $t=0$ on the marked region. The shaded area illustrates the region where the field significantly deviates from $v$. Right panel: calculation of the expectation value of the projector $P_-$.  }
\label{instanton}
\end{center}  
\end{figure}

To compute the expectation value, we use a semiclassical limit. We thus need to find solutions to the classical equations of motion
\be 
\partial_\mu\partial^{\mu} \phi =V'(\phi)\;,
\ee
with the appropriate boundary conditions. In this limit, such path integral is computing a configuration where the field goes to $+v$ at infinity in every direction. To achieve that, and the boundary condition at $t=0$, the field should take the value $\phi(\vec{x})=0$ at $t=0$ inside the ball, and grows positively as we move away from $t=0$ in any direction of time. This is a configuration with a non-trivial field around $t=0$ that has the form of an instanton interpolating between $-v $ and $v$ in the time direction, but where we change the sign of the negative part of the trajectory. This is still a solution to the equations of motion because of the action of the twist (see the left panel in figure \ref{instanton}). The action is the same as the one of the instanton.\footnote{We can as well compute $\langle v|\tau \, g |-v\rangle =\langle v|\tau|v\rangle $ where $g$ is the global group operation. In this case, the twisted boundary condition seats outside $R$ and there is nothing in $R$. We have then to interpolate between $-v$ and $v$, so the calculation in this form gives directly the instanton.} Thus, we are computing an instanton corresponding to the tunneling from one vacuum to the other inside the spherical region. This has finite action if we keep the volume large but finite. Analogously, we are computing an overlap $\langle \tau \rangle\equiv \langle v\vert\tau\vert v\rangle_V =\langle -v\vert v\rangle_V$ of the two vacua in the region of volume $V$. In the large volume limit, this transition amplitude is just originated from a translation-invariant solution, for which
\be 
\frac{d^2  \phi}{dt^2} =V'(\phi)\;,
\ee
which is an instanton in one dimension. One can alternatively think of it as a domain wall. Call the corresponding one-dimensional action of this one-dimensional instanton $S_I$, which, however, has $d-1$ dimensions in energy. This is the usual instanton action of a non-relativistic degree of freedom $\phi$ in a double-well potential $V(\phi)$ (see \cite{Coleman:1985rnk} for specific examples and general features). We are ignoring subleading corrections from fluctuations around the saddle point. The total action has a factor of the volume, and we get
\be 
\langle \tau \rangle\sim e^{-S_I V}\;.
\ee
This allows us to compute the coefficient of the volume term of the expectation value of the twist, which does not depend on the shape of the region, as far as this region is large enough. As opposed to the conformal scenario, the leading coefficient of the exponent can be explicitly computed. 
The entropic order parameter in the twist algebra is then given by
\be 
S_{\{\tau\}} (\omega| \omega\circ E_\tau) \sim  e^{-2\,S_I \,V}\;.\label{dfdf}
\ee
We remind that the factor of $2$ in the exponents appears because the correction to formula \eqref{entryt} comes at second order.

We should be able to find a volume scaling using the certainty relation and the intertwiner relative entropy as well, connecting \eqref{11100} with \eqref{sir2}. To find such a contribution, we need to understand how to choose our intertwiner. First, we define a homogeneously smeared operator $\phi^{A}$ over a region $A$. Doing its spectral decomposition, we define projectors on the space of positive and negative eigenvalues of $\phi^A$, namely $P^{A}_+$ and $P^{A}_-$ respectively. Since the global symmetry acts as $\tau_A \phi^{A} \tau_A ^{-1}=-\phi^{A}$, we then have
\be 
\tau_A P^{A}_{\pm}\tau_A^{-1}=P^{A}_{\mp}\,.
\ee
Associated with these projectors there is a charged operator $V^{A}=P^{A}_+ -P^{A}_-$. This transforms as $\phi$ itself.  There are similar projectors $P^{B}_\pm$, associated with homogeneous smearing of the scalar field for an outside region $B$. With those, we can find an analogous charged operator $V^{B}$ in that region. With these charged operators we define the following intertwiner ${\cal I}\equiv V^{A}V^{B}$, which satisfies ${\cal I}^2=\mathbf{1}$. Considering the region $B$ large enough and $\langle \phi\rangle=v>0$, we can set $\langle P^B_-\rangle=0$, $\langle P^B_+\rangle=1$. Then, the probabilities of the projectors $P^\pm=(\mathbf{1}\pm {\cal I})/2$ are $ \langle P^+ \rangle= 1-\langle P^A_- \rangle/2$ and $\langle P^- \rangle= \langle P^A_- \rangle/2$.  

To compute this expectation value, we again turn to the path integral. Now, we are in the situation of the right panel of figure \ref{instanton}. The value of the field is negative inside the region at $t=0$ because of the insertion of the projector, and the classical solution prefers to sit at $\phi=-v$ there. The solution is now formed by two consecutive instantons taking us from $v$ to $-v$ and again to $v$ in the time direction. Then, we have 
\be
p\sim e^{- 2\, S_I \, V_A}\;.
\ee
We could have obtained this result also from the approximation $\langle v|P^A_-|v\rangle\sim \langle v|-v\rangle_A \langle -v|v\rangle_A\sim |\langle v|\tau_A|v\rangle|^2$.
 
The relative entropy in the intertwiner algebra is found to be
 \be
 S_{\{\mathcal{I}\}} (\omega| \omega\circ E_\mathcal{I})\sim \log (2) - k\, e^{- 2\, S_I \, V_A}\,,
\ee 
where $k$ includes subleading factors depending on the size of the region. From this, we get an upper bound to the twist relative entropy. The best bound follows by enlarging $A$ to cover most of the region. Combining this upper bound with the lower bound arising from \eqref{dfdf}, we get for the twist order parameter
\be
 S_{\mathcal{F}(S)\vee \{\tau\}\vee G}\left(\omega|\omega\circ E_\tau\right)\sim e^{- 2\, S_I \, V}\,.\label{tirso}
\ee
This gives exactly the value of exponent $c$ in \eqref{sir2}.

For computing an upper bound to the intertwiner order parameter, we could as well have followed the calculation at the end of section \eqref{improving}, considering many regions of size $V_A$ small compared to $V$. We have to insert the probability $p= \langle P^A_- \rangle/2$ in formula \eqref{vvvvv}. For $N$ such regions $A$ inside the ball $R$ and covering it, from \eqref{vvvvv} we get
\be 
\log(2)- S_{\{\mathcal{I}\}} (\omega| \omega\circ E_\mathcal{I})\sim e^{- N\, S_I \, V_A }\sim e^{- S_I \, V } \;.
\ee
We get a worse upper bound than \eqref{tirso}, but still shows the volume law is obligatory from the existence of multiple uncorrelated intertwiners.

\subsubsection{Summary}

The entropic order parameters clearly distinguish between the phases of QFT in these scenarios. The ordinary operator order parameters are usually studied in the limit of sharp localized unsmeared operators. This means point-like, string-like, surface-like operators, etc. To compare with these operator order parameters we have analyzed the relative entropies for ``thin'' geometries. For example, we consider two balls far away separated from each other for the intertwiner order parameter and a large thin shell for the twist order parameter. For these geometries, the symmetry broken phase has a constant law for the intertwiner relative entropy and specific exponential decay with the volume of the enclosed regions for the twist parameter. In contrast, in the symmetric phase, this behavior is not possible. In the symmetric phase, if the charged fields become gapped, the intertwiner parameter decays to zero exponentially with the distance at large distances  (an ``area law'') and the twist parameter has a constant law. This is the dual behavior of the SSB phase. The conformal regime is intermediate in the sense that none of the order parameters displays an ``area law'' nor a constant law.
 
It is not so surprising that the physics of these phases are captured by the relative entropies because they are related to the expectation values of the associated order-disorder operators. What is interesting is that the entropic approach,  due to the certainty relation, relates, in a quantitative manner, the characterization of the phases in terms of the order or the disorder parameters. The present approach shows they are dual to each other, the duality relation given by the certainty relation \eqref{cglobal}.

In this sense, it is quite clear that it would not be possible to have area-area behavior for both order/disorder parameters. Such putative phase conflicts with the certainty relation. Indeed, we have seen that the area behavior of one parameter is always tied to the constant behavior of the other. This is because to fulfill the certainty relation where one parameter decays exponentially with the area, an area worth of the dual operators with independent and approximately constant expectation values is necessary. It would be interesting to prove these interrelations more rigorously.

\subsection{Gauge symmetry} \label{miga}

We consider the algebra of a simple ring (solid torus) $R$, which contains non-contractible one dimensional loops. Its complement $R'$ contains non contractible $S^{d-3}$ surfaces. The group of non-local operators is Abelian.  For $d=4$, these two complementary ``rings'' have the same topology once we compactify the space at infinity.
 
Let us first consider $d>4$, where the ring and its complement have different topologies. In analogy with the case of global symmetries, we have two possible acceptable algebras for the same region $R$ (other algebras containing only subgroups of the non-local operators can be considered as well). Here, the relevant algebra is the one containing the non-contractible Wilson loops and the additive operators in $R$. Dually, for $R'$, the relevant algebra is the one containing the non-breakable t' Hooft loops and the additive operators in $R'$. To ease the notation, and to highlight the type of non-local operators and algebra a given region contains, in this section, we rename the different algebras as
\begin{align}
{\cal A}(R) & :={\cal A}_{\textrm{add}}(R)\,, & \hspace{-15mm} {\cal A}(R') & :={\cal A}_{\textrm{add}}(R')\,,\\
{\cal A}_{W}(R) & :={\cal A}_{\textrm{add}}(R)\vee\{W\}\,, & \hspace{-15mm} {\cal A}_{T}(R') & :={\cal A}_{\textrm{add}}(R')\vee\{T\}\,. \nn
\end{align}
In $d>4$, notice that ${\cal A}_{W}(R)\equiv {\cal A}_{\textrm{max}}(R)$ and ${\cal A}_{T}(R')\equiv {\cal A}_{\textrm{max}}(R')$.

Any element $x\in{\cal A}_{W}(R)$ can be decomposed as
\be
x= \sum_{r} x_{r}\, W_r\,,
\ee
for a fixed set $\{ W_r\}$ of non locally generated Wilson loop operators going along the ring $R$, where $r$ are the irreducible representations of the corresponding Abelian group $Z$ (the center of the gauge group for pure gauge theories). The elements $x_{r}$ belong to the additive algebra ${\cal A}(R)$.  We can now use the conditional expectation that kills the Wilson loops previously studied
\be
E_{W}:  {\cal A}_{W}(R)\rightarrow {\cal A}(R)\,,\hspace{1cm} E_{W}(x)=x_1\,.
\ee
Then, we have two natural states on ${\cal A}_{W}(R)$. The first one is the vacuum $\omega$, and the second is $\omega\circ E_{W}$. A ``magnetic'' order parameter (since it measures magnetic fluctuations) is then given by 
\be
S_{{\cal A}_{W}(R)}(\omega|\omega\circ E_{W})\,.
\ee

Analogously, any operator $y\in{\cal A}_{T}(R')$ can be written as
\be
y=\sum_{z} y_{z}\, T_{z}\,,  
\ee
where the subindex $z$ runs over $Z$. We can again define a conditional expectation that eliminates the non-trivial t' Hooft loops
\be 
E_{T}:  {\cal A}_{T}(R')\rightarrow {\cal A}(R')\,,\hspace{1cm}E_{T}(y)=y_1\,.
\ee
The ``electric'' order parameter is then
\be
S_{{\cal A}_T (R')}(\omega|\omega\circ E_{T})\,.
\ee
Both of these order parameters vanish for theories having only global symmetries, which do not have duality problems for rings. In the same way, the order parameters for global symmetries vanish for theories containing only gauge sectors. The order parameters only detect their associated symmetries.

Considering this scenario, the observations done until this moment can be condensed in the following complementarity diagram
\bea
{\cal A}_{W}(R) & \overset{E_{W}}{\longrightarrow} & \mathcal{A}(R)\nonumber \\
\updownarrow\prime\! &  & \:\updownarrow\prime\\
{\cal A}(R') & \overset{E_{T}}{\longleftarrow} & {\cal A}_{T}(R')\,,\nonumber 
\eea
with an associated entropic certainty relation
\be 
S_{{\cal A}_{W}(R)}\left(\omega|\omega\circ E_{W}\right)+S_{{\cal A}_T(R')}\left(\omega|\omega\circ E_{T}\right)=\log \vert Z\vert\;.
\ee

Let us now consider the case of $d=4$, where $R$ and $R'$ have the same topology. Both are conventional rings once we compactify the space at infinity. In other words, 't Hooft loops are one dimensional loops. This implies that the maximal algebra for $R $,  namely $({\cal A} (R'))'$,\footnote{We remember that we are calling ${\cal A}(R)$ to the additive (minimal) algebra.} contains both non-local Wilson and 't Hooft loops, all based on $R$. We call this maximal algebra
\be 
{\cal A}_{WT}(R):= ({\cal A} (R'))'={\cal A}(R)\vee \{W_r\}\vee \{T_z\}\;.
\ee
These two sets of non-local loops can be chosen such they commute within each other, and we can expand a general element $x\in {\cal A}_{WT}(R)$ as
\be
x=\sum_{z,r} x_{z,r}\, T_z W_r\,.
\ee
We can also define a new conditional expectation $E_{WT}:{\cal A}_{WT}(R) \rightarrow {\cal A}(R)$ satisfying $E_{WT}(x)=x_{1,1}$. In this case, the complementarity diagram reads
\bea
{\cal A}_{WT}(R) & \overset{E_{WT}}{\longrightarrow} & \mathcal{A}(R)\nonumber \\
\updownarrow\prime\! &  & \:\updownarrow\prime\\
\mathcal{A}(R') & \overset{E_{WT}'}{\longleftarrow} & {\cal A}_{ WT}(R')\,,\nonumber 
\eea
with an associated entropic certainty relation
\be 
S_{{\cal A}_{WT}(R)}\left(\omega|\omega\circ E_{WT}\right)+S_{{\cal A}_{ WT}(R')}\left(\omega|\omega\circ E_{WT}'\right)=2\log  |Z|\;.
\ee

As explained in section \ref{nonabelian}, charged fields can break the group formed by non-local Wilson and t' Hooft loops  $\{W_rT_z\}$ to subgroups, which may not have this particular product structure. The above still applies to such a scenario just by taking the conditional expectations that kill all remaining non-local operators, and where $2 \log|Z|$ is replaced by the order of the group of non-local operators. In the above scenario, or more generally when the group of non-local operators has a subgroup, we can choose other non-local algebras associated with such subgroups and define their corresponding relative entropies (order parameters). We describe these parameters in the case of a $\{W_rT_z\}$ group.

There are other possible choices of algebras for $R$. One is ${\cal A}_{W}(R):={\cal A}(R)\vee \{W_r\}$, which contains the non-local Wilson loops but excludes the non-local t' Hooft loops. Equivalently, we have ${\cal A}_{T}(R):={\cal A}(R)\vee \{T_z\}$, which contains the non-local t' Hooft loops but excludes non-local Wilson loops. This leads to two other natural order parameters, $S_{{\cal A}_{W}(R)}(\omega|\omega\circ E_{W})$ and $S_{{\cal A}_{T}(R)}(\omega|\omega\circ E_{T})$. The associated complementarity diagrams are
\be
\begin{array}{ccccccc}
{\cal A}_{ W}(R) & \overset{E_{W}}{\longrightarrow} & \mathcal{A}(R) & \,\hspace{1.5cm} \, & {\cal A}_{ T}(R) & \overset{E_{T}}{\longrightarrow} & \mathcal{A}(R) 
\\
\updownarrow\prime\! &  & \:\updownarrow\prime &\, \hspace{1.5cm} \,& \updownarrow\prime\! &  & \:\updownarrow\prime
\\
\mathcal{A}_W(R') & \overset{E_T'}{\longleftarrow} & {\cal A}_{ WT}(R')\,, & \,\hspace{1.5cm}\, & \mathcal{A}_T(R') & \overset{E_{W}'}{\longleftarrow} & {\cal A}_{ WT}(R')\,.
\end{array}
\ee
The certainty relations read in this case
\bea
S_{{\cal A}_{W}(R)}(\omega|\omega\circ E_{W})+S_{ {\cal A}_{WT}(R')}(\omega|\omega\circ E_{ T}')=\log |Z|\,,\\
S_{{\cal A}_{T}(R)}(\omega|\omega\circ E_{T})+S_{ {\cal A}_{ WT}(R')}(\omega|\omega\circ E_{ W}')=\log |Z|\,.
\eea

We can show that these relative entropies are not all independent. We will adopt here a simplified notation for the relative entropies detailing only the content of the non-local operators of the two algebras connected by the conditional expectation. For example, we define $S_{WT,0}(R):= S_{{\cal A}_{WT}(R)}\left(\omega|\omega\circ E_{WT}\right)$ and $S_{WT,T}(R):= S_{{\cal A}_{WT}(R)}\left(\omega|\omega\circ E_{W}\right)$.  Using this notation, the certainty relations described above are 
\bea
S_{W,0}(R)+S_{WT,W}(R')=S_{T,0}(R)+S_{WT,T}(R')\!\!\!&=&\!\!\!\log |Z| \,,\\
S_{WT,0}(R)+S_{WT,0}(R')\!\!\!&=&\!\!\!2 \log|Z| \,.
\eea
Using the conditional expectation property \cite{petz2007quantum}, we obtain
\be
S_{WT,0}(R)=S_{W,0}(R)+S_{WT,W}(R)=S_{T,0}(R)+S_{WT,T}(R)\,. \label{doos}
\ee
These relations combined give, for example, the symmetry relations 
\be
S_{T,0}(R)+S_{W,0}(R')=S_{W,0}(R)+S_{T,0}(R')\,.
\ee
Curiously, from these observations, it follows that all asymmetries between parameters corresponding to Wilson loops and the ones corresponding to t' Hooft loops are equal
\be
S_{W,0}(R)-S_{T,0}(R)=S_{W,0}(R')-S_{T,0}(R')= S_{WT,T}(R)-S_{WT,W}(R)=S_{WT,T}(R')-S_{WT,W}(R')\,.
\ee
If there was a form of duality under the interchange of non-additive electric and magnetic loops, all these quantities would be zero. Finally, it is worth mentioning that these relative entropies satisfy certain inequalities. From monotonicity of the relative entropy, we have
\be
S_{WT,0}(R)\ge S_{W,0}(R), S_{T,0}(R)\,,\hspace{.6cm}
S_{WT,T}(R)\ge S_{W,0}(R)\,,\hspace{.6cm}
S_{WT,W}(R)\ge S_{T,0}(R)\,.
\ee
These relations and \eqref{doos} imply 
\be
S_{WT,0}(R)\ge S_{W,0}(R)+S_{T,0}(R)\,.
\ee
This last inequality also follows from the commuting square property of the conditional expectations $E_W$, $E_T$, and $E_{WT}=E_W\circ E_T=E_T\circ E_W$, and their respective algebras \cite{petz2007quantum}.

\subsubsection{Bounds from Wilson and 't Hooft operator expectation values}

The algebra generated by an independent set of non-local 't Hooft loops is isomorphic to the algebra of an Abelian group $Z$, whereas the algebra generated by an independent set of non-local Wilson loops is isomorphic to the algebra of its (irreducible) representations. Therefore, the formulas for the relative entropies for these algebras are the ones obtained for twist and intertwiners, but specializing for Abelian groups. In particular, the algebra of a fixed set of 't Hoof loops is
\be
T_{z_1} T_{z_{2}}=T_{z_{1}z_2}\,,
\ee
where the $z_1,z_2 \in Z$. All these operators are unitary. Again, it is convenient to define projectors labeled by irreducible representations
\be
P_r:=|Z|^{-1}\, \sum_z \chi_r^*(z) T_z \,, \hspace{.6cm}P_r P_{r'}=\delta_{r,r'}\, P_{r}\,, \hspace{.6cm}\sum_ r P_r=\mathbf{1}\, .\label{centryW}
\ee
The conditional expectation is analogous to the twist conditional expectation \eqref{tw_ce}
\be 
E_{T}(T_z)=\delta_{z,1}\,,\hspace{.6cm}
E_{T}(P_r)=\frac{\mathbf{1}}{|Z|}\,.
\ee
Any state $\omega$ in this Abelian algebra is determined  by the probabilities of the different sectors 
\be
q_r=\langle P_r \rangle\,. 
\ee
The relative entropy in the algebra of the t' Hoof loops becomes
\be  
S_{\{T_z\}} (\omega| \omega\circ E_T)= \log |Z|+\sum_r q_r \log q_r =\log|Z|-S_T\,,
\ee
where $S_T$ is the von Neumann entropy over the t' Hooft loop algebra. We have $S_{\{T\}}(\omega| \omega\circ E_\tau)\in [0,\log Z]$.

For the Wilson loops, the situation is similar but replacing representations by elements of the group and vice versa. The minimal projectors of the Algebra are then
\be
P_{z}:=|Z|^{-1}\,\sum_r   \chi_r^*(z)   W_r \,,
\ee
with probabilities
\be
q_{z}=\langle P_{z}\rangle\,.
\ee
The conditional expectation kills all non-trivial Wilson loops,
\be
E_W (W_r)=\delta_{r,1}\,,\hspace{.6cm}
E_W(P_{z})=\frac{\mathbf{1}}{|Z|} \,.
\ee
The relative entropy in the algebra of Wilson loops  becomes
\be
S_{\{W_r\}}(\omega|\omega\circ E_W)=\log |Z|+\sum_{z} q_{z} \log q_{z}= \log|Z|-S_W \,,
\ee
where $S_W$ is the von Neumann entropy on the Wilson loop algebra.

Reducing the certainty relation to the subalgebras of non-local Wilson and 't Hooft loops (using monotonicity of the relative entropy) and using the previous expressions, we derive an uncertainty relation for the von Neumann entropies
\be
\log |Z|\le S_W +S_T\,.
\ee

\subsubsection{Ring order parameter for the Maxwell field}

In this section, we compute upper and lower bounds for a ring relative entropy for the free Maxwell field. We find the behavior of this relative entropy to be surprisingly well determined by these bounds.

\begin{figure}[t]
\begin{center}  
\includegraphics[width=0.5\textwidth]{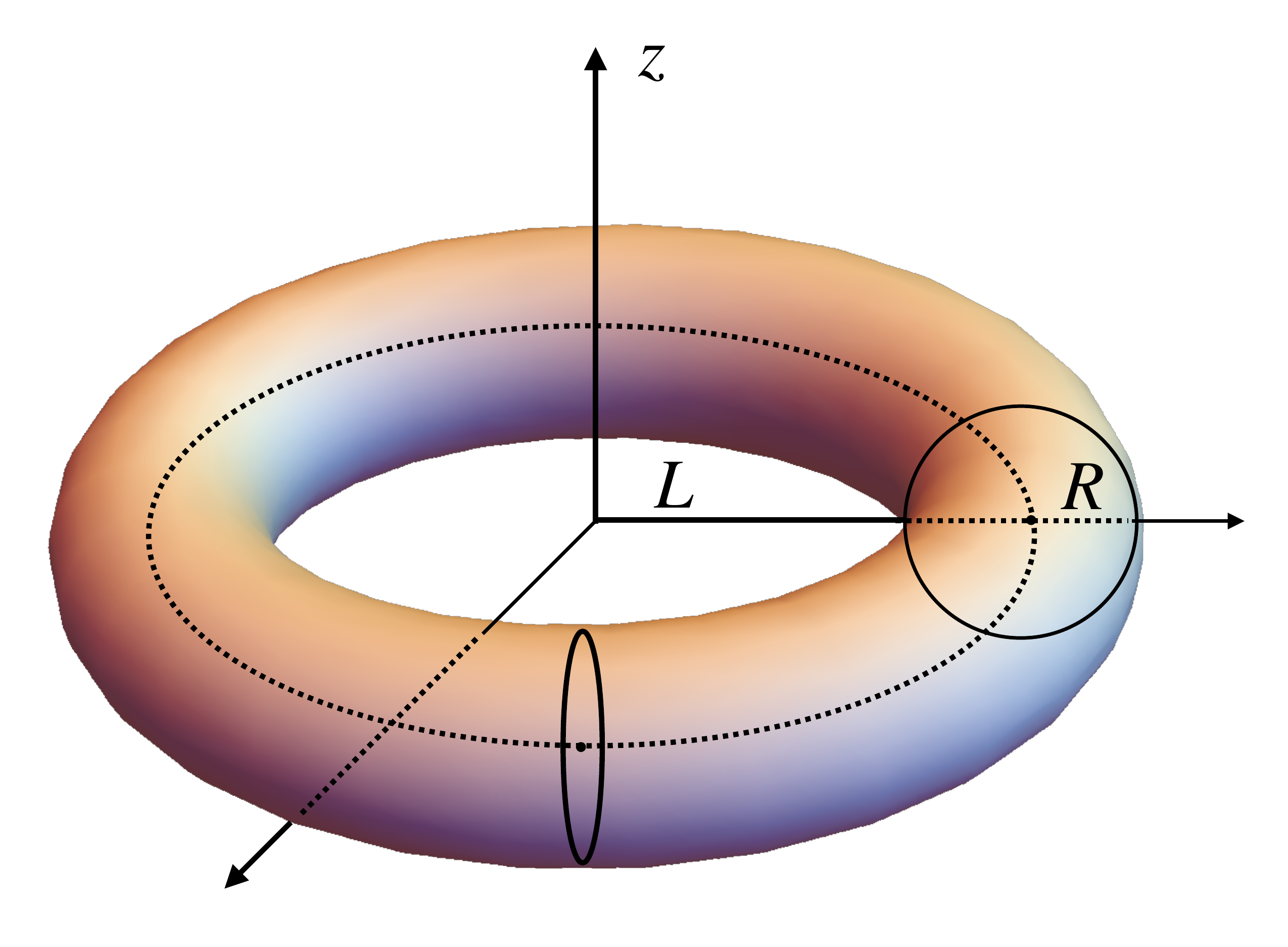}
\captionsetup{width=0.9\textwidth}
\caption{A ring formed by the revolution around the $z$ axes of a disk $D$ of radius $R$, such that the inner radius of the ring is $L$.}
\label{toro}
\end{center}  
\end{figure}

Let us take a ring formed by the revolution around the $z$ axes of a disk $D$ of radius $R$, such that the inner radius of the ring is $L$ (see figure \ref{toro}). In this case, ${\cal A}_{\textrm{add}}(R)$ is the additive algebra of the electric and magnetic fields inside the ring.\footnote{The radius of the torus $R$ should not be confused with the name of the ring-like region itself.}  For the Maxwell field, the symmetry group of the non-local operators in a ring forms an infinite non-compact group ${\mathbb R}^2$ of electric and magnetic charges, and the order parameters arising from the comparison of  ${\cal A}_{\textrm{max}}(R)$ and ${\cal A}_{\textrm{add}}(R)$ are divergent for any $R$. Therefore, we consider the bigger algebra for $R$ a subalgebra of the maximal one. This is obtained by adding to ${\cal A}_{\textrm{add}}(R)$ a closed group of Wilson loops corresponding to charges that are integer multiples of some fixed charge $q>0$. We call this algebra ${\cal A}_{W_q}(R)$. We will be interested in the relative entropy $S_{{\cal A}_{W_q}}(\omega|\omega\circ E_W)$ for this choice of algebra, where $E_W$ eliminates the non-additive Wilson loops. In the following, we adopt the simplified notation of the previous section and call simply ${\cal A}(R)$ to the additive algebra ${\cal A}_{\textrm{add}}(R)$ and ${\cal A}_{WT}(R)$ to the maximal one ${\cal A}_{\textrm{max}}(R)$.

The complementarity diagram is then
\bea
{\cal A}_{W_q}(R) & \overset{E_W}{\longrightarrow} & {\cal A}(R)\nonumber \\
\updownarrow\prime\! &  & \:\updownarrow\prime\\
\mathcal{A}_{W,T_g}(R') & \overset{E_T}{\longleftarrow} & {\cal A}_{WT}(R')\,\nonumber 
\eea
The algebra $\mathcal{A}_{W,T_g}(R')$ contains all non-local Wilson loops, and the non-local t' Hooft loops with charges integer multiples of $g:=2\pi/q$. These are the ones that commute with the Wilson loops in ${\cal A}_{W_q}(R)$. The difference between ${\cal A}_{W_q}(R)$ and  ${\cal A}(R)$ is a group $\mathbb{Z}$ of Wilson loops, while the difference between ${\cal A}_{WT}(R')$ and $\mathcal{A}_{W,T_g}(R')$ is its dual group $U(1)$ of t' Hooft loops. The dual conditional expectation, denoted by $E_T=E'_W$, eliminates the t' Hooft loops with non-integer multiple of the minimal magnetic charge $g$.  Because these dual groups contain an infinite number of elements, the index is infinite. It will turn out that the relative entropy $S_{{\cal A}_{W_q}(R)}(\omega|\omega\circ E_W)$, arising from a discrete group, is finite, though it can take arbitrarily large values depending on the geometry. The complementary relative entropy $S_{{\cal A}_{WT}(R')}(\omega|\omega\circ E_T)$ diverges, as it should be due to the divergence on the index in the certainty relation.\footnote{The divergence in the relative entropy, due to a continuous group of non local operators, appears for $S_{{\cal A}_{W_q}(R)}(\omega|\omega\circ E_W)$ as a $-\log(q)$ behaviour in the limit of a continuous group $q\rightarrow 0$.} However, we will be able to use a subtracted form of the certainty relation to get an upper bound to  $S_{{\cal A}_{W_q}(R)}(\omega|\omega\circ E_W)$ from the expectation values of the t' Hooft loops.

A choice of smeared Wilson loops can be written in cylindrical coordinates, appropriate for the ring geometry, as
\be   
W_{q n}= e^{i \, q\,  n \int_D dr\, dz\, \alpha(r,z) \int_0^{2\pi} d\varphi \,r\, (\hat{\varphi}\cdot A(r,z,\varphi))}\,.\label{qrt}
\ee
We have used smearing functions localized at $t=0$. To quantize the charge, we have to impose
\be
\int_D dr\, dz\, \alpha(r,z)=1\,.
\ee
These loop operators form a group $\mathbb{Z}$ that gives rise to the following ring algebra
\be
{\cal A}_{W_q}={\cal A} \vee \{W_{qn}\}\,.
\ee
This algebra is independent of the smearing functions $\alpha$ since changes in $\alpha$ can be produced by additive operators in the ring. 

\subsubsection*{Lower bound}

Let start by computing a lower bound to this relative entropy. We evaluate the relative entropy in the subalgebra $W_q$ generated by the non-local Wilson loops \eqref{qrt} without additional additive operators. By monotonicity of relative entropy, we have
\be
S_{W_q}(\omega|\omega\circ E_W)\le S_{{\cal A}_{W_q}}(\omega|\omega\circ E_W)\,.
\ee
The algebra $\{ W\}$ of Wilson loops is Abelian, and it can be represented as the pointwise multiplicative algebra of functions on $k\in [-\pi,\pi)$ by the identification
\be
W_{qn}\leftrightarrow f_n(k)=e^{i k n}\,. 
\ee
The probability density in the $k$-space, corresponding to the state $\omega$, is then given by the relation
\be
\langle W_{qn}\rangle=\int_{-\pi}^\pi dk\, P(k)\, e^{i k n}\,.\label{ev}
\ee
On the other hand, being exponential operators in a free theory, the Wilson loops have expectation values 
\be
\langle W_{q n}\rangle= e^{-\frac{1}{2} q^2 n^2\,  \langle \Phi_B^2\rangle } \,,
\ee
where $\Phi_B=\int_D dr\, dz\, \alpha(r,z) \int_0^{2 \pi} d\varphi \,r\, (\hat{\varphi}\cdot A(r,z,\varphi))$ is the magnetic flux operator. Then, using Poisson summation formula for inverting \eqref{ev}, we get
\be
P(k)=\sum_{n=-\infty}^\infty \frac{e^{-i n k}}{2\pi} e^{-\frac{1}{2} q^2 n^2 \langle \Phi_B^2\rangle}=\frac{e^{-\frac{k^2}{2 q^2 \langle \Phi_B^2\rangle}}}{ \sqrt{2 \pi} \sqrt{q^2 \langle \Phi_B^2\rangle}}\Theta_3\left(\frac{i k \pi}{q^2 \langle \Phi_B^2\rangle}, e^{-\frac{2 \pi^2}{q^2 \langle \Phi_B^2\rangle}}\right)\,,\label{pp}
\ee
where $\Theta_3$ is the elliptic function. This gives the probability distribution corresponding to the vacuum state. The state $\omega\circ E_W$ on the algebra $W_q$ of Wilson loops just gives zero expectation value to all Wilson loops except for $n=0$, which corresponds to the identity operator. Then, the probability distribution in the $k$-space corresponding to the state $\omega\circ E_W$ is the uniform distribution $Q(k)=(2 \pi)^{-1}$. Then, the relative entropy is
\be
S_{W_q}(\omega|\omega\circ E_W)=\int_{-\pi}^\pi dk\, P(k) \log(P(k)/Q(k))=\int_{-\pi}^\pi dk\, P(k) \log(2 \pi P(k))\,.
\ee
This depends through \eqref{pp} on the smearing function $\alpha$ and the "elementary" charge $q$. Later, we will analyze this dependence in more detail.

A simplification of these expressions can be obtained in the limits of large and small $q^2\langle \Phi_B^2\rangle$. For $q^2\langle \Phi_B^2\rangle \ll 1$, we can convert the sum \eqref{pp} into an integral, and get
\bea
P(k)\!\!\! &\sim & \!\!\!\frac{e^{-\frac{k^2}{2 q^2 \langle \Phi_B^2\rangle}}}{\sqrt{2 \pi q^2 \langle \Phi_B^2\rangle}}\,,\label{pova}\\
S_{W_q}(\omega|\omega\circ E_W)\!\!\! & \sim & \!\!\!\frac{1}{2}\left(\log\left(\frac{2 \pi}{q^2\langle \Phi_B^2\rangle }\right)-1)\right)\,,\hspace{1cm} q^2\langle \Phi_B^2\rangle\ll 1\,.\label{yuyo}
\eea
Note that in the limit of a non compact Wilson loop group $q\rightarrow 0$, this relative entropy diverges logarithmically with the charge as $S_{W_q}(\omega|\omega\circ E_W)\sim -\frac{1}{2} \log q^2$, and the same happens for the ring order parameter $S_{{\cal A}_{W_q}}(\omega|\omega\circ E_W)$. 

\begin{figure}[t]
\begin{center}  
\includegraphics[width=0.55\textwidth]{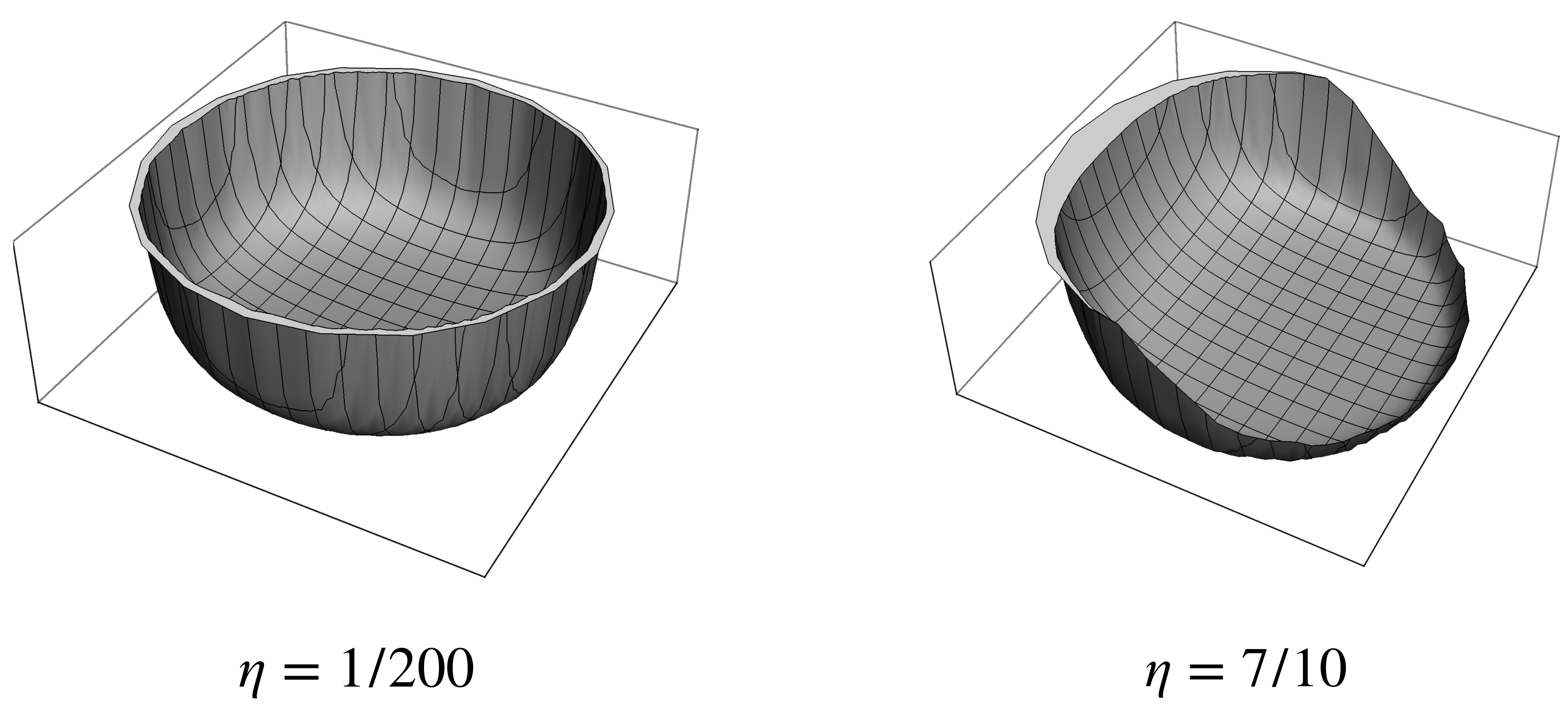}
\captionsetup{width=0.9\textwidth}
\caption{Three dimensional plot of the smearing function $\alpha$ that minimizes the flux. In this example, $\eta=1/200$ (left) and  $\eta=7/10$ (right).}
\label{perfiles}
\end{center}  
\end{figure}

In the opposite limit, $q^2\langle \Phi_B^2\rangle\gg 1$, only the first terms in the sum \eqref{pp} give a non-negligible contribution, and we get
 \bea
P(k)\!\!\! & \sim&\!\!\!   \frac{1}{2\pi}+\frac{\cos(k)}{\pi} e^{-\frac{1}{2} q^2\langle \Phi_B^2\rangle}\,,\\
S_{W_q}(\omega|\omega\circ E_W) \!\!\! & \sim&\!\!\!   e^{- q^2\langle \Phi_B^2\rangle}\,,\hspace{3.3cm} q^2\langle \Phi_B^2\rangle\gg 1\,.\label{yuyo2}
\eea
The best lower bound is obtained for the largest relative entropy for the subalgebra. This corresponds to a smearing function in which $\langle \Phi_B^2\rangle$ is minimal, producing the largest difference between the $\omega$ and $\omega \circ E_W$ expectation values. These latter are zero for non-trivial Wilson loops. To solve this problem, we first express $\langle \Phi_B^2\rangle$ in terms of $\alpha$. From \eqref{qrt}, writing the $(r,z)$ coordinates as a vector $u\in D$, we have
\bea
\langle \Phi_B^2\rangle \!\!\!&=&\!\!\! \alpha\cdot K\cdot \alpha = \int_D d^2u\, d^2u'\, \alpha(u) K(u,u') \alpha(u')\,,\\\label{sss}
K(u,u') \!\!\!&=&\!\!\! \frac{r r'}{2\pi}\int_0^{2 \pi} d\varphi\, \frac{\cos(\varphi)}{(z-z')^2+r^2+r'^2-2 r r' \cos(\varphi)}\nonumber\\
&=&\!\!\! \frac{1}{2}\left(\frac{(z-z')^2+r^2+r'^2}{\sqrt{((z-z')^2+r^2+r'^2)^2-4 r^2 r'^2}}-1\right)\,,
\eea
where in \eqref{sss} we have used  the correlator of the vector potential in Feynmann gauge   
\be
\langle A_i(x) A_j(0)\rangle=\frac{1}{(2 \pi)^2} \frac{\delta_{ij}}{x^2}\,. 
\ee 
Now, from \eqref{sss}, we see that finding $\alpha$ such that $\langle \Phi_B^2\rangle$ is minimal corresponds to minimizing $\alpha\cdot K \cdot \alpha$ subject to the constraint $\alpha\cdot {\bf 1}=1$, where ${\bf 1}(u)=1$ is the function that is identically $1$ on the disk $D$. The solution is
\bea
\alpha \!\!\! &=&\!\!\!  \frac{K^{-1} \cdot {\bf 1}}{{\bf 1}\cdot K^{-1}\cdot  {\bf 1}}\equiv \frac{\int_D d^2u\,  K^{(-1)} (u,u') }{\int_D d^2u\, d^2u' \, K^{(-1)}(u,u')} \,, \label{micro1}  \\
\langle \Phi_B^2\rangle \!\!\!&=&\!\!\!  \alpha\cdot K \cdot \alpha= ({\bf 1}\cdot K^{-1}\cdot  {\bf 1})^{-1}=\left(\int_D d^2u\, d^2u' \, K^{(-1)}(u,u')\right)^{-1}\,.\label{micro}
\eea
These depend on the ring parameters $L$ and $R$ only through the cross-ratio
\be
\eta=\frac{R^2}{(R+L)^2} \in (0,1)\,, \label{cross-ratio}
\ee
which determinss the geometry of the ring (see appendix \ref{confor}). Consequently,  the lower bound $S_{W_q}(\omega|\omega\circ E_W)$ will also be a function of the cross ratio $\eta$ (and $q$). 
 We have computed  numerically the smearing function \eqref{micro1} discretizing the kernel in a square lattice with site labels $(i,j) \leftrightarrow (r,z)$. In these coordinates, the ring $(R,L)$ is given by the set of points $(i,j)$ such that  $(i - L - R)^2 + j^2 \leq R^2$ with $L\leq i \leq L+2R $ and $-R\leq j \leq R$. Alternatively, for the ring $(R,\eta)$, we have $(i - \frac{R(1-  \sqrt{\eta})}{\sqrt{\eta} }- R)^2 + j^2\leq R^2$ with $\frac{R(1-  \sqrt{\eta})}{\sqrt{\eta} }\leq i \leq \frac{R(1-  \sqrt{\eta})}{\sqrt{\eta} }+2R$, and $-R\leq j \leq R$. As it shown in figure \ref{perfiles},  $\alpha$  evolves from being rotationally invariant but mostly concentrated on the boundary for small $\eta$ (thin ring), to a crescent moon concentrated on the inner left boundary of the disk for $\eta\sim 1$.
 
The relative entropy can be solved analytically in terms of $\eta$ for the opposite regimes $L/R\ll 1$ and $L/R\gg1$.  Let us start by considering the limit of thin rings $\eta\ll 1$. In this case, the expression of the kernel $K$ is rotational and translational invariant
\be
K(u,u')\sim \frac{L}{2} \frac{1}{|u-u'|}\,.\label{kuu}
\ee
Moreover, in this regime, $K\cdot \alpha$ is proportional to the Coulomb potential, and therefore, the condition $K\cdot \alpha=\textrm{const}$ means $\alpha$ is proportional to a charge density on a conductor disk.  The solution to this problem can be obtained for squeezed ellipsoids using oblate spherical coordinates and then taking the limit of the disk. The solution is
\be
\alpha= \frac{1}{2 \pi R \sqrt{R^2- \tilde{r}^2}}\,, \label{alpha}
\ee
where $\tilde{r}$ is the radial coordinate in the disk. This is in agreement with the profile shown in figure \ref{perfiles} for $\eta=1/200$. From  \eqref{kuu} and \eqref{alpha}, we see that the flux that gives the best lower Wilson loop bound for large $L/R$ and fixed $q$ satisfies
\be
\langle \Phi_B^2\rangle=\frac{\pi}{4}\frac{L}{R}\,.\label{fluxb}
\ee
This gives an exponentially small relative entropy 
\be
 S_{W_q}(\omega|\omega\circ E_W)  \sim  e^{- q^2 \frac{\pi}{4}\frac{L}{R}} \sim e^{- q^2 \frac{\pi}{4}\eta^{-1/2}}\,, \hspace{.6cm} \eta q^{-4}\ll 1\,, \eta \ll 1 .\label{con}
\ee

Now we study the opposite regime  $L/R\ll 1$. For this geometry corresponding to $\eta\sim 1$, it is convenient to consider the complementary region of the ring. This is a thin ring with cross-ratio $\tilde{\eta}=1-\eta$ corresponding to the ratio $\tilde{L}/ \tilde{R}\gg 1$ of the new radius of the torus.  Then, in this limit, the original geometry of a ring with $L/R\ll 1$ turns out to approach a much simpler one given by the complement of a large tube of radius $\tilde{R}$.  The smear function in this setup is translation invariant along the $z$ direction, and the kernel $K$ is the same as the one already found above. Moreover, translation invariance implies that the equivalent problem cannot depend on $z$ but only on the radial coordinate $r$
\bea
\alpha(r,z) \!\!\!&=&\!\!\! \frac{\tilde{\alpha}(r)}{2 \pi \tilde{L}}\,,\hspace{.7cm} \textrm{with } \int_{\tilde{R}}^\infty dr\, \tilde{\alpha}(r)=1\,,     \\
\langle \Phi_B^2\rangle \!\!\!&=&\!\!\! (2 \pi\tilde{L})^{-1}\int dr\, dr'\, \tilde{\alpha}(r) \tilde{K}(r,r') \tilde{\alpha}(r') \,,\label{flux2} \\
\tilde{K}(r,r') \!\!\!&=&\!\!\! \int_{-\infty}^\infty dz\, K(z,r,r')\,. 
\eea
We could not solve this limit analytically. However, employing dimensional arguments,  the flux giving the best lower bound in this limit has to be proportional to $\tilde{R}/\tilde{L}$
\be
\langle \Phi_B^2\rangle =c \,\frac{\tilde{R}}{\tilde{L}}\,, 
\ee
where the constant $c$ can be evaluated numerically by inverting a discretized version of the kernel $\tilde{K}$. We get 
\be
c\simeq \frac{1}{\pi}\sim\, 0.318\,\,.
\ee
The relation between new and original variables can be obtained from the cross-ratio in this limit (see appendix \ref{confor})
\be
 1-\eta=2\frac{L}{R}=\tilde{\eta}=\frac{\tilde{R}^2}{\tilde{L}^2}\,.\label{etab}
\ee
Finally, from \eqref{yuyo}, \eqref{fluxb} and \eqref{etab}, the best lower bound is
\bea
 S_{W_q}(\omega|\omega\circ E_W)  \!\!\!&\sim&\!\!\! \frac{1}{2}\left(\log\left(\frac{\sqrt{2}\pi}{q^2 c}\sqrt{\frac{R}{L}}\right)-1\right)\label{compari}\\
 &=& \!\!\! \frac{1}{2}\left(\log\left(\frac{2\pi}{q^2 c}(1-\eta)^{-\frac{1}{2}}\right)-1\right)\,,\,\,\,\,\, q^4 (1-\eta)\ll 1, (1-\eta)\ll 1 \,.\nonumber
\eea
From the above equation, the relative entropy increases logarithmically for wide rings $R/L\rightarrow \infty$.

\subsubsection*{Upper bound}

Having found a lower bound, let us now proceed with an upper bound. Such an upper bound can be found by considering the dual algebra of t' Hooft loops in the complement ring. Regularizing the continuous group by discrete subgroups, and taking the limit afterward, we can still use the certainty relation to put an upper bound based on the statistics of the t 'Hooft loops on $R'$. The divergences of the index and the complementary relative entropy $S_{{\cal A}_{WT}}(\omega|\omega\circ E_T)$ cancel to each other when we use an algebra of t' Hooft loops to get upper bounds.  We then get (see appendix \ref{subgroups})
\be
S_{{\cal A}_{W_q}}(\omega|\omega\circ E_W)\le S_{T}-S_{T_g}=: \Delta S\,,\label{ub}
\ee
where $S_T$ is the entropy in the full algebra of t' Hooft loops, and $S_{T_g}$ is the one in the algebra of loops with magnetic charges, which are integer multiples of
\be
g=\frac{2 \pi}{q}\,,
\ee   
due to the Dirac quantization condition. Note that the upper bound \eqref{ub} is not (at least in an evident way) a relative entropy, but a difference of entropies on continuous Abelian algebras.  This is the result of applying the entropic certainty and uncertainty relations restricted to the case of a subgroup\footnote{We are only considering discrete charges multiples of $q$.} of the total symmetry gauge group (see appendix \ref{subgroups}). To calculate the entropy in $T$, we note that this algebra is represented as the one of complex-valued functions on the full real line by the identification $T_x \leftrightarrow e^{i x k}$ with $x\in \mathbb{R}$. The probability density, corresponding to the state $\omega$, has then a formula analogous to \eqref{pova}. Substituting the sum by an integral in \eqref{pp}, we get 
\be
P(k)= \frac{e^{-\frac{k^2}{2 \langle \Phi_E^2\rangle}}}{\sqrt{2 \pi  \langle \Phi_E^2\rangle}}\,,
\ee
giving an entropy $S_T$
\be
S_T=-\int_{-\infty}^{+\infty} dk\, P(k) \log P(k)=\frac{1}{2}\left(1+ \log(2 \pi \langle \Phi_E^2\rangle )\right)\,.\label{unoz}
\ee

Regarding the entropy $S_{T_g}$, the calculation follows the same line as the lower bound one. Note the t' Hooft  loops in $T_g$ are represented as $e^{i g k n}$, $k \in g^{-1}[-\pi, \pi)$. A calculation analogous to the one in the previous section gives the probability density
\be
Q(k)=\frac{e^{-\frac{k^2}{2 \langle \Phi_E^2\rangle}}}{\sqrt{2 \pi \langle \Phi_E^2\rangle}}  \Theta_3\left(\frac{i k \pi}{g \langle \Phi_E^2\rangle},e^{-\frac{2 \pi^2}{g^2\langle \Phi_E^2\rangle}}\right).\label{doz}
\ee
The entropy is then
\be
S_{T_g}=-\int_{-\pi/g}^{\pi/g}dk\, Q(k)\log Q(k)\,.\label{lll}
\ee
The upper bound is then given by $S_T-S_{T_g}$ using equations \eqref{unoz} and \eqref{doz}. This is a function of $g \equiv (2\pi)/q$ and $\langle \Phi_E^2\rangle$. Again, we have to use the best loop smearing with the smallest $\langle \Phi_E^2\rangle$ to get the lowest entropy difference and the best upper bound. By electromagnetic duality, this is given by the same function used for the magnetic flux, but evaluated in a complementary cross-ratio
\be
\langle \Phi_E^2\rangle(\eta)= \langle \Phi_B^2\rangle(1-\eta)\,.\label{fluxbe}
\ee

Let us then compute the limits of wide and thin loops. The limit $g^2\langle \Phi_E^2\rangle\ll 1$ allows us again to convert the sums into integrals and integrate over the real line in \eqref{lll}. The leading order in $S_{T_g}$ exactly cancels $S_T$. The  entropy difference is given by the following integral
\be
\Delta S= -2\int_{\pi/g}^\infty dk\, 	\frac{e^{-\frac{k^2}{2 \langle \Phi_E^2\rangle}}}{\sqrt{2 \pi  \langle \Phi_E^2\rangle}} \log\left(\frac{e^{-\frac{k^2}{2 \langle \Phi_E^2\rangle}}}{\sqrt{2 \pi  \langle \Phi_E^2\rangle}}\right)\sim \frac{\sqrt{2\pi}}{\sqrt{g^2 \langle \Phi_E^2\rangle}} e^{- \frac{\pi^2}{2 g^2 \langle \Phi_E^2\rangle }}\,. 
\ee
Replacing $\langle \Phi_E^2\rangle\sim c R/L$ and $g=2\pi/q$, we get 
\be
\Delta S= \frac{1}{2} \frac{\sqrt{q^2 L}}{\sqrt{2 \pi c R}} e^{- \frac{  q^2 L }{8   c R }}\,, \hspace{1cm} \frac{L}{R}\gg 1\,, q^2\frac{L}{R}\gg 1\,,
\ee
which is compatible with \eqref{con} because $c\ge (2 \pi)^{-1}$. This confirms, in this regime, that the relative entropy has a perimeter law
\be 
S_{{\cal A}_{W_q}}(\omega|\omega\circ E_W)\sim e^{- x\, q^2 \frac{L}{R}}\,,\hspace{.5cm}  0.39 \sim (8 c)^{-1}\le x\le \pi/4 \sim\,  0.78\,\,.
\ee 
 
\begin{figure}[t]
\begin{center}  
\includegraphics[width=0.70\textwidth]{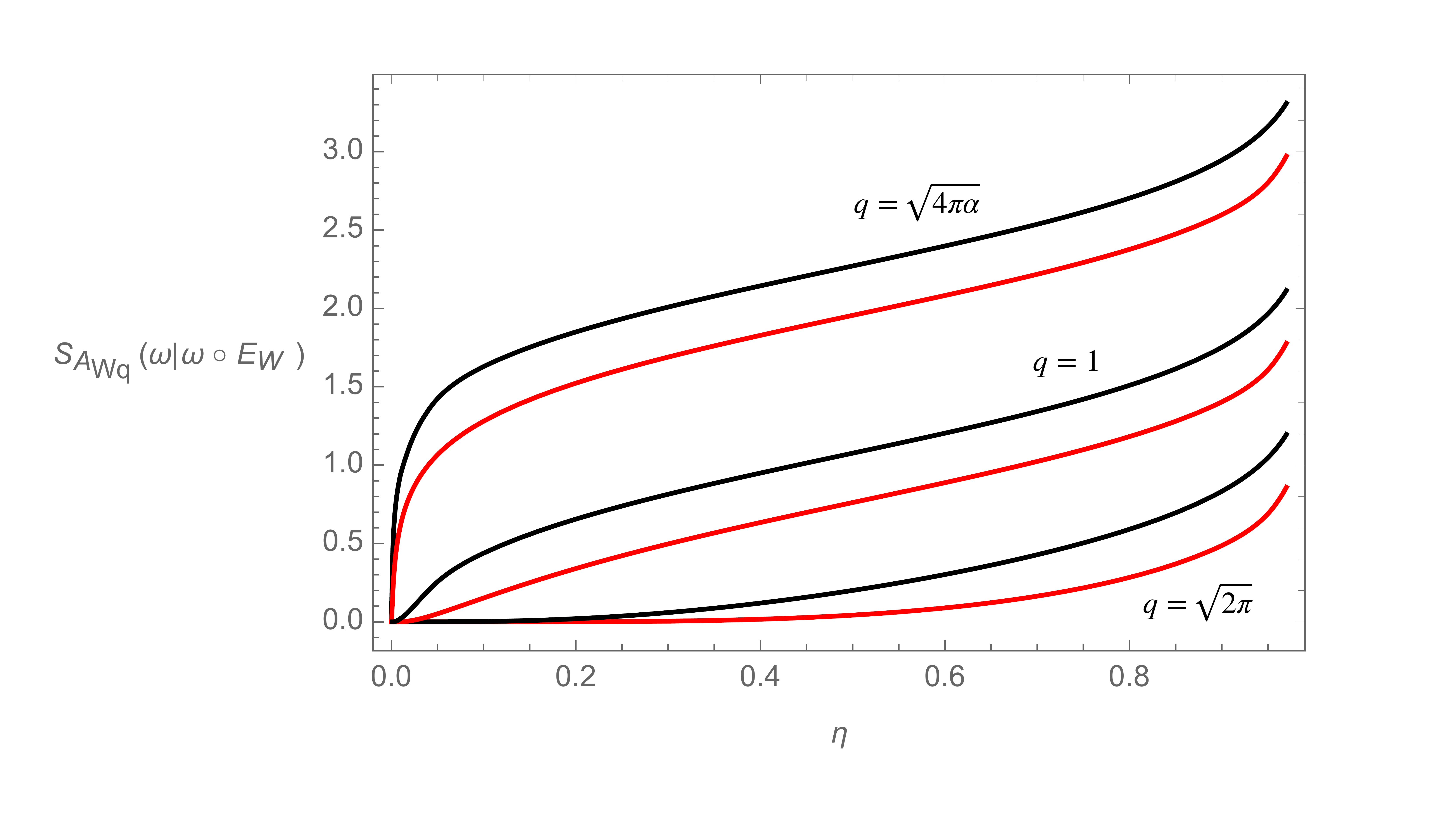}
\captionsetup{width=0.9\textwidth}
\caption{Relative entropy upper (black) and lower (red) bounds for different charges, using numerical evaluation of the function $\langle \Phi_B^2\rangle(\eta)$ (see text). The lowest pair of curves correspond to the case of equal electric and magnetic charges $q=g=\sqrt{2\pi}$. The upper pair to the electron charge. At $\eta=1$, the curves diverge logarithmically. Approaching $\eta=0$, they go to zero exponentially fast. This is not seen in the two upper curves because it happens for quite a small $\eta$.}
\label{reldownup}
\end{center}  
\end{figure}   

For $g^2\langle \Phi_E^2\rangle \gg 1$, we get $Q(k)\sim \frac{|g|}{2\pi}$ up to exponentially small terms. Then, we get
\be
\Delta S=S_T+\log(g/(2\pi))=\frac{1}{2}\left(1+ \log( g^2 \langle \Phi_E^2\rangle/(2 \pi) )\right)\,.
\ee
From \eqref{fluxb} and \eqref{fluxbe}, we get $\langle \Phi_E^2\rangle=\frac{\pi}{4} \sqrt{\frac{R}{2L}}$ and
\be
\Delta S=\frac{1}{2}\left(\log\left( \frac{ \pi^2}{2\sqrt{2} q^2}\sqrt{\frac{R}{L}}\right)+1\right)\,, \hspace{1cm} \frac{R}{L}\gg 1\,, q^{-4}\frac{R}{L}\gg 1\,.
\ee
This is compatible with \eqref{compari}. 

The upper and lower bounds give a surprisingly precise determination of the ring order parameter in this limit,  
\be
S_{{\cal A}_{W_q}}(\omega|\omega\circ E_W)\sim \frac{1}{4}\log\left(\frac{R}{q^4 L}\right)+ \kappa\,,  \hspace{1cm} \frac{R}{L}\gg 1\,, q^{-4}\frac{R}{L}\gg 1
\ee
with 
\be
0.81\sim \frac{1}{2}\log\left(\frac{\sqrt{2} \pi}{c e }\right)\le \kappa\le \frac{1}{2}\log\left(\frac{\pi^2 e}{2\sqrt{2}}\right)\sim 1.12\,.
\ee
We conclude that, while the relative entropy in the ring can be very large for thick rings, it always differs from the one on the best Gaussian Wilson loop algebra by less than half a bit.

The numerical calculation of the lower and upper bounds on intermediary regimes is shown in figure \ref{reldownup} for different charges. The upper pair of curves corresponds to the electron charge.

 \subsubsection{Ring order parameters in CFT's}
 
For a CFT there are no scales and the order parameters are dimensionless conformally invariant functions of the geometry. One can take advantage of the conformal symmetry and consider the toroidal rings previously described for the Maxwell field. We will consider $d=4$ in this section. The relative entropies must be a function of the cross-ratio $\eta$ (see equation \eqref{cross-ratio}). The complementary ring has a cross-ratio $1-\eta$, as described in appendix \ref{confor}. All relative entropies are increasing functions of $\eta$ due to monotonicity. 

For small width $R$ and large size $L$, all relative entropies associated with the ring should vanish exponentially fast in the perimeter
\be
S\sim e^{-c\,\frac{L}{R}}\,,\label{ghk}
\ee
matching the behavior of the loop operator expectation values. Indeed, the loop operators in this ring have a perimeter law, and this gives a lower bound. An upper bound follows from the certainty relation. Using a perimeter worth of small loops, wrapped around the thin ring and well separated between themselves, we can show a perimeter law for an upper bound for thin loops. This follows from the ideas described in section \ref{improving} about how to improve bounds with the help of uncorrelated dual operators. However, we cannot produce stronger upper bounds using larger dual loops crossing the bulk of the ring. This is because these large dual loops expectation values must decay fast with the size and have non-trivial correlations in the conformal case, preventing the direct application of the ideas in \ref{improving}.   

Accordingly, for large width, we expect the parameters to saturate the bound and approach $\log |Z|$ in the finite group case. From the certainty relation, we expect
\be
S\sim \log |Z|- k\, e^{-c\,\frac{L}{R}}\,.
\ee
For a continuous Abelian group $U(1)^n$, the order parameter involving the elements of the group is divergent for any $\eta$ because the group is continuous. It has to be so to match the certainty relation with divergent index too. However, the dual group of representations $Z^*$ is infinite and discrete for a compact group $Z$. We expect that the corresponding relative entropy to behave as in the case of $n$ Maxwell fields studies in the previous section.  In the limit of wide rings $R/L\gg 1$, we then have
\be 
S_{WT,0}(R)\sim \frac{n}{2}\log(R/L)\,, \hspace{.5cm}S_{W,0}(R),S_{T,0}(R),S_{WT,T}(R),S_{WT,W}(R)\sim \frac{n}{4}\log(R/L) \,.
\ee

The case of finite groups and conformal symmetry is quite special, given the tendency of gauge theories based on non-Abelian Lie groups to confine. It is achieved with a special balance of gauge and matter degrees of freedom. The matter fields should transform accordingly the adjoint representation of the gauge group to preserve the generalized symmetry of the center of the gauge group. A famous case is $\mathcal{N}=4$ $SU(N)$ SYM theories.    

Given the duality relation between $\eta$ and $1-\eta$, there are some peculiar features in the conformal case for finite groups. The certainty relation gives
\be
S_{WT,0}(\eta)+S_{WT,0}(1-\eta)=2 \,\log |Z|\,.  
\ee
In particular, we have
\be
S_{WT,0}(1/2)=\log|Z|\,.\label{lz}
\ee
Both relations are rather surprising. The relative entropy for a fixed cross-ratio should be dependent on the dynamics that set the expectations values of the non-local operators and their relation with the rest of the algebra. The relation \eqref{lz} tells that, for example, in conformal SYM theories this relative entropy does not depend on the coupling constant.

Likewise, from the other properties studied in section \ref{miga}, we obtain
\be
S_{W,0}(1/2)+S_{WT,W}(1/2)=S_{T,0}(1/2)+S_{WT,T}(1/2)=\log|Z|\,.
\ee
The first equation is easily understood if there is an analogous to electromagnetic duality symmetry. However, these equations do not seem to require some form of duality equating t' Hooft and Wilson loops. 

\subsubsection{Confinement and Higgs phases}

The duality between the confinement and Higgs phases was transparently argued in 't Hooft's work \cite{tHooft:1977nqb}. In such a work, he defined the dual disorder parameters, latterly called 't Hooft loops. They were defined by their simple commutation relations with the order parameters (Wilson loops). It was argued that, although the confinement phase might be difficult to approach given strong coupling issues, the dual physics can be studied in the Higgs phase at weak coupling.

In the confinement phase, the ``electric'' charges (the quarks) are confined, and this can be measured by the area law of the Wilson loop. In the Higgs phase, the magnetic charges are the ones confined, and the 't Hooft loops are the operators displaying area law. The 't Hooft loop, a ``disorder'' parameter, is the natural order parameter for the Higgs phase. Although the physics of both phases is similar (or dual), the Higgs phase can be approached semi-classically.

In this section, we want to analyze what we expect for the entropic order parameters in these phases. This provides a different perspective to the physics already known, where new light is shed by the certainty relation. This relation is valid at any coupling, and it relates the physics of order and disorder parameters. In the confinement-Higgs scenarios, it relates the physics of Wilson and 't Hooft loops.

\begin{figure}[t]
\begin{center}  
\includegraphics[width=0.45\textwidth]{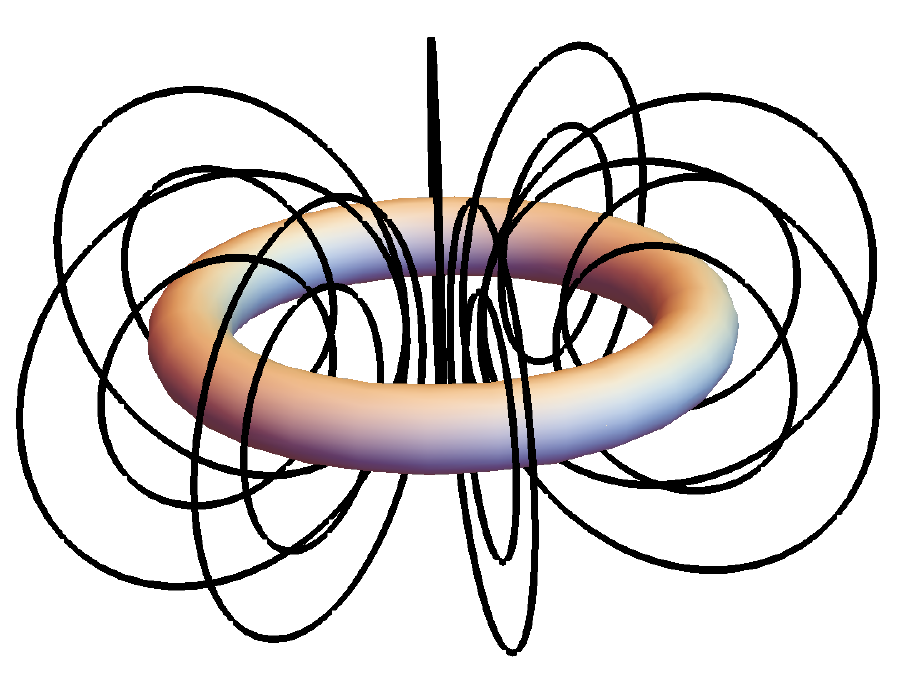}
\captionsetup{width=0.9\textwidth}
\caption{A ring interlocked by thin loops. A number of thin loops proportional to the area can be placed keeping them separated at a fixed distance.}
\label{madeja}
\end{center}  
\end{figure}

As described for global symmetries, spontaneous symmetry breaking implies that the expectation value of the intertwiner should go to a constant, and the certainty relation then implies that the twist expectation values decay exponentially with the volume. In such a phase, the intertwiner factorizes into the non-vanishing product of one-point functions of the charged operators. A volume worth of constant intertwiners can be used to induce a volume law for the twist expectation value. This is very different from the conformal scenario, where the intertwiner expectation value shows some decay typical of a conformal field theory, and the twist expectation value decays with the area of the boundary.
 
For gauge symmetries, we have a very similar picture. Both confinement and Higgs phases are expected to become gapped. We will focus on the Higgs phase, where we have semi-classical control. But the same (or dual) behavior is expected in the confinement phase. The gauge field then becomes massive.  This has the consequence that the loop operators become uncorrelated and, further, as shown below, the Wilson loop displays an approximately constant behavior. This implies through the certainty relation that the dual loop (the t' Hooft loop in the Higgs phase) is bound to display an area law. This is because we can place an area worth of constant uncorrelated Wilson loops crossing the sectional area of the t' Hooft loop as it is shown in figure \ref{madeja}.    
 
In the literature, this phenomenon is interpreted as a symmetry breaking of the ring generalized symmetry \cite{Gaiotto:2014kfa}. However, loops are almost always considered as line operators, without width. In this limit, the loop operator dual to the one displaying area law shows a perimeter law in general, as it happens in the conformal scenario, rather than a constant one. This constant behavior of loop operators is invoked in the literature to make a parallel to the case of spontaneous symmetry breaking of a global symmetry. It is argued that the loops can be dressed by local operators to convert the perimeter law into a constant behavior, while this cannot be done in the case of an area law. But by the very same means, a constant behavior can be induced as well for any loop operator having perimeter law, including the conformal case. However, for our purposes, this is unsatisfactory. The reason is that we need the loop operators to be real (smeared) operators satisfying a group law. This tells us that they cannot be dressed arbitrarily. There is no way to dress a loop in the conformal regime to have constant behavior while keeping the group fusion rules. Otherwise, the corresponding relative entropy will not have the perimeter law dependence on $R/\epsilon$ that it has. This is not the case of a massive field. The dressing is here replaced by looking at operators that are smeared in a ring of a certain width. To get a constant law, we need to widen the ring size and be away from the conformal phase. 

To test this behavior, we study the Higgs phase. In this scenario, the gauge field appearing in the Wilson loop has become massive, and this should lie at the root of the expected constant scaling. We thus consider a Wilson loop for a massive vector field. This is of course the case of Wilson loops inside a superconductor, which is a specific simple scenario of the Higgs phase. In the massive case, the Wilson loop typically shows a perimeter law. We want to show that with transversal smearing in a size larger than the scale of the inverse mass, we can do better and obtain a Wilson loop whose expectation value is (almost) constant, independent of the perimeter. 

\begin{figure}[t]
\begin{center}  
\includegraphics[width=0.40\textwidth]{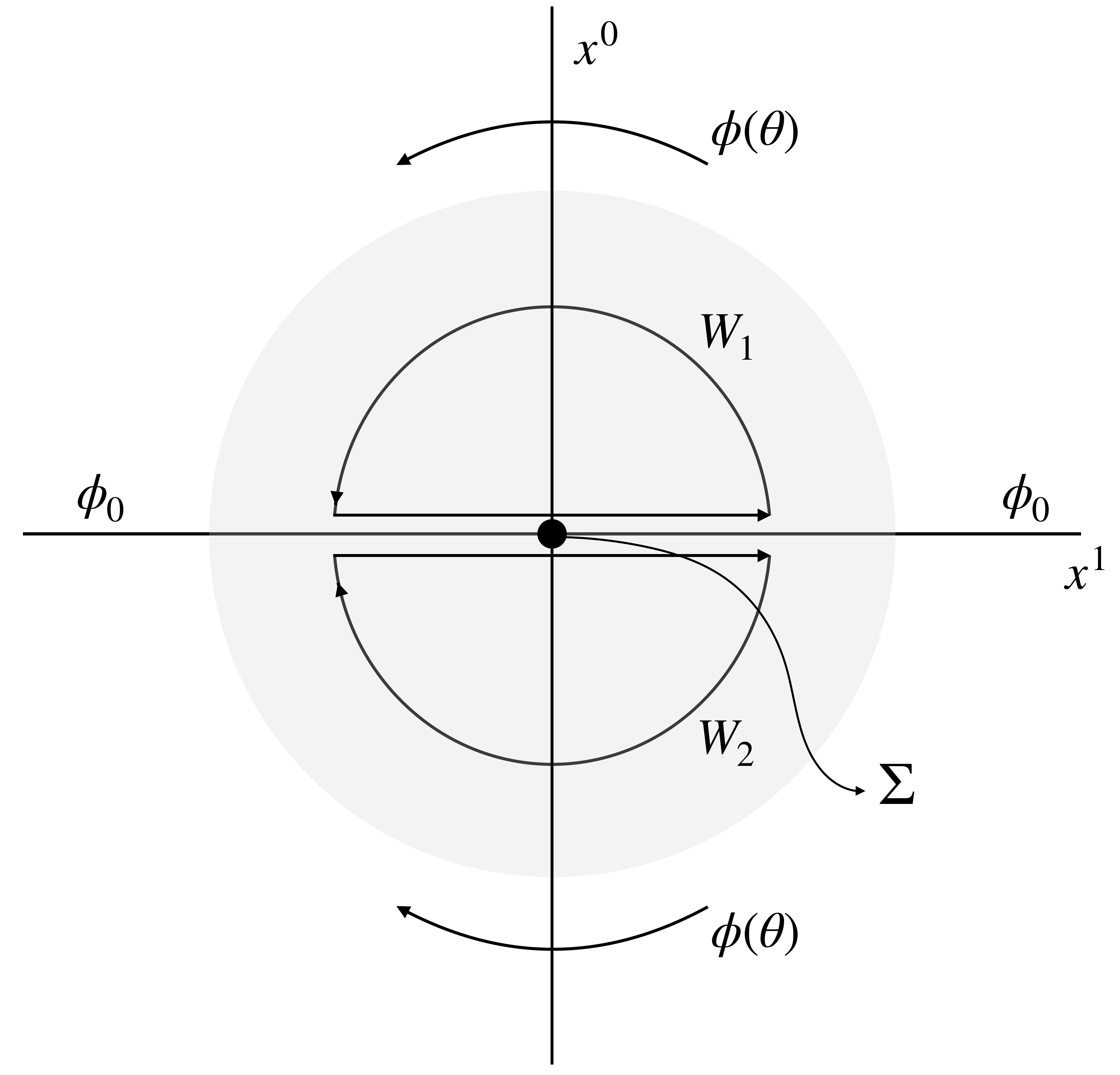}
\captionsetup{width=0.9\textwidth}
\caption{The vortex-instanton configuration giving the t' Hooft loop expectation value.}
\label{vin}
\end{center}  
\end{figure}

We have to compute the expectation value of a Wilson loop in the classical regime
\be
\langle W \rangle =\langle e^{i\int d^4x\, J(x)\cdot A(x)}\rangle
=e^{-\frac{1}{2}\int d^4x\, d^4y\, J^\mu(x)\Delta_{\mu\nu}(x-y) J^\nu(y)}=: e^{-V}\,, \label{vvv}
\ee
where the Wightman correlator of a massive vector field is
\be
\Delta_{\mu\nu}(x)=\int \frac{d^4p}{(2\pi)^3}\, e^{-i p\cdot x} \delta(p^2-m^2) \theta(p^0) \left(-g_{\mu\nu}+\frac{k_\mu k_\nu}{m^2}\right)\,.
\ee
Since, for the moment, we are interested in the perimeter term, we take an operator invariant in the $\hat{z}$ direction with a current in that direction given by
\be
J(x)= \alpha(y)\, \hat{z}\,,   
\ee
where we have written $y=(x^0,x^1,x^2)$. We also need to normalize the charge of the operator $W$ accordingly to
\be
\int dy\, \alpha(y)=1\,,\label{lala}
\ee
and impose $\alpha(y)$ to has its support inside the causal development of a region of size $R$ in the coordinates $x^1$ and $x^2$.

Plugging this into \eqref{vvv}, and considering a  tube of large length $L$, the leading linear $L$ dependence of the exponent $V$ becomes
\be
V=\frac{1}{2} L \int d^3 y_1 \, d^3 y_2\, \alpha(y_1)\, \alpha(y_2)\, G(y_1-y_2)\,,\label{per}
\ee
with
\be
G(y)=\int_{-\infty}^\infty dx^3\, \Delta_{33}(y,x^3)=\int \frac{d^3p}{(2\pi)^2} \delta(p^2 -m^2) \theta(p^0) e^{-i p\cdot y}\,.  
\ee
Writing the smearing functions in momentum space, we obtain from \eqref{per} the perimeter law
\be
V= \frac{1}{2} L \int \frac{d^3 p}{(2\pi)^2} \, \delta(p^2 -m^2) \theta(p^0) |\tilde{\alpha}(p)|^2 \,. \label{mass}
\ee
The condition \eqref{lala} on $\alpha$ gives the constraint $\tilde{\alpha}(0)=1$. The mass shell in \eqref{mass} is separated from the point $p=0$, and the Fourier transform of the smearing function can be chosen to be exponentially small outside $p\sim R^{-1}$.\footnote{It cannot has compact support because the Fourier transform of a function of compact support is an entire function.} Therefore, we can device a smearing function such that the coefficient of the perimeter goes to zero exponentially fast for $m R\gg 1$. Note that a pure spatial smearing function $\alpha(y)=\delta(t) \beta(x^1,x^2)$ is not allowed for this exponential suppression, and we can get a power law suppression at most. One may doubt this improvement from spatial smearing to spacetime smearing given the fact that the fields at $t\neq 0$ can still be written at $t=0$ using the equations of motion. However, in writing the fields at $t=0$, our loop operator will also contain a term of the conjugate momentum of $A$ (the electric field) at $t=0$ in the exponent. This is what achieves the additional suppression of the exponent and the increase in the expectation value of the loop.  This is an instance of improvement produced by the locally generated operators on the expectation value of a non locally generated one.   

We conclude that the coefficient of the perimeter law for a massive field can be made exponentially small for loops wider than the mass scale. We can have an almost constant law for wide enough Wilson loops. The gap then implies uncorrelated loops, and an area law for the t' Hooft loop should arise from the certainty relation.

Let us see directly how this area law appears in the classical regime in the Higgs phase. The t' Hooft loop is a singular gauge transformation of the center of the gauge group on a surface $\Sigma$ of area $A_\Sigma$ and boundary in a ring $R$. For simplicity, let us think in a group $\mathbb{Z}_2$, where we have only one (non-trivial) t' Hooft loop. This is the case of a spontaneously broken $SU(2)$ gauge theory. To break it without converting the Wilson loops in the fundamental representation into local operators in the ring, we need to couple the gauge fields with adjoint Higgs fields. More than one Higgs is necessary to break the symmetry completely. We will not enter into the details of the model building. We use the Euclidean path integral to compute the expectation value of the t'Hooft loop. Far from the surface $\Sigma$, and at both sides of it at $x^0=0$, the Higgs fields stay into their vacuum values such that they are continuous at large distances outside the loop. So we expect a Higgs configuration that remains in its vacuum value at $x^0=0$ at spatial infinity, deviating from it only near the $\Sigma$. We are interested in the area term, so we choose a large loop and neglect the boundary effects. Then, in this large surface limit, the configuration of interest is only dependent on the coordinates $x^0$ and $x^1$, where $x^1$ is the coordinate perpendicular to $\Sigma$. The t' Hooft loop sits at $(x^0,x^1)=(0,0)$ in this plane. The two Wilson loops $W_1$ and $W_2$ in the fundamental representation in figure \ref{vin} pass through the t' Hooft loop at $x^0=x^1=0$, closing above and below $x^0=0$. They get a factor $-1$ (the central element in the gauge group $SU(2)$) due to the t' Hooft loop insertion. Then, the configuration is such that the circulation $\mathcal{P} e^{i \int_{x^1>0}^{x^1<0} dx^\mu A_\mu}$ of the gauge field, on the upper plane and far from $x^0=0$, is $-1$. The same, in a time reflected manner, happens on the lower plane. The classical solution is a ``vortex-instanton'' that exists because of the insertion of the t' Hooft loop. The Higgs field, as usual for these vortex configurations, accompanies the rotation of the gauge field such as to minimize the action. It rotates, from $x^1\gg 0$ to $x^1\ll 0$, an angle $2\pi$ in the group parameter. But, it ends up at the same vacuum value $\phi_0$ because a $2\pi$ rotation is the identity on the adjoint representation. The full configuration has the same action as a $4\pi$ rotation vortex. The instanton action $S_I$ is a quantity with two dimensions of energy (it is an action density over the surface). We get
\be
\langle T \rangle \sim e^{-A_\Sigma\, S_I}\,.           
\ee
This gives a lower bound on the corresponding relative entropies\footnote{Small expectation values affect the relative entropy quadratically.}
\be
S_{WT,W}(R), S_{T,0}(R) \gtrsim  e^{-2\, A_\Sigma \, S_I}\,. \label{lowi}
\ee
This also gives an upper bound to the relative entropy corresponding to Wilson loops in the wide complementary ring.

\begin{figure}[t]
\begin{center}  
\includegraphics[width=0.40\textwidth]{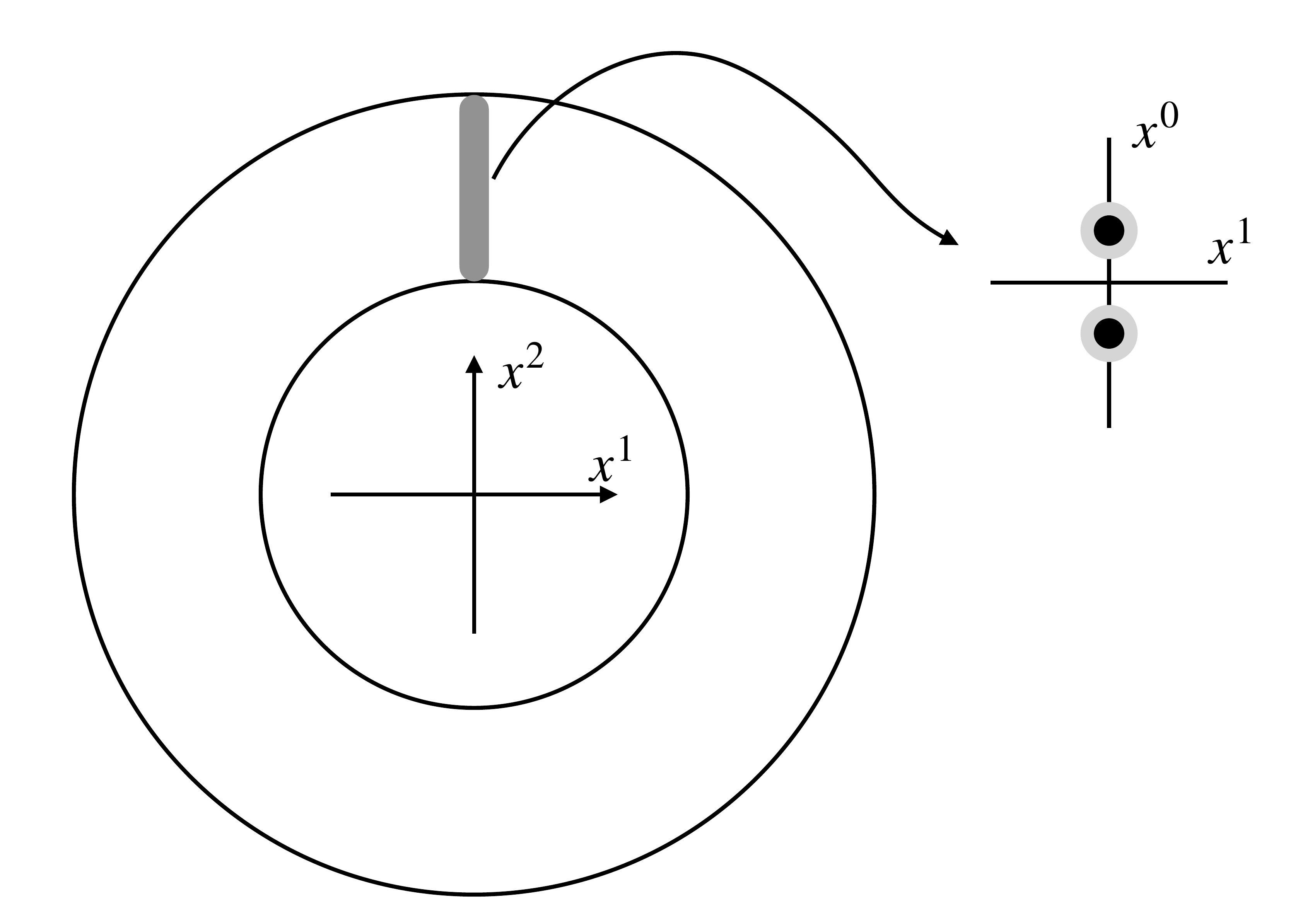}
\captionsetup{width=0.9\textwidth}
\caption{Field configuration computing the expectation value of the projector over negative values of the Wilson loop. The Wilson loop is interlocked with the t' Hooft loop previously discussed. In the plane $x^1,x^2$ it occupies the region between the two circles in the picture. It also extends in the $x^3$ direction (perpendicular to the figure plane) such that its $(x^2,x^3)$ cross-section fits inside the t' Hooft loop. The shaded area in the $x^1,x^2$ plane and the two spots in the $x^0,x^1$ plane show the position of the non-trivial action in the classical solution contribution to the expectation value of the projector over negative eigenvalues of the Wilson loop. The extension over the $x^3$ direction is not shown.}
\label{dospuntos}
\end{center}  
\end{figure}

Now we compute an upper bound. For that, we need to understand how the sufficiently wide Wilson loop in the fundamental representation approaches maximal expectation value. We follow the same route as for the case of intertwiners in section \ref{ssb}. The Wilson loop can be decomposed into orthogonal projectors $P_\pm=(\mathbf{1} \pm W)/2$  (we are using $W^2=\mathbf{1}$ for $\mathbb{Z}_2$). We have $\langle P_+ \rangle \sim 1 $ and $\langle P_-\rangle =p\ll 1$. The expectation value of the projector $P_-$ follows again by inserting it into the path integral. The gauge field is then constrained to produce a $2\pi$ rotation in the gauge group at $x^0=0$ along the path of the loop. 
This rotation has to be completed to a full $4\pi$ rotation in returning back through negative or positive times. Therefore, we have two vortices of $4\pi$ rotation, one after the other in the time direction, which have the same classical action as the one previously discussed (see figure \ref{dospuntos}). They can be positioned anywhere along the path of the loop, but the contribution is quite concentrated around a cross-section of the loop of area $A$. We then get
\be
p\sim e^{-2 A \, S_I}\,.      
\ee
The entropy on the algebra of $W$, for small $p$, behaves as $S_{\{W\}}\sim -p \log (p)$. Therefore, taking the transversal area $A\sim A_\Sigma$ to be as wide as possible for a loop interlocked with the t' Hooft loop,  we get the leading exponential behavior of the upper bounds 
\be
S_{WT,W}(R), S_{T,0}(R) \lesssim  e^{-2\, A_\Sigma \, S_I}\,.
\ee
This is nicely consistent with \eqref{lowi} and gives the exact coefficient of the area in the exponent of the entropic order parameter. The same calculation shows that wide rings have a Wilson loop order parameter going exponentially fast to $\log(2)$. 

\section{Remarks on RG flows of order parameters}
\label{remarks}

Having studied the order parameters in different phases, the main challenge becomes to understand the running of these parameters with the scale. A related objective would be to arrive at some conclusions about the possible realizations of a certain symmetry in the IR and UV. In this final section, we make a few comments on these questions.

For any type of symmetry, the analysis of spontaneous symmetry-breaking scenarios is typically phrased as follows. At low energies, the symmetry might be broken. This is signaled by some order parameters approaching a constant value and the dual order parameter decaying exponentially with some characteristic exponent. At high energies, where we approach some conformal fixpoint, the symmetry is restored. This is signaled again by the behavior of the order/disorder parameters. The question we want to comment on concerns the transition between the different phases through the RG flow.

regions. For scaling regions, the terminology of phases and parameters is simplified considerably. Under scaling, the relative entropy corresponding to a region $R$ of certain topology might show different behaviors. A first case is when it either goes to zero or tends to $\log|G|$ (or the logarithm of the order of a subgroup) as we scale $R$ to infinity, and this happens independently of the precise shape of $R$. A second possibility is that the limit value is some value in the range $(0,\log |G|)$, and it depends on the conformal geometry of $R$. In the first case, one of the symmetries (order vs disorder) is unbroken but the dual symmetry is broken. In the second case, which is the conformal one, we are forced to associate with it the idea that none of the symmetries is broken nor unbroken in the present sense.

In the same line, scaling non-local operators\footnote{In order to associate unique operators to spacetime regions we can use the standard construction described in section \ref{standard}.} leads, in the symmetry breaking scenario, to expectation values $1$ or $0$ in the large scaling limit. This is independent of the details of the shape. We just need to scale the characteristic length of the region to infinity. Operators with expectation value $1$ correspond to the unbroken symmetry. They form a group because $a_1|0\rangle\sim a_2|0\rangle\sim |0\rangle$ implies $\langle 0| a_1 a_2| 0\rangle=1$. In the conformal case, it leads to operators with intermediate expectation values, which depends on the conformal geometry of the region.
 
These observations suggest that the breaking of asymmetry is tied to a gap. At least, it seems tied to the absence of correlations between the relevant non-local operators. This is simply because unitary operators, which their expectation values are saturated to $1$, have zero connected correlations between themselves.  The same will happen for operators with zero expectation value due to the commutation relations. If $\langle a_1\rangle =\langle a_2\rangle=0$ for non local operators seated at spatially separated regions $R_1$ and $R_2$, we can use an operator $b$ commuting with $a_2$ but not with $a_1$ and satisfying $\langle b\rangle \sim 1$, to show
\be
\langle a_1 a_2 \rangle =   \langle b\, a_1 a_2 \rangle=  \chi_b(a_1)\langle a_1 a_2 \,b \rangle=\chi_b(a_1)\langle a_1 a_2 \rangle=0\,.
\ee

The usual terminology in terms of area/perimeter laws for line operators is a bit more cumbersome, especially because we can enlarge line operators in only one direction. For example, the improvement from perimeter to constant law for a loop operator occurs as we increase the width. It holds only in certain cases and for loops that are wide enough. Further, this constant law does not persist without a perimeter term for exponentially large loops of constant width. The reason for these nuisances is that line operators are UV and IR operators at the same time, according to the two widely different scales involved in their geometry.

From the present perspective of the RG flow, one would like to prove that some phase are realized in the IR by connecting this phase with the departure of conformal behavior in the UV. For example, to prove confinement, it may not be necessary to compute the area law of Wilson loops in the IR, but to address the question of which UV behavior leads inevitably to this area law if that were possible. This approach would of course be especially appealing, given that for asymptotically free gauge theories, the UV is under perturbative control.

In investigating the change of the order parameters with the scale, it is not possible to use directly monotonicity of relative entropy for scaling regions. The reason is that, for non-trivial topologies, none of the two scaled regions will be contained in the other one. In the scaling limit, the order parameters are ordered by the inclusion ordering on conformally equivalent shapes (i.e. ordered by the value of the cross-ratio in the special ring shapes used in the paper). This ordering is trivially realized in the case of broken symmetries, where the scaling limit of the relative entropy is independent of the shape.  

Even if the simple monotonicity property is not enough to obtain a UV-IR connection, the heuristic ideas around entropic order parameters in this paper suggest that there is indeed a tendency for the increase of asymmetry between dual order parameters as we move to larger regions. If some non-local operators have larger expectation values than the complementary ones, say $\langle a\rangle> \langle b \rangle$, this seems to seed still larger expectation values for the operators $a$ in larger regions (and smaller ones for the operators $b$),  through the certainty relation. Though it is not clear how to keep under control the effect of correlations, it suggests that the expectation of some connection between the UV and IR asymmetries may not be hopeless. With the purpose to illustrate further this point, in the next subsection we construct an (admittedly quite crude) toy model. But first, let us end this section discussing briefly some special entropic order parameter that exists only for global symmetries.

For the case of global symmetries, we can describe this tendency to symmetry breaking in more precise terms by using the theory ${\cal F}$ containing charged operators. In this case, there is a simpler relative entropy than the ones studied so far concerning the geometry of two balls. It is the relative entropy in just one topologically trivial region introduced in \cite{Casini:2019kex}, and mentioned in section \ref{gs},
\be
S_{{\cal F}(R)}(\omega|\omega\circ E_\psi)\,. \label{rerere}
\ee
The reason this quantity has not been much discussed above is that it is not easily generalizable to the case of gauge symmetries. In any case, it was studied at sufficient length in \cite{Casini:2019kex}.
 
This quantity will be trivially zero if there is no SSB, just because the two compared states are identical in this scenario. However, in an SSB scenario, this relative entropy will be non-zero at all scales. Furthermore, by monotonicity, it is always an increasing function of the region $R$. This leads to the following conclusion. Even if the relative entropy goes to zero at the UV, however, small the deviation from zero for small regions, will not go down again as we move to the IR scaling the region. More interestingly, it cannot remain at a small value. To observe this we notice that for this relative entropy, the role of the intertwiners in the two-ball order parameter discussed above is played by the charged operators inside the ball. If there is a charged operator with a non-zero expectation value we can take many copies of this operator separated by large distances between each other, such that they are statistically independent. Using the results of section \ref{improving}, we conclude that this will make the relative entropy grow until it reaches $\log|G|$ if the symmetry is completely broken. If the symmetry is only partially broken to a subgroup $H\subset G$, it will tend to $\log (|G|/|H|)$ instead. This argument exposes the general idea described above. Once a global symmetry is broken in the UV, notwithstanding the size of breaking, there is no way back. The symmetry will be completely broken in the IR.\footnote{However, we remark that this does not mean there could not be {\sl new} effective symmetries appearing at the IR. If we attempt to extend this effective (local) symmetry at the IR to the UV, it will probably get badly non-local, or highly broken, such as to make these relative entropies of regions either ill-defined or divergent.} The same could be said about an explicit breaking, driven by a small perturbative relevant operator in the UV.

From the operatorial point of view, detecting SSB in the UV is also easy in the ${\cal F}$ theory: it corresponds to a non-zero expectation value of a charged operator. However, this expectation value is non-perturbative, and then not easy to understand in the perturbed UV theory. The same remark applies to the relative entropy \eqref{rerere}. Even if this relative entropy shows the irreversibility of the SSB phenomenon, there is no clear indication on how to exploit it in the UV.

If we fix attention to operators in the neutral model ${\cal O}$, a signal of SSB for small regions is difficult to separate from UV fluctuations. Correlators of a charged-anti-charged neutral operator (intertwiner) will be almost conformal, and concavity properties for short distances do not tell if this will end up as a constant, a conformal, or a massive case, as we move to large distances (see appendix \ref{twopoint}). However, once it has set to a constant expectation value, by reflection positivity, the expectation value of the intertwiner will not start decaying again.

\subsection{Toy model for the RG flow}

This model is intended to illustrate more concretely the instability of the asymmetry between dual order parameters, and how it drives symmetry breaking. We think in a simple case of a $\mathbb{Z}_2$ symmetry. We have the non-local surface operators $a$ and $b$ of dimensions $k$ and $d-2-k$ respectively. To describe RG flows, we fix a way of scaling a specific region $R$. Any such (sufficiently symmetric) region $R$ can be characterized by two length scales $r$ and $\epsilon$. For example, for $k=1$, we have loops, $r$ is the radius of the loop, as a one-dimensional object, while $\epsilon$ is the width of the loop. We will fix $\epsilon$ to be sufficiently small and scale $r$ from $\epsilon$ to infinity.\footnote{This does not mean $r$ needs to be in the IR nor $\epsilon$ in the UV scales.} Thus, we take generalized thin loops. Notice the dual region of a thin loop is not a thin loop.

Ideally, we would want to find the relative entropies for the algebras of the two types of thin loops associated to the dual non local operators as functions of $r$. We call these relative entropies $S_{\textrm{thin},a}(r)$ and $S_{\textrm{thin},b}(r)$. They are not dual to each other because they are not based on complementary regions. We call $S_{\textrm{thick},a}(r)$ and $S_{\textrm{thick},b}(r)$ to the relative entropies in the complementary regions. We also call $S_{ {a}} (r)$ and $S_{ {b}} (r)$ to the relative entropies on the (finite dimensional) thin loop algebras $\{1,a\}$ and $\{1,b\}$, excluding further additive operators.  While the exact computation seems out of reach, the  certainty relation together with monotonicity of relative entropy, gives
\bea
S_{ {a}} (r) \!\!\!&\leq &\!\!\! S_{\textrm{thin},a}(r) = \log (2) -S_{\textrm{thick},b}(r) \,, \\
S_{ {b}} (r) \!\!\!&\leq &\!\!\! S_{\textrm{thin},b}(r) = \log (2) -S_{\textrm{thick},a}(r) \;.
\eea

For $r/\epsilon \gg 1$, we can let many dual thin loops cross the original thin loop, as in figure \ref{madeja}. We can then use the results in section \ref{improving} to estimate upper bounds for the right-hand-side of these equations using a large number of decoupled dual thin loops, under the simplified assumption that we can neglect correlations between the thin loops. To obtain the best upper bound we need to locate as many thin loops as we can. The optimal configuration is the one where we place increasingly big thin loops to fill the original one (see figure \ref{madeja}).  Writing lengths in units of $\epsilon$, we can use formula \eqref{saddle11} to obtain an inequality with the form
\bea\label{RGineq11}
S_{ {a}} (r) \!\!\!&\leq &\!\!\! S_{ {a}}^0\,e^{- \int_0^r dr'\, (r-r')^{k}\, b (r)}\,,\nonumber\\
S_{ {b}} (r) \!\!\!&\leq &\!\!\!   S_{ {b}}^0 \,e^{- \int_0^r dr'\, (r-r')^{d-2-k}\, a (r)}\;,
\eea
where $S_{ {a}}^0$ and $S_{ {b}}^0$ are constants and we have defined the function $a (r)$ (respectively $b (r)$) to be the classical relative entropy between two probability distributions: the first one is the homogeneous distribution $(1/2,1/2)$, and the second one is $p_a (r)$ (respectively $p_b (r)$) associated with the generalized thin loops of radius $r$ (the probability of the two projections $(\mathbf{1}\pm a)/2$ (respectively $(\mathbf{1}\pm b)/2$)). For thin loops, the probabilities $p_a (r)$ and $p_b (r)$ are near the distribution $(1/2,1/2)$ and we can approximate  
\be
 S_{ {a}} (r)\sim a (r)\,, \hspace{1cm}
 S_{ {b}} (r)\sim b (r)\;.
\ee
This gives a pair of coupled inequalities involving only two unknown functions
\bea\label{RGineq1}
a (r) \!\!\!&\leq &\!\!\!  a_0\,e^{- \int_0^r dr'\, (r-r')^{k}\, b (r)}\nonumber\\
b  (r) \!\!\!&\leq &\!\!\!    b_0 \,e^{- \int_0^r dr'\, (r-r')^{d-2-k}\, a (r)}\nonumber\;.
\eea
In these equations, $a_0$ and $b_0$ are the boundary conditions as $r\rightarrow 0$. The content of these uncertainty relations can be further analyzed by the related equations
\bea\label{eqab}
\tilde{a}(r) \!\!\!&= &\!\!\! a_0 \,e^{- \int_0^r dr'\, (r-r')^{k}\, \tilde{b}(r)}\,,\\
 \tilde{b}(r) \!\!\!&= &\!\!\! b_0 \,e^{- \int_0^r dr'\, (r-r')^{d-2-k}\, \tilde{a}(r)}\,.  
\eea
This pair of coupled integral equations can be easily solved numerically. Notice that, if in some interval of the radius $r$ one of the functions is approximately constant, then the other function decays exponentially fast to zero with the expected scaling with $r$. For example, for $d=4$ and $k=1$, the case of pure gauge theories in four dimensions, if one of the functions remains constant, then the other decays with an area law. The behavior shows similarities for different dimensions and $k$. To be more explicit, we can take $d=2$ and $k=0$, where we can analytically solve the previous equations (which are equivalent to the differential equations $\tilde{a}' = \tilde{b}'=-\tilde{a} \tilde{b}$) 
\bea
\tilde{a}(r) \!\!\!&=&\!\!\! a_0\frac{(a_0-b_0)}{a_0-b_0\,e^{(b_0-a_0) r}}\,,\\
\tilde{b}(r) \!\!\!&=&\!\!\! b_0\frac{(b_0-a_0)}{b_0-a_0\,e^{(a_0-b_0) r}}\,.
\eea 
If $a_0>b_0$, the limit for $r\rightarrow \infty$ gives $\tilde{a} \rightarrow a_0-b_0$ constant and $\tilde{b} \rightarrow 0$ exponentially fast (area law). If $a_0<b_0$, the opposite happens. There is a tendency of the RG flow to fall in one of the two possibilities and never come back, and the outcome only depends on data for small loops. The IR fate is controlled in this simplified toy model by the order between the dual entropic parameters. For other cases, to find the dual constant/area behavior for each of the parameters, we need to input from the start an asymmetry between the initial conditions. If that asymmetry is given, the outcome of the equations is a long period in which one of the functions remains constant and the other decays with the appropriate dual scaling.

The main drawback of this toy model lies in the fact that in the UV, the assumption that non-local operators are uncorrelated is invalid.  Related to this comment, an interesting behavior in the previous case of $d=2$, $k=0$ arises if we take the limit $(a_0-b_0)\ll a_0$. In this scenario, there is a regime, when $r\ll (a_0-b_0)^{-1} $, obeying a power law\be
b(r)\sim a(r)\sim \frac{1}{r}\,,
\ee  
pointing to a phase transition with some universal behavior when we cross the critical initial conditions $a_0 =b_0$. The precise functional decay in this regime should again not be taken seriously since, in the conformal scenario, we cannot use the approximation of uncorrelated thin loops. 
 
\section{Conclusions}

In the description of a QFT in terms of algebras and regions, some basic relations are structurally natural. One of them is Haag duality, expressing that a region should contain all admissible operators allowed by causality. The other is additivity, expressing that operators in a region should be generated by local degrees of freedom in the same region. However, these properties are not required by the consistency of the theory. Only sufficiently complete theories should satisfy all these properties.  In this paper, we have put forward the idea that the algebraic origin of symmetries in QFT is most naturally framed as the violation of these properties in regions with specific topology.

This point of view seems to be fruitful. To start with, considering algebras constructed additively, it is the case that different classes of symmetries correspond to violations of duality for regions with different topologies. We thus can see generalized symmetries as tied to violations of duality for regions with non-trivial homotopy groups $\pi_0$, $\pi_1$, $\pi_2$, etc. These violations correspond respectively to global, local, and generalized symmetries, which are thus treated on the same footing. The focus on these simple properties of the net of algebras allows us to describe these symmetries without appealing to topological non-trivial spaces, excited states, or superselection sectors. 

One key consequence that arises when taking duality as the fundamental starting point is that whenever duality is violated for a region with a non-trivial $\pi_k$, then duality is also violated for the complementary region, which has non-trivial $\pi_{d-2-k}$. Besides, the operators that violate duality in the complementary region are in one-to-one correspondence with those in the original region, and the commutation relations between both algebras are uniquely determined. This provides a unified perspective on order/disorder parameters. These can be just defined as the operators which are responsible for the violation of the duality condition. They necessarily are not locally generated in the region in question, and everything else follows from this. For global symmetries, we have intertwiners and twists. For local symmetries, we have (unbreakable) Wilson and 't Hooft loops, and the commutation relations that arise in our construction are exactly those enforced by the original definition in 't Hooft's seminal work \cite{tHooft:1977nqb}. Generalized symmetries follow similar patterns. In this light, the Dirac quantization condition, together with its generalizations, follows when enforcing causality on a possible completion of a net of algebras showing violations of duality.

Regarding gauge symmetries, being not physical symmetries, its true meaning has become a recurring theme in QFT. From the present perspective, the breaking of duality for ring-shaped regions is an unambiguous physical remnant of the gauge symmetry. As we have shown, it is also related to a good definition of confinement order parameters. Indeed, loop order parameters satisfying area law necessarily need to violate duality in a ring. In this precise sense, the conventional confinement order parameters imply a violation of duality. In turn, this means that the inclusion of algebras ${\cal A}(R)\subseteq ({\cal A}(R'))'$ is not saturated, and entropic order parameters suggest themselves from the non-trivial inclusion.

The last part of the paper has been devoted to study the properties of entropic order parameters in several cases of interest. In this context, there are some aspects to highlight. The first one is the use of the entropic certainty relation. This quantitatively relates the physics of order and disorder parameters in complementary regions. We have confirmed such a prediction in different cases. The certainty relation gives a useful and geometrical picture of the origin and relations between the different laws followed by dual order parameters. A constant law for one order parameter forces an area law for the dual one. Area law for both parameters, or even area and perimeter laws, would be forbidden: the fluctuations in both parameters were high enough to prevent saturation of the certainty relation.

However, it is fair to say that we feel we have not yet understood how to profit from these relations in full force. For example, though it is known that area-area laws for complementary parameters should be forbidden, and we see a compelling heuristic reason for this, we could not prove this, in the present approach, in a rigorous manner. Including some information on the correlations between loops would be important for further progress. In the same lines, our approach shows the importance to understand the behavior of wide loops, which are dual to thin loops. These latter have been the focus of almost all past efforts.  A simple heuristic reasoning suggests that the change of a loop with the size is seed by the behavior of the dual loop in a self-consistent manner. A further understanding of this self-consistency is important to have a clearer picture of the RG flow on the order parameters.      

\section*{Acknowledgements} 
This work was partially supported by CONICET, CNEA, and Universidad Nacional de Cuyo, Argentina. The work of H. C. and J. M. is partially supported by an It From Qubit grant by the Simons Foundation.

\appendix

\section{Intertwiners with simple fusion rules} \label{app2}

In this appendix, we show how to construct intertwiners satisfying the simple fusion rules of section \ref{global}, and how the algebra of intertwiners and invariant twists can be embedded in a $|G|\times |G|$ matrix algebra. 

We start from the existence of charge creating operators in ball $R_1$ corresponding to the regular representation of the group \cite{Casini:2019kex,Longo:1994xe}. These operators $\{V_g\}_{g\in G}$ satisfy
\be
\sum_ {g\in G} V_g V_{g}^\dagger=\mathbf{1}  \,,\hspace{1cm}
V_g^\dagger V_{g'}=\delta_{g,g'}\,,\hspace{1cm}
\tau_h^\dagger V_g \tau_h= V_{h^{-1}g}\,.
\ee
In particular, we have orthogonal projectors
\be
T(g):=V_g V_{g}^\dagger \,, \hspace{1cm} \sum_{g\in G} T(g)=\mathbf{1} \,,\\
\ee
forming a basis for the regular representation of the group
\bea
T(g)T(g')\!\!\!&=&\!\!\!\delta_{g,g'}\, T(g)\,,  \label{system}\\
\tau_h^\dagger T(g) \tau_h \!\!\!&=&\!\!\! T(h^{-1}g)\,.\label{system1}
\eea
We decompose this regular representation, with basis vectors $T(g)$, into irreducible representations. This is achieved by  
\be
U^{i,j}_{r}:=\sum_{g\in G} R_r^{ij}(g)\, T(g)\,,\label{inver}\\
\ee
where $R_r^{ij}$ are the matrices of the irreducible representation $r$. In this way, for each fixed $j=1\,\cdots\, d_r$, the operators $U^{i,j}_{r}$ transform as a vector of the representation $r$ in the index $i$, i.e.
\be
\tau_h^\dagger U^{i,j}_{r} \tau_h =\sum_{g\in G} R_r^{ij}(g)\, T(h^{-1} g)=\sum_{g\in G} R_r^{ij}(h g)\, T(g)=R_r^{ik}(h)\sum_{g\in G} R_r^{kj}(g) T(g)=R_r^{ik}(h) U^{k,j}_r\,.
\ee
In the decomposition of the regular representation into irreducible representations, the representation of type $r$ appears $d_r$ times (one for each value of the index $j$). Because the operators $T(g)$ are Hermitian, we have for the conjugate representation $\bar{r}$
\be
U^{i,j}_{\bar{r}}= (U^{i,j}_{r})^\dagger\,. 
\ee

Now we consider the following operators based on two disjoint balls $R_1$ and $R_2$,
\be
{\cal I}_r:= \sum_{i,j=1}^{d_r} U^{i,j}_{r}(R_1) (U^{i,j}_{r}(R_2))^\dagger= \sum_{g, h\in G} \chi_r(g h^{-1})\, T_{R_1}(g)\, T_{R_2}(h)\,.\label{b10}
\ee
It is immediate that these operators are invariant under global group transformations (acting on both $R_1$ and $R_2$ simultaneously), and they are formed by linear combinations of operators with charge $r$ in $R_1$ and $\bar{r}$ in $R_2$. Therefore, they are intertwiners of representation $r$. Using equation \eqref{system} and 
\be
\chi_{r_1}(g)\chi_{r_2}(g)=\sum_{r_3} n_{r_1 r_2}^{r_3} \,\chi_{r_3}(g)\,,
\ee
where $ n_{r_1 r_2}^{r_3}$ is the fusion matrix of the group representations, we get intertwiners that close the fusion (Abelian) algebra
\be
{\cal I}_r {\cal I}_{r'}= \sum_{r''} n_{r r'}^{r''}{\cal I}_{r''}\,.
\ee
We also have from \eqref{b10}
\be
{\cal I}_{\bar{r}}=({\cal I}_r)^\dagger\,,
\hspace{.8cm}
{\cal I}_1=1\,.
\ee
It is clear from \eqref{tt}, \eqref{215}, \eqref{system}, and \eqref{system1} that these intertwiners and the invariant twists belong to the finite dimensional algebra of invariant operators of the form
\be
\sum_{a,b,c\in G} f(a,b,c) \, T_{R_1}(a)\, T_{R_2}(b)\, \tau_c\,,
\ee
where $f(a,b,c)=f(g a, g b, g c g^{-1})$ for all $g\in G$ imposes invariance under the global group. This algebra has dimension $|G|^2$, and the invariant twists and intertwiners form Abelian subalgebras. 

\section{Gauge field on a lattice}\label{apg}

We consider pure gauge fields on the lattice based on a compact group $G$. We take a square lattice and think in terms of a finite group for simplicity. The basic variables (at fixed time) are elements $U_{(ab)}\in G$ of the gauge group $G$ assigned to each oriented link $l=(ab)$ joining neighbor lattice vertices $a$ and $b$. The link $\bar{l}=(ba)$ with the reverse orientation is assigned the inverse group element $U_{\bar{l}}=U_{(ba)}=U_{(ab)}^{-1}=U_l^{-1}$, and hence it does not correspond to a different independent variable. The gauge transformations correspond to variables $g_a \in G$ attached to the vertices $a$ of the lattice. The gauge transformation law is 
\be
U_{(ab)}^\prime=g_a U_{(ab)} g_b^{-1}=: U_{(ab)}^{g}\,.\label{hi}
\ee 

Consider the vector space ${\cal V}$ of all complex wave functionals $|\Psi\rangle:= \Psi[U]$, where $U=\{U_{l}\}$ 
is an assignation of group elements to all links. The inner product in ${\cal V}$ is defined as
\be
\langle \Psi_1|\Psi_2\rangle=\sum_{U_{1}\in G} ... \sum_{U_{N_L}\in G} \Psi_1[U]^* \Psi_2[U]\,,\label{scalar1}
\ee
where $U_l$ is the variable corresponding to the link $l=1,...,N_L$, for a lattice with $N_L$ links. The gauge transformation at a vertex $a$ is implemented by a unitary operator $C_a^g$ in ${\cal V}$
\be
(C_a^g \Psi)[U]=\Psi[U']\,,
\ee
with the gauge transformed variables $U'$ given by \eqref{hi} with $g_x=g\, \delta_{a,x}$ ($x$ any vertex). Gauge transformations based at different points commute with each other. 

The physical Hilbert space is the subspace ${\cal H}\subset {\cal V}$ of gauge invariant functionals
\be
C_a^g\Psi=\Psi\,, \hspace{.7cm} \forall a \textrm{ vertex},g\in G \,.
\ee    
The subspace ${\cal H}$ is also a Hilbert space with the scalar product \eqref{scalar1}. Gauge invariant operators form a subalgebra ${\cal A}\subset {\cal B}({\cal V})$ of the algebra ${\cal B({\cal V})}$ of all operators in ${\cal V}$
 \be
 {\cal A}:=\{X\in {\cal B}({\cal V}), \, (C_a^g)^{-1} X C_{a}^g=X,\, \forall a \textrm{ vertex},g\in G\}\,.
\ee 
The subalgebra ${\cal A}$ maps ${\cal H}$ in itself. This is formalized employing a conditional expectation 
\be
E:{\cal B}({\cal V})\rightarrow {\cal A}\,,\hspace{.8cm} E(X):=\prod_{ a} \frac{1}{|G|}  \sum_{g \in G} (C_a^g)^{-1} X C_a^g\,.
\ee
This conditional expectation acts locally in the lattice. The gauge invariant gauge transformation operators are
\be
\tilde{C}_a^{c}=E(C_a^g)=\frac{1}{n_{c}}\sum_{h\in c}C_a^h\,,   
\ee
where $c\subset G$ is a conjugacy class of $G$ and $n_c$ is its number of elements, form a set of gauge invariant constraint operators in ${\cal A}$ labeled by the conjugacy classes of $G$. These generators commute with all the elements of ${\cal A}$ and generate the center of this subalgebra. All $\tilde{C}_a^{c}$ act as the identity on ${\cal H}$ giving constraints analogous to the Gauss law. A representation of ${\cal A}$ where the global center is not mapped to the identity contains external charges.

\subsection{Local generators of the algebra}

We now want to construct a set of generators of ${\cal A}$ to understand the local properties of the gauge invariant operator algebra. A complete set of local operators for each link in ${\cal B}({\cal V})$ is defined by
\be
(\hat{U}^{(r)\, ij}_l \Psi)[U]:=U^{(r)\,ij}_l \Psi[U]\,,
\ee
where $U^{(r)\,ij}_l:=R_r^{ij}(U_l)$ is the numerical value of the matrix element $i,j$ corresponding to $U_l\in G$ in the  representation $r$. These operators are the analogous to the position operators in the description of a quantum system in terms of wave functions. The analogous to the momentum operator are labeled by elements $g$ of $G$,
\be
(L_{l}^g \Psi)[U_1,...,U_{L_N}]=\Psi[U_1,...,g U_{l},...,U_{L_N}]\,.
\ee
We have $(L_{l}^g)^\dagger =L^{g^{-1}}_{l}$ and $L^{g_1}_l L^{g_2}_l=L^{g_1 g_2}_l$. Furthermore, these operators form a unitary representation of the group.\footnote{An operator $R_l^g$ analogous to $L_l^g$ but acting on the right, $(R_{l}^g \Psi)[U_1,...,U_{L_N}]=\Psi[U_1,...,U_{l} g,...,U_{L_N}]$, can be simply written 
$R_l^g=L_{\bar{l}}^{g^{-1}}$. $R_l^{g_1}$ and $L_l^{g_2}$ commute for all $g_1,g_2 \in G$.} Gauge transformations are implemented by $C_a^g=\prod_b L^g_{(ab)}$. 
 
Operators $\hat{U}^{(r)\,ij}_l$ for different $r$ (and $l$) clearly commute. However, they are not gauge invariant. A gauge invariant operator can be constructed with products of link operators in a closed oriented line. These are the well-known Wilson loop operators
 \bea
 W^r_\Gamma \!\!\!&:=&\!\!\! \hat{U}^{r}_{(a_1 a_2)}\hat{U}^{r}_{(a_2 a_3)}...\hat{U}^{r}_{(a_k a_1)}\,, \\
 (W^r_\Gamma \Psi)[U] \!\!\!&=&\!\!\! \chi_r (U_{(a_1 a_2)}U_{(a_2 a_3)}...U_{(a_k a_1)}) \Psi[U] = \chi_r(U_\Gamma)\,,
 \eea
where $\Gamma={a_1 a_2 ... a_k a_1}$ is an oriented closed path made by links in the lattice, the matrix indices are contracted, and $\chi_r$ is the character of the representation $r$. These magnetic operators are then labelled by closed lines and group representations. We have $(W^r_\Gamma)^\dagger= W^{\bar{r}}_\Gamma=W^r_{\bar{\Gamma}}$, where $\bar{\Gamma}$ is $\Gamma$ with the reversed orientation.  
$W_\Gamma^r$ is unitary only if the representation is one dimensional. Wilson loops for elementary plaquettes will be called plaquette operators. 

Wilson loops (at $t=0$), together with the constraint operators $\tilde{C}_a^{c}$, form a maximal commuting subalgebra of gauge-invariant operators. Then, the physical Hilbert space ${\cal H}$, where all constraints are trivial, is the linear span of polynomials on Wilson loops \cite{Giles:1981ej}. These polynomials can be thought of as polynomials of the Wilson loop operators acting on the trivial state $\Psi[U]=1$. Different bases are also useful. For example, it is possible to decompose the products $\hat{U}_l^{r_1}\cdots \hat{U}_l^{r_k}$ for the same link $l$ into a linear combination of the $\hat{U}_l^{r}$ for different $r$ using the Glebsch-Gordan decomposition. This shows that the wave function is at most linear in each of the $\hat{U}_l^{r}$. This gives place to the spin network representation \cite{Baez:1994hx}.

Gauge invariant local electric operators can be defined for each link $l$ and conjugacy class $c$ of the group as
\be
E_l^{c}:= \sum_{h\in c} L_l^h\propto E(L_l^g)\,. \label{elec-cong}
\ee
A general gauge invariant electric operator (for the link $l$) can be constructed by linear combinations of these operators. More specifically, for any element $k=\sum_{g} k_{g} g $, $k_g=k_{h g h^{-1}} \in \mathbb{C}$ (for all $h\in G$), of the center of the group algebra, the electric operator
\be
E_l^{k}:= \sum_{g\in G}  k_g \,L_l^g\,. \label{gauge-inv-ele}
\ee
is gauge invariant. It is immediate that $E_l^k$ (and in particular $E_l^c$) commutes with all Wilson loops not passing through the link $l$ and with all other electric variables based on any link. However, it does not commute with Wilson loops passing through the link $l$. Using the minimal (orthogonal) projectors of center of the group algebra
\be
P_r=\frac{d_r}{|G|} \sum_{g \in G} \chi_r^*(g) g\,,\hspace{.7cm} P_r\, P_{r'}=\delta_{rr'}P_r\,,\hspace{.7cm}\sum_r P_r=1\,, 
\ee
which are labeled by irreducible representations, we can construct electric projectors for any link $l$ of the lattice
\be
E_l^{r}:=\frac{d_r}{|G|} \sum_{g\in G} \chi_r^*(g)  \,L_l^g\,,\hspace{.4cm}(E_l^r)^\dagger=E_l^r\,,\hspace{.4cm}E_{\bar{l}}^r=E_{l}^{\bar{r}}\,,\hspace{.4cm} E_l^r E_l^{r'}=\delta_{r r'} E_l^r\,,\hspace{.4cm} \sum_r E^{r}_l=\mathbf{1}\,.
\ee
It is immediate that these operators are also gauge invariant since they are of the form \eqref{gauge-inv-ele}, where $k_g= d_r \chi_r^*(g)/|G|$.

Let us decompose the wave function in different terms according to the representation assigned to the link $l$
\be
\Psi[U]=\sum_{r,ij} U_l^{(r) ij} \, f^r_{ij}[U]= \sum_{r,ij} R_r^{ij}(U_l) \, f^r_{ij}[U]\,, \label{ds}  
\ee
where the wave functions $f^r_{ij}[U]$ do not depend on the variable $U_l$ of the site $l$. Each term of this decomposition is gauge invariant since
\be
E_l^r \Psi[U]= U_l^{(r) ij} \, f^r_{ij}[U]\,. \label{ds1} 
\ee

Let us choose the representation of the algebra in the Hilbert space of gauge-invariant vectors ${\cal H}$. All gauge constraints are then set to the identity, i.e. $\tilde{C}_a^{c}\equiv 1$, and the global algebra representation has a trivial center. With the elementary operators described above, we can form local algebras attached to subsets of links of the lattice by taking the algebras generated by the magnetic plaquette and electric link operators which can be formed in the subset. These algebras will be additive by definition. Subalgebras assigned to disjoint lattice subsets are mutually commuting. We now have to understand the interplay between duality and additivity properties for algebras of different regions. 

\subsection{Additivity and duality}

To a given subset $R \subset \{1,\cdots,N_L\}$ of the lattice we have assigned an additive algebra ${\cal A}_{\textrm{add}}(R)$ generated  by all plaquettes and electric link operators in $R$. Our first task is to understand whether a Wilson loop with path $\Gamma\subset R$ belongs to ${\cal A}_{\textrm{add}}(R)$, that is, whether it can be generated additively inside the region. We will show first that it can be generated additively if the path $\Gamma$ is contractible inside $R$. This is elementary for Wilson loops of a one dimensional representation (for example, any irreducible representation of Abelian groups). In that case, for two loops $\Gamma_1 l$ and $\bar{l}\Gamma_2 $, sharing the link $l$ with opposite orientation, we have
\be
\chi_r(U_{\Gamma_1} U_l) \chi_r(U_{l}^{-1}U_{\Gamma_2} )=\chi_r(U_{\Gamma_1 \Gamma_2}).   
\ee
Thus, bigger loops can be produced by multiplying smaller ones. Provided that for any contractible loop $\Gamma$ inside the region $R$ there is a surface inside $R$ with boundary $\Gamma$, contractible Wilson loops will be a product of plaquette operators in the region.      

To do the same job for non-Abelian representations, we need to use some electric generators (inside $R$) to sew the plaquette magnetic operators. We have for two loops sharing the link $l$
\bea
&&\sum_{r'} E_l^{r'}\, W^r_{(\Gamma_1 l)}  W^r_{(\bar{l} \Gamma_2)} E_l^{r'} \Psi[U]= \sum_{r'} E_l^{r'}\, \chi_r(U_{\Gamma_1} U_l)  \chi_r(U_{l}^{-1} U_{\Gamma_2})\, E_l^{r'} \,\sum_{r''} U_l^{(r'') ij} \, f^{r''}_{ij}[U]\label{sew}\\
&=&\sum_{r'} E_l^{r'}\, \chi_r(U_{\Gamma_1} U_l)  \chi_r(U_{l}^{-1} U_{\Gamma_2})\, U_l^{(r') ij} \, f^{r'}_{ij}[U]=\sum_{r'} E_l^{r'}\, U^{(r) mn}_{\Gamma_1} U_l^{(r)nm}  U_{l}^{(r)ts\,*} U_{\Gamma_2}^{(r)ts}\, U_l^{(r') ij} \, f^{r'}_{ij}[U]\nonumber\\
&=&d_r^{-1}\,\sum_{r'} E_l^{r'}\, U^{(r) mn}_{\Gamma_1}  U_{\Gamma_2}^{(r)nm}\, U_l^{(r') ij} \, f^{r'}_{ij}[U]
=\frac{1}{d_r}\, W^r_{(\Gamma_1\Gamma_2)} \Psi[U]\,.\nonumber
\eea
In passing from the second to the third line we have used the fact that in the Klebsch-Gordan decomposition of $U_l^{(r)nm}  U_{l}^{(r)ts\,*}$ only the component proportional to the identity can keep $U_l^{(r') ij}$ into the representation $r'$, as is required by the projector $E_l^{r'}$. We can then replace $U_l^{(r)nm}  U_{l}^{(r)ts\,*}$ by the term proportional to the identity in the Clebsch-Gordan decomposition, namely $(r n,\bar{r} t|1) (1|r m, \bar{r} s)=d_r^{-1}\, \delta_{nt}\delta_{ms}$. Therefore, we get
\be
W^r_{(\Gamma_1\Gamma_2)}=d_r\,\sum_{r'} E_l^{r'}\, W^r_{(\Gamma_1 l)}  W^r_{(\bar{l} \Gamma_2)} E_l^{r'} \,.
\ee
Then, Wilson loops along $\Gamma$ can be generated by plaquette and electric operators lying on a surface bounded by $\Gamma$. 

Now, the question remains whether a Wilson loop based on $\Gamma$ can be generated additively inside a region $R$ containing $\Gamma$ but not containing any surface $\Sigma$ with $\partial \Sigma=\Gamma$. Let us take a simple region $R$ with the topology of a ring $S^1\times \mathbb{R}^{d-2}$, and consider a loop $\Gamma$ that winds around the $S^1$ direction once. We will also ask $R$ to be wide enough to contain a one dimensional closed strip of plaquettes bounded by $\Gamma$ on one side and another loop $\tilde{\Gamma}$ on the other side. $\tilde{\Gamma}$ is just displaced a link with respect to $\Gamma$. We take plaquette operators of the representation $r$ with boundaries in $\Gamma$ and $\tilde{\Gamma}$, and sew them in order to produce locally, after the final plaquette is added, the operator $W_\Gamma^r W_{\tilde{\Gamma}}^{\bar{r}}$. Sewing plaquettes to  $W_{\tilde{\Gamma}}^{\bar{r}}$ we can displace it laterally to finally obtain, by local operations in $R$, $W_\Gamma^r W_\Gamma^{\bar{r}}$ based on the same loop $\Gamma$.

For two generic Wilson loops based on the same path, we have the fusion rule
\bea
(W^{r_1}_\Gamma W^{r_2}_\Gamma \Psi)[U]\!\!\!&=&\!\!\!\chi_{r_1}(U_\Gamma)\chi_{r_2} (U_\Gamma) \Psi[U]\label{eeea}\\
\!\!\!&=&\!\!\! \sum_{r_3}  n_{r_1 r_2}^{r_3} \,\chi_{r_3}(U_\Gamma)\,  \Psi[U]= \sum_{r_3}  n_{r_1 r_2}^{r_3} \, (W^{r_3}_\Gamma \Psi)[U]\,,\nonumber 
\eea
where  $n^{r_3}_{r_1 r_2}=n^{r_3}_{r_2 r_1}$ are the fusion matrices of the group representations. Therefore, we obtain that the following operator is locally generated
\be
W_\Gamma^r W_\Gamma^{\bar{r}}=\sum_{r'} n_{r \bar{r}}^{r'} W_\Gamma^{r'}\,.\label{topo}
\ee
We want to select only one Wilson loop in this linear combination. Notice that, in the operation of sewing two loops along a link in equation \eqref{sew}, we have started with two loops in different representations, the result would have vanished. Then, we can sew the operator \eqref{topo} with a plaquette operator in representation $r'$ sharing a link with $\Gamma$ to obtain a Wilson loop in representation $r'$ along a curve deformed from $\Gamma$ in only one plaquette. We can move this back to the loop $\Gamma$ sewing a plaquette again. Then, we finally conclude that we can locally generate in the ring $R$ any Wilson loop $W_\Gamma^{r'}$ such that $n_{r \bar{r}}^{r'}\neq 0$ for some representation $r$.

Loops winding $n$ times with $n>1$ along the $S^1$ direction of the region $R$ can be deformed using local operators to wind $n$ times along the same line $\Gamma$.  This corresponds to $\chi_r(U_\Gamma^n)$. As a function of $U_\Gamma$, this is a class invariant function. It can be decomposed into a linear combination of characters with coefficients that depend on the group. Then, it can be decomposed into elementary loops of different representations winding just once along $\Gamma$. If we have a product of two loops along curves winding once, it can be locally transformed into a product of loops for the same path $\Gamma$, which can be also decomposed into elementary loops of different irreducible representations. Then, when seeking to understand the non locally generated operators, we only need to worry about the case of simple loops winding once.

The locally generated loops can fuse, and the result is also locally generated. Locally generated loops form a subalgebra of the fusion algebra. For an Abelian group we have $n_{r \bar{r}}^{r'}=0$ for any $r' \neq 1$, and non contractible loop cannot be formed additively in this way. To see the structure of the loops that are not locally generated, and to prove they are such, we have to discuss t' Hooft loops, which are certain combinations of electric operators. 

For a $z$ in the center $Z$ of the group $G$, the class $[z]$ consist of a single element $z$, and according to \eqref{elec-cong}, $E^z_l=E^{[z]}_l=L^z_l$ is gauge invariant. For any $(d-2)$-dimensional surface $\Sigma$ in the dual lattice and each element  $z\in Z$, we define the t' Hooft operator       
\be
T_{\Sigma}^{z}:= \prod_{l\perp \Sigma} E_l^{z}\,.\label{thh}
\ee
This is analogous to the electric flux through $\Sigma$. Recalling that $C_a^{z}\equiv 1$ in the ${\cal H}$ representation, we can multiply these operators in a volume $A$ of the lattice to get
\be
1=\prod_{a\in A} C_a^z=\prod_{l\perp \partial A} E^z_l\,.  
\ee
We have used here the fact that as $z$ commutes with all elements of the group, the action of the gauge transformations on the interior links of $A$ cancel. Therefore, the electric ``flux'' corresponding to $z$ vanishes on any closed surface. In consequence, the flux is the same operator for two surfaces with the same boundary. This means that the t' Hooft operator corresponding to $z$ is independent of the precise surface $\Sigma$ in the definition, and it depends only on the boundary $\partial \Sigma$. For later convenience, we call this $(d-3)$ dimensional closed surface $\Gamma'$. The outcome is that we have t' Hooft loop operators $T_{\Gamma'}^z$, for any $z\in Z$ and any $(d-3)$ dimensional closed surface $\Gamma'$. We also have $(T_{\Gamma'}^z)^\dagger=T_{\Gamma'}^{z^{-1}}=(T_{\Gamma'}^z)^{-1}$, i.e., these operators are unitary.

We could have used the electric fields $E_l^{c}$ for any conjugacy class $c$ in \eqref{thh}, but if $g\notin Z$ this operator does not commute with local operators along the surface, and therefore, it is not an operator that can be thought of as to be localized along the boundary of $\Sigma$.

For a one dimensional loop $\Gamma$ interlocked with $\Gamma'$ (winding number one), it is not difficult to see that 
\be
T_{\Gamma'}^z W_\Gamma^r= \phi_r(z)\, W_\Gamma^r T_{\Gamma'}^z\,,\hspace{.7cm}\phi_r(z):=\frac{\chi_r(z)}{d_r} \,.\label{tito}
\ee
This uses the irreducibility of the representation $r$, which implies, through Schur's lemma, that $z$ is represented inside the loop as a matrix proportional to the identity for $z\in Z$. Then $\chi^r(U z)=  \phi_r(z) \chi^r(U)$. The value  $\phi_r(z)$ is a phase which corresponds to one of the (one dimensional) representations of $Z$, and we have $\phi_r(z)= \phi_r(z^{-1})^*= \phi_{\bar{r}}(z)^*$. A similar calculation shows that a t' Hooft loop commutes with all Wilson loops with a zero winding number with $\Gamma'$. This is because they cross the same number of times the surface $\Sigma$ in opposite directions giving factors $\phi_r(z)$ and $\phi_r^*(z)$ an equal number of times. Another way to see this is that, for zero winding between $\Gamma$ and $\Gamma'$, we can deform $\Sigma$ to lie outside the support of $\Gamma$.

Choosing a $\Gamma'$ in the complement of the region $R$ and interlocked with $\Gamma$, we conclude that Wilson loops along $\Gamma$ for a representation where $\phi_r(z)\neq 1$ for some $z\in Z$ cannot be locally generated in the ring $R$. Any locally generated operator must commute with the t' Hooft loop, and this operator does not commute with $W_\Gamma^r$ if $\phi_r(z)\neq 1$.

Then, we have two sets of representations, both of them closed under fusion. One is formed by the representations generated by the fusion of $r \bar{r}$ for all $r$, which we have shown that give locally generated WL. The other set is the one formed by the representations that are trivial (proportional to the identity) in the center of the group. The complement of this set (i.e., the representations that are non-trivial on the center) has been shown to provide WL that cannot be locally generated.

Any representation $r$ restricted to the center of the group can be put into a diagonal form, where the diagonal elements are proportional to phases. The conjugate representation $\bar{r}$ is given by the conjugate matrices, and $r\bar{r}$ is proportional to the identity. Therefore the first set of representations is included in the second. We knew this already from the above reasoning about the locally generated representations. We will now show how these two sets coincide.               

We need to introduce the adjoint representation $D_\textrm{Ad}(g)$ of a group. This is a $|G|$-dimensional representation in the group algebra (viewed as $|G|$-dimensional vector space) given by the adjoint action, defined as 
\be
D_\textrm{Ad}(g)\, \left(\sum_{h\in G} b_h h \right)=\sum_{h\in G} b_h\,  g h g^{-1}\,.
\ee
The group algebra is isomorphic to a space of block-diagonal matrices $\bigoplus_r M_{d_r\times d_r}$, where an element of the group is represented on the block  $r$ by the corresponding irreducible representation. The adjoint action reduces to the adjoint action of $D_r(g)$ on each block, which is equivalent to the tensor product representation $r \bar{r}$. Therefore, the character of the adjoint representation is $\chi_\textrm{Ad}(g)=\sum_r \chi_r(g)\chi_{\bar{r}}(g)$, and it contains all irreducible representations that can be formed by fusion of $r\bar{r}$. On the other hand, if we view the adjoint action on a basis $|h\rangle$ given by the elements of the group, the adjoint action produces a permutation of the basis elements and we have $D_\textrm{Ad}(g)_{h h'}=\delta_{h, g h' g^{-1}}$. Then, the character is also $\chi_\textrm{Ad}(g)=\sum_h \delta_{h, g h g^{-1}}$, i.e. $\chi_\textrm{Ad}(g)$ is the number of elements of the group that commute with $g$. The elements of the group that commute with $g$ form a subgroup, which is called the centralizer of $g$ and is denoted by $C(g)$. Then, $\chi_a(g)=|C(g)|$. It is immediate that $\chi_\textrm{Ad}(z)=|G|$ for $z\in Z$ and $\chi_\textrm{Ad}(g)<|G|$ for $g\notin Z$. In consequence, $\lim_{n\rightarrow \infty} \frac{\chi_a(g)^n}{|G|^n} = \theta_Z(g)$, the characteristic function of the center, which equals $1$ for elements on $Z$ and  $0$ otherwise. Fusing the adjoint representation with itself $n$ times we get the character $\chi_a(g)^n$, and it follows that, for $n$ large enough, this will contain any irreducible representation which is constant on the center of the group. This proves the statement that the fusion algebra generated by $r \bar{r}$ for all $r$ coincides with one of the characters which are constant on the center.\footnote{One may wonder if the irreducible representations contained in the adjoint without further fusing it with itself (that is, representations in $r\bar{r}$ for all $r$) is enough to obtain all representations constant in the center. This is known as Roth's conjecture \cite{roth1971conjugating}. It does hold in most groups but some counterexamples show that further fusing is sometimes necessary \cite{pena2017lie,heide2013conjugacy}.} We denote by $\Xi_1$ the set of Wilson loop operators corresponding to this set of characters.   

This proves that all Wilson loops which commute with the t' Hooft loops are locally generated in the ring. Those that do not commute with at least one t' Hooft loop are not locally generated. Further, any character $\varphi_s$ of $Z$ (irreducible representation of $Z$) induces a representation of $G$ whose character evaluated on $Z$ is proportional to $\varphi_s$. Therefore, for every t' Hooft loop there is a Wilson loop that does not commute with it. In other words, t' Hooft loops are not locally generated in the complement $R'$ of $R$.  

The character $\chi_r$ of each irreducible representation of $G$ when evaluated on $Z$ is proportional to the character $\varphi_s$ of one irreducible representation of $Z$. Then, Wilson Loops $W^r_\Gamma$ are divided into equivalence classes $\Xi_s$ according to the characters $\varphi_s$, $s=1,\cdots,|Z|$ of $Z$. The class of the identity is the one formed by locally generated loops $\Xi_1$. Further, $\Xi_1 \Xi_s=\Xi_s$, and the different classes select Wilson loops that are locally transformable to each other. The dual sets of non locally generated operators in $R$ and $R'$ are then labeled respectively by $z\in Z$, and $s\in Z^*$,  $Z^*$ is the group of irreducible characters (representations) of $Z$.

\begin{figure}[t]
\begin{center}  
\includegraphics[width=0.5\textwidth]{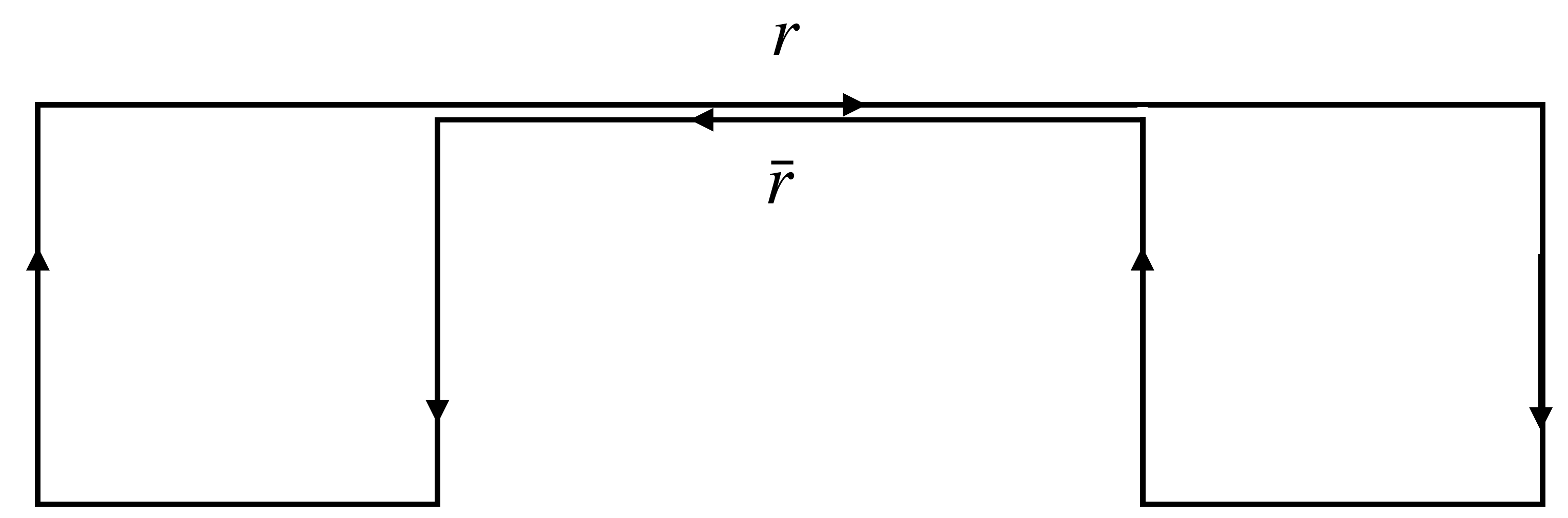}
\captionsetup{width=0.9\textwidth}
\caption{A Wilson loop of representation $r$ in the lattice formed by two plaquettes joined by a line transited in the two opposite directions. This acts as a Wilson line for the representations arising from decomposing $r\otimes \bar{r}$.}
\label{wl}
\end{center}  
\end{figure}   

As we mentioned before, in a gauge theory in the continuum, the adjoint Wilson loop can be broken into pieces by the Wilson lines \eqref{indices} formed with the curvature as charged fields. In the lattice, this type of Wilson line is represented by two plaquettes joined by a segment that is passed in the two opposite directions (see figure \ref{wl}). The plane of the plaquettes represents the indices of the curvature in \eqref{indices}. The adjoint representation on the Lie algebra is the adjoint representation of the group discussed above when we look at elements near the identity.  
 
In fact, in analogy to the case of finite groups, for a Lie group the adjoint representation generates all the representations that arise from fusing $r \bar{r}$. As an example, for the group $SU(2)$, the adjoint representation of spin $1$ generates all integer spin representations, and the same is true for the fusion $j\times j$ for any $j$. Only half-integer representations are not locally generated. The center of the group is isomorphic to $\mathbb{Z}_2$, and it is formed by the identity and the $2\pi$ rotation. The corresponding t' Hooft loops commute with the Wilson loops of integer spin representations. For $SO(3)$,  Wilson loops of all representations are locally generated, and the center is trivial.
  
\subsection{Algebra and maximal nets}

Summarizing, as dual non-local operators in a (pure) gauge theory we have Wilson loops $W_s$ labeled by the characters (irreducible representations) $\chi_s \in Z^*$ of the center $Z$ of the gauge group, and t' Hooft loops labeled by elements of $Z$. For $d=4$, the set of non-local operators in a ring is doubled since both types of loops live in regions with topology $S^1 \times \mathbb{R}^2$. Hence, in this case, the group of non-locally generated operators group is $Z\times Z^*$.  

Explicit construction of Wilson loop operators $W_s$ follows by choosing one representative of each class $\Xi_s$ and applying the same construction as in section \ref{allgg}. The commutation relations between Wilson and t' Hooft loops follow from \eqref{tito}
\be
 W_s T_z= \chi_s(z) T_z W_s\,.  
\ee
A maximal local net satisfying duality requires choosing a set of non-local operators for $R$ and $R'$, which cannot be enlarged without violating additivity. A relevant case is $d=4$, where both types of operators live in the same topology. In this case, we can form dyons labeled by a pair $(z,s)$, where $z\in Z$ and $s\in Z^*$. The set of chosen dyons should be closed under fusion and conjugation
\bea 
(z_1,s_1)(z_2,s_2)\!\!\!&=&\!\!\!(z_1 z_2,s_1 s_2)\,,\label{fus} \\
(z,s)*\!\!\!&=&\!\!\!(z^{-1},s^{-1})\,\label{conju} 
\eea
where $s^{-1} \equiv \bar{s}$. Locality implies the generalized Dirac-Zwanziger quantization condition
\be
\chi_{s_2}(z_1)=\chi_{s_1}(z_2)\;, \label{dsz}  
\ee
for any pair of dyons. This is automatically consistent with (\ref{fus}-\ref{conju}). As in the Maxwell case, there could be several solutions for a maximal set satisfying these conditions, including taking all the electric charges $\{ (1,s):  s\in Z^*\}$ or all the magnetic monopoles $\{ (z,1):  z\in Z\}$. Several examples are worked out for the centers of Lie groups in \cite{Aharony:2013hda}. We do not know the classification of solutions to this problem for general finite Abelian groups. 

\section{Perimeter law for additive operators}
\label{perimeter}
We want to show that the vacuum expectation values of operators in the additive algebra of a ring-like region, with topology $S^1 \times \mathbb{R}^{d-2}$, decrease at most with a perimeter law 
\be
\langle {W}\rangle\ge k\, e^{-\mu R }\;, \label{19}
\ee
for large radius $R$, where $k$ and $\mu$ are constants that depend on the operator. This would prohibit an area law, the expected behavior of confinement order parameters, for additively generated operators. However, for this statement to make sense, we have to further qualify the operator ${W}$ living in a ring on how it is supposed to depend on the size of the ring. First, we have to construct a sequence of operators $W_R$ for different ring radius $R$ in such a way that they represent ``the same type of operator'' but for different ring regions. Otherwise, we can obtain any behavior $\langle W_R\rangle$ as a function of $R$ by multiplying the operator by an arbitrary function of the radius. We also need to increase the size of the ring keeping the cross-section constant. It would also be convenient to use only operators with positive expectation values.

We first discuss a simple case of correlators of operators localized in two balls. This corresponds to the order parameters of global symmetries, instead of gauge symmetries.

\subsection{Two point correlators}  \label{twopoint}

We look at the behavior of the vacuum correlator of a pair of operators localized in two balls when the distance between the two balls $R$ goes to infinity. We conveniently take operators $O_i(x):=e^{i P x} O_i e^{-i P x}$, with $O_i$  ($i=1,2$) fixed operators localized in a ball. We define
\be
f(R):= \langle O_1(-\vec{R}) O_2(\vec{R})\rangle \,,
\ee
where $\vec{R}=R \hat{n}$ with $\hat{n}$ a unit vector.\footnote{For large $R$, the leading behavior of $f(R)$ will not depend on $\hat{n}$.} The clustering property in QFT implies 
\be
 \lim_{R\rightarrow \infty} f(\vec{R}) = \langle O_1\rangle \langle  O_2\rangle\,.\label{18}
\ee
Looking at the behavior of $f(R)$ with $R$, we have already selected a sequence of operators for different distances by translating fixed operators in the same ball. The norm of the operators are also fixed. Several improvements of \eqref{18} are well known. For example, the unitarity bound $|f(R)-f(\infty)|\le c \,R^{-(d-2)}$ for conformal models, and  $|f(R)-f(\infty)|\le c\, e^{-m_0 R}$ for gapped ones, where $m_0$ is the mass gap. If $f(\infty)=0$, these give upper bounds on the asymptotic behavior of $|f(R)|$. 

However, on the contrary, in an inequality such as \eqref{19}, we intend a lower bound. Indeed, there is also a lower bound associated with the clustering of operators. To see this, we have to select operators with positive expectation values such that $f(R)$ could not be zero or change sign. This is easily done by taking $O_1$ and $O_2$ to be CRT conjugate operators
\be
O_1(-\vec{R})=J O_2(\vec{R}) J =: \overline{O_2(\vec{R})}\,,
\ee
with $J$ the CRT operator, which coincides with the Tomita-Takesaki reflection for the Rindler wedge and the vacuum state. In the above equation, we used a CRT reflection with respect to some wedge region $\mathcal{W}$ such that $O_2(\vec{R})$ is inside $\mathcal{W}$ for all $R>0$.\footnote{It is not difficult to see that we can always take a wedge region satisfying this property.} By wedge reflection positivity, or CRT positivity (see for example \cite{Casini:2010bf})
\be
 f(R)= \langle O(\vec{R}) \overline{O(\vec{R})}\rangle > 0\,.
\ee 
This cannot vanish for a non-trivial operator. In a massive theory and provided that $f(\infty)=0$, this will decay exponentially fast. We will now show that it cannot decay faster than exponentially in any theory.

Writing 
\be
f(R)=:e^{-V(R)}\;,
\ee 
wedge reflection positivity tell us that\footnote{In addition, $f(R)$ is a completely monotonic function \cite{Blanco:2019gmt}.} 
\be
V''(R)\le 0\,.\label{sd}
\ee
The slope $V'(R)$ cannot be negative for any $R$. Otherwise, if $V'(R_0)<0$ for some $R_0$, \eqref{sd} would imply a slope always negative for $R>R_0$, leading to increasingly negatives $V(R)$ and a violation of the clustering property. Hence, $V(R)$ is increasing and concave. This means that there is a limit for the slope
\be
V'_\infty=\lim_{R\rightarrow \infty} V'(R)\,\ge 0.   
\ee

If $V'_\infty>0$ is positive, we have an ``area law'' for large radius
\be
f(R)\sim e^{- R\, V'_\infty} \, ,\hspace{.3cm} R \rightarrow \infty.
\ee
If $V'_\infty=0$, we may have different situations. A perimeter law would be  $f(R)\sim \textrm{const.}= |\langle O\rangle|^2$. An intermediate case is the conformal case $V(R)\sim \log(R)$. The potential cannot increase faster than the ``area'', in analogy with the result for loops. 

For an orbifold, the intertwiner, which is a non locally generated operator in the two balls, has a constant law if the symmetry is spontaneously broken, and area law or some milder logarithmic law if the symmetry is not broken. Hence, it is a good order parameter for symmetry breaking.  However, an operator locally generated (not an intertwiner) can also have both area and constant law. This is a difference with what we expect for operators on rings, where locally generated operators (hence non-good for order parameters) have a potential that increases at most as the perimeter. This is, in part, because an operator in the two balls may be locally generated for a particular orbifold, but it might also be an intertwiner for other groups.  A way to regain the expected behavior of locally generated operators is to take $O$ as a CRT positive operator, $O=Q\bar{Q}$. This prevents $\langle O\rangle=0$ and leads to a constant law. In this case, $O(\vec{R}) \overline{O(\vec{R})}$  can never be an intertwiner for any orbifolding since it is the product of non charged operators.
  
\subsection{Constructing additive ring operators}

In analogy with the previous discussion, we limit the  operator order parameters to be CRT positive operators, i.e., operators of the form
\be
 W = \sum_{ij} \lambda_{ij} O_i \overline{O_j}\,,
\ee
with $\lambda_{ij}$ a positive definite matrix and $O_i$ localized in the right wedge. We could also take limits of these types of operators. In particular, this gives   $\langle W \rangle > 0$. We also impose $W$ to be localized in a ring and invariant under the rotation symmetry of the ring (see below). In principle, this should allow us to select a ``cross-section'' for the operator from which to construct the sequence of operators for different ring radius. Both these properties (rotation symmetry and CRT positivity) can be also imposed on non locally generated loops. Nevertheless, we only consider the case of locally generated operators in this appendix.

To begin with, we can divide the angular span of the ring in $N$ pieces,  and consider operators $O^{i_1}_1$ localized in the first of these angular sectors. With the aim to construct an operator invariant under the subgroup of rotations $\mathbb{Z}_N$, we consider operators in other angular sectors $O_k^{i_k}$, $k=0,\cdots,N-1$ which are rotations of angle $2\pi k/N$ of $O^{i_1}_1$ along the ring. A $\mathbb{Z}_N$ invariant operator is of the form
\be
W=\sum_{i_1,\cdots,i_{N-1}} \Lambda_{i_1,\cdots,i_{N-1}}\prod_k O_k^{i_k}\;,
\ee
where $\Lambda_{i_1,\cdots,i_{N-1}}$ is invariant under cyclic permutations. We can write this in "tensor network form" 
\be
\Lambda_{i_1,\cdots,i_{N-1}}=T_{i_1}^{a_1 a_2} T_{i_2}^{a_2 a3}\cdots T_{i_N}^{a_N a_1}\;,
\ee
where the indices $a_i$ run over a sufficiently large set and are contracted. In this way, we get a representation in a ``Wilson loop form''
\be
W=O^{a_1 a_2}_1 O^{a_2 a3}_2\cdots O^{a_N a_1}_N\;,
\ee
as a path ordered product of matrices of operators with $O_i^{a b}=T_{l}^{a b} O_{i}^l$. Each $O^{ab}_i$ is obtained from rotating $O^{ab}_1$. 

Now we impose CRT positivity which requires that $ O_N^{ab}=\overline{O_1^{ba}}$, where the modular conjugation is respect to the plane separating the two. It is convenient to impose $O_1^{ab}=\overline{O_1^{ba}}$, with respect to the plane bisecting the first angular sector. These relations, together with rotation symmetry, imply that the operator $W$ is CRT positive. We also have that all partial operators
\be
W_{k_1,k_2}=O^{a_1 a_2}_{k_1} O^{a_2 a_3}_{k_1+1}\cdots O^{a_n a_1}_{k_2}
\ee
are also CRT positive with respect to the plane dividing them in two.  

The same reflection positivity argument used in section \ref{confina-additivo} shows that these operators have concave potential. We then have 
\be
 \langle {W}_{-k,k'} \rangle  \langle {W}_{-k',k}\rangle  \le  \langle {W}_{-k,k} \rangle  \langle {W}_{-k',k'} \rangle\,,
\ee
with the convention of mod $N$ classes for the position indices. Then, writing  $\langle W_{k_1,k_2} \rangle=e^{-V(k_2-k_1)}$, and for $k,k'>0$, we have 
\be
 2 V(k_1+k_2)\ge V(2 k_1) +V(2 k_2)\,.
\ee
Call the difference  $ V(1)-V(0)=\mu$. If $\mu\le 0$, the function $V(k)$ is upper bounded by $V(0)$. If $\mu\ge 0$ we have
\be
V(k)\le \mu \,k +V(0)\,.
\ee 
Therefore,
\be
\langle W \rangle \ge e^{- \mu N-V(0)}\,, 
\ee
for some $\mu\ge 0$, and the value of $\mu$ depends only on the expectation value of two adjacent operators. 

Then, we can use the same matrix operator as a seed for creating operators in larger circles using rotations in larger circles. When $N$ and $R$ are large, and $R/N=c$ fixed, the two adjacent operators may have to be readjusted for larger circles by a very tiny contrary rotation of the reflected operators. In the limit, the value of $\mu$ converges, and we get the lower bound
\be 
\langle W \rangle(R) \ge e^{- 2 \pi \, (\mu/c) R -V(0)}\,, \hspace{4mm}R \rightarrow \infty\,. 
\ee

\subsection{Wilson type operators}

The CRT positive, rotational invariant, and additive operators form an algebraically closed convex cone. To get a result at this level of generality we should analyze how to take limits of $\mathbb{Z}_N$ symmetric to rotational symmetric operators.  We do not pursue the mathematical details of this construction any further. Rather, we analyze the case of Wilson type operators of the form
\be
W=\textrm{Tr} \, \mathcal{P} e^{i\oint ds \, A(s)}\,,
\ee
which are suggested both by the standard presentation of Wilson loops and by the preceding discussion. Here $\mathcal{P}$ is the path order, $A(s)$ is the rotation of $A(0)$, and $A(0)$ is a matrix of fields smeared in the direction perpendicular to the loop. $A(0)$ is the cross-section from which we can obtain a sequence of operators for rings of different radius by using rotations for different circles. CRT positivity is then reduced to
\be
A_{ab}(0)=-\overline{A_{ba}(0)}\,.  
\ee
Define the partial operators 
\be
W(s_1,s_2,\epsilon)=\textrm{Tr} \, \mathcal{P} e^{i\int_{s_1-\epsilon}^{s_2+\epsilon} ds \, A(s) \alpha(s)}\,,
\ee
where $\alpha(s)$ is a smearing function equal to one inside the interval $(s_1,s_2)$, zero outside $(s_1-\epsilon,s_2+\epsilon)$ and smooth everywhere. The shape of the smearing function between one and zero is the same for any interval $(s_1,s_2)$. The concavity holds for the expectation values of these operators. To get the perimeter law for $W$, we then only have to take care of the step going from an almost closed loop to a closed one. Between the closed loop and loops with one gap and two gaps, there is also an inequality from CRT by reflecting in some plane that does not pass through the gap. Given that the loops with one and two gaps satisfy perimeter laws with the same perimeter coefficient, it follows the perimeter law for the full closed loop.  

If $A(s)$ is a gauge field in some representation where the loop cannot be broken by charged fields, the construction does not work. The partial operators are not gauge-invariant. If we fix the gauge, several problems appear. If we do not, the partial operators have zero expectation value. Notice that nothing goes wrong, except that we cannot use the calculation to put a useful lower bound on the closed loop. 

\section{Conformal transformations of the ring} \label{confor}

Consider a ring ${\cal R}$ formed by rotating around the $z$ axis a circle of radius $R$, centered around the point $z=0, x=r_0$, in the plane $z,x$. Written in cylindrical coordinates, the surface of the ring is given by the equation
\be
(r-r_0)^2+z^2=R^2\,.
\ee
The intersection with the plane $x,y$ is the circular corona with inner radius $L:=r_0-R$ and outer radius $r_0+R$. In Cartesian coordinates, the surface of the ring corresponds to the quartic equation
\be
 (L^2 + 2 L R + x^2 + y^2 + z^2)^2-4 (L + R)^2 (x^2 + y^2)=0\,. \label{torus_cart}
\ee

A natural parameter describing the geometry of this ring in the conformal case is the cross-ratio $\eta$ between the four points $x=(-L-2 R, -L, L, L+2R)$,  of the intersection of the $x$ axis with the surface of the ring
\be
\eta=\frac{R^2}{(R+L)^2}\,.  
\ee
We have $\eta\in (0,1)$, with the limit $0$ corresponding to thin rings and $1$ to thick ones. 

We want to show that the geometry of the complement ${\cal R}'$ of the ring is conformally equivalent to another ring with cross-ratio $\eta'=1-\eta$, that is 
\be
{\cal R}'(1-\eta)\sim {\cal R}(\eta)\,.
\ee  
From the relation
\be
\frac{R'^2}{(R'+L')^2}=\eta' \equiv 1-\eta=\frac{L(L + 2 R)}{(R+L)^2}\,,
\ee
we have $R'/L'\sim 2/(R/L)^2$ for small $R/L$, and $R'/L'\sim \sqrt{2L/R}$ in the limit of large $R/L$. 

To obtain the conformal transformation on $\mathbb{R}^{3}$ that maps the interior of the torus with the exterior, we use the (conformal) stereographic projection that maps the whole euclidean plane (plus the infinity) onto the unit sphere\footnote{This transformation can be thought as the restriction of a transformation from $\mathbb{R}^{4}$ to $\mathbb{R}^{4}$ which is a (conformal) inversion respect to a sphere of radius $2$ centered on $\left(0,0,0,1\right)$.}
\be
\varphi:\mathbb{R}^{3}\cup\{\infty\}\leftrightarrow S^{3}\subset\mathbb{R}^{4}\,,\label{stereo}
\ee
where $\varphi(0,0,0)=\left(0,0,0,-1\right)$ and $\varphi(\infty)=\left(0,0,0,1\right)$. On the sphere, a torus surface can be described by the equation\footnote{Equation \eqref{torus_sphere} automatically implies that the other two coordinates satisfy the equation $x_{3}^{2}+x_{4}^{2}=1-\alpha$.}
\be
x_{1}^{2}+x_{2}^{2}=\alpha\,,\quad0<\alpha<1\,.\label{torus_sphere}
\ee
A natural way to parametrize it is
\bea
x_{1}(s_{1}) \!\!\!& = &\!\!\! \sqrt{\alpha}\cos(s_{1})\,,\\
x_{2}(s_{1}) \!\!\!& = &\!\!\! \sqrt{\alpha}\sin(s_{1})\,,\\
x_{3}(s_{2}) \!\!\!& = &\!\!\! \sqrt{1-\alpha}\cos(s_{2})\,,\\
x_{4}(s_{2}) \!\!\!& = &\!\!\! \sqrt{1-\alpha}\sin(s_{2})\,,
\eea
where $s_{1},s_{2}\in(-\pi,\pi]$. Applying the inverse of the stereographic projection \eqref{stereo} to the above parametrization, we can explicitily check that the transformed surface in $\mathbb{R}^{3}$ satisfies the equation \eqref{torus_cart} for
\be
L_{\alpha}=\frac{2\sqrt{1-\alpha}}{\sqrt{\alpha}+1}\quad\textrm{ and }\quad R_{\alpha}=\frac{2\alpha}{\sqrt{(1-\alpha)\alpha}}\,.\label{LR_can}
\ee
This torus has cross-ratio $\eta_{\alpha}=1-\alpha$. In the same way, we can check that the interior of the torus is the region on the sphere described by the inequality
\be
x_{1}^{2}+x_{2}^{2}\leq\alpha\,,
\ee
whereas the exterior region is then described by the opposite inequality. The advantage of describing the torus on the sphere is that we immediately see that the exterior of a torus is also a torus because it is described by the region
\be
x_{3}^{2}+x_{4}^{2}\leq1-\alpha\,.\label{torus2}
\ee
In the same way, using the inverse of another stereographic projection $\tilde{\varphi}:\mathbb{R}^{3}\cup\{\infty\}\leftrightarrow S^{3}\subset\mathbb{R}^{4}$ on the plane $x_{1}=0$,\footnote{This stereographic projection maps $\tilde{\varphi}(0,0,0)=(-1,0,0,0)$ and $\tilde{\varphi}(\infty)=(1,0,0,0)$.} the region \eqref{torus2} is mapped into a torus in $\mathbb{R}^{3}$ with dimensions $L_{1-\alpha}$ and $R_{1-\alpha}$, and cross ratio $\eta_{1-\alpha}=\alpha$.

Then, the transformation $T=\tilde{\varphi}^{-1}\circ\varphi$ maps the plane $\mathbb{R}^{3}\cup\{\infty\}$ onto itself, it is conformal by definition, and it maps the interior region of a torus with dimensions $(L_{\alpha},R_{\alpha})$ (cross ratio $\eta_{\alpha}=1-\alpha$) onto the exterior region of a torus with dimensions $(L_{1-\alpha},R_{1-\alpha})$ (cross ratio $\eta_{1-\alpha}=\alpha$). We were able to explicitly compute  this (orientation preserving) conformal transformation $\tilde{x}=T(x)$. It reads
\bea
\tilde{x}^{1} \!\!\!& = &\!\!\! \frac{8(x_{1}-2)}{(x_{1}-4)x_{1}+x_{2}^{2}+x_{3}^{2}+4}+2\,,\label{conf1}\\
\tilde{x}^{2} \!\!\!& = &\!\!\! \frac{8x_{3}}{(x_{1}-4)x_{1}+x_{2}^{2}+x_{3}^{2}+4}\,,\\
\tilde{x}^{3} \!\!\!& = &\!\!\! \frac{8x_{2}}{(x_{1}-4)x_{1}+x_{2}^{2}+x_{3}^{2}+4}\,.\label{conf3}
\eea
Due to the conformal invariance of the problem, it is important to remark that the above transformation has to be considered as a transformation that maps the equivalence class of torus interiors with cross-ratio $\eta$ onto the equivalence class of torus exteriors with cross-ratio $1-\eta$. Moreover, given any torus with dimensions $(L,R)$, it is no longer true that the transformation (\ref{conf1}-\ref{conf3}) sends it to a torus. This transformation only works if it is applied to the representative within the conformal class having the dimensions \eqref{LR_can}, where $\eta=1-\alpha$. If we start with any generic torus $(L,R)$ in the conformal class of cross-ratio $\eta$, we have first to apply a dilatation $x\mapsto\lambda x$ with
\be
\lambda=\frac{2}{\sqrt{L(L+2R)}}\,.
\ee
After this transformation, we obtain a torus in the same conformal class satisfying \eqref{LR_can}, and it is to this new torus that we have to apply (\ref{conf1}-\ref{conf3}).

\section{Relative entropy of a subgroup of non local operators}\label{subgroups}

Let $G$ be an Abelian group of non-local operators and $H\subseteq G$ a subgroup. In this appendix, we study an upper bound to the relative entropy $S_{A \vee H}(\omega|\omega\circ E_H)$, which "compares" the state $\omega$ with the one resulting from erasing the information on the expectation values of the operators in $H$. Let $\tilde{G}:=G^*$ and $\tilde{H}:=H^*$ be their dual groups. We have the complementary diagram 
\be
\begin{array}{ccc}
A\vee H & \stackrel{E_H}{\longrightarrow} & A \\
\downarrow '& & \downarrow ' \\
A'\vee \widetilde{G/H} & \stackrel{E_{\tilde{H}}}{\longleftarrow} & A'\vee \tilde{G} 
\end{array}
\ee   
Here $\widetilde{G/H}$ is a subgroup of $\tilde{G}$ formed by all the characters which are the identity over $H$ (and hence the operators in this subgroup commute with the operators in $H$). The conditional expectations are 
\bea
E_H(x)=\frac{1}{|G|} \sum_{\tilde{g}\in \tilde{G}} \tilde{g} \, x\, \tilde{g}^{-1}\,,\\
E_{\tilde{H}}(x)= \frac{1}{|H|} \sum_{h\in H} h \, x \, h^{-1}\,. \label{app_ce}
\eea 
An upper bound follows from the certainty relation
\bea
 S_{A \vee H}(\omega|\omega\circ E_H)\!\!\!&=&\!\!\!\log |H| -S_{ A'\vee \tilde{G}}(\omega|\omega\circ E_{\tilde{H}})\le \log |H| -S_{ \tilde{G}}(\omega|\omega\circ E_{\tilde{H}})\nonumber \\
\!\!\!&=&\!\!\! \log|H|-S_{\tilde{G}}(\omega\circ E_{\tilde{H}})+S_{\tilde{G}}(\omega)\,. \label{app_upb}
\eea
The algebra of the group $\tilde{G}$ can be represented as the algebra of complex-valued functions on $G$ through the  characters $\tilde{g} \mapsto \chi_{\tilde{g}}(\cdot)$. The probabilities $p_g$ are obtained from the $\tilde{g}$ expectation values as
\be
p_g= \frac{1}{|G|}\sum_{\tilde{g}} \chi_{\tilde{g}}(g)^*\, \omega(\tilde{g})\,. 
\ee
The entropy is then
\be
S_{\tilde{G}}(\omega)=-\sum_{g\in G} p_g \log p_g\,.
\ee
The state $\omega\circ E_{\tilde{H}}$ in $\tilde{G}$ is computed using the explicit form of the conditional expectation
\be
\omega\circ E_{\tilde{H}}(\tilde{g})=\omega\bigg( \frac{1}{|H|} \sum_{h\in H} h \tilde{g} h^{-1}\bigg)=\frac{1}{|H|}\, \sum_{h\in H} \omega(\tilde{g})\, \chi_{\tilde{g}}(h)\,. 
\ee
The probabilities $q_g$ of this new state are then given by
\be
q_g=\frac{1}{|G||H|} \sum_{\tilde{g},h} \chi_{\tilde{g}}(g^{-1} h)\, \omega(\tilde{g})\,.
\ee
For each element $g\in G$, there are $|H|$ elements $g H$ with equal probability $q_g$. Then, for each element $x\in G/H$ (which we identify with a representative in $G$) we can define a probability 
\be
\tilde{q}_x:= \frac{1}{|G|} \sum_{\tilde{g}\in \tilde{G},h\in H} \chi_{\tilde{g}}(x^{-1} h)\, \omega(\tilde{g})=\frac{|H|}{|G|} \sum_{\tilde{g}\in \widetilde{G/H}} \chi_{\tilde{g}}(x)^*\, \omega(\tilde{g})\,,\hspace{.6cm} \sum_{x\in G/H} \tilde{q}_x=1\,.
\ee
Calling 
\be
S_{\widetilde{G/H}}(\omega)=-\sum_{x\in G/H} \tilde{q}_x \log(\tilde{q}_x)\,,
\ee   
we get 
\be
S_{\tilde{G}}(\omega\circ E_{\tilde{H}})=\log |H|+S_{\widetilde{G/H}}(\omega)\,.
\ee
This gives for the upper bound \eqref{app_upb}
\be
S_{A \vee H}(\omega|\omega\circ E_H)\le S_{\tilde{G}}(\omega)-S_{\widetilde{G/H}}(\omega)\,.
\ee

\bibliography{EE}{}
\bibliographystyle{utphys}

\end{document}